\documentclass[prd,notitlepage,nofootinbib,superscriptaddress]{revtex4-1}
\usepackage{amsmath}
\usepackage{amsfonts}
\usepackage[utf8]{inputenc}
\usepackage{graphicx}
\usepackage{hyperref}
\usepackage{slashed}
\usepackage{subcaption}
\usepackage{mathrsfs}
\usepackage{array}
\usepackage{amsmath,empheq}
\usepackage{booktabs}
\usepackage{multirow}
\usepackage{float}
\usepackage[capitalise]{cleveref}
\newcolumntype{P}[1]{>{\centering\arraybackslash}p{#1}}
\hypersetup{colorlinks, linkcolor = [rgb]{0,0.0,0.75}, citecolor = [rgb]{0,0.0,0.75}, urlcolor = [rgb]{0,0.0,0.75}}

\DeclareMathOperator{\Tr}{Tr}
\newcommand{\MSbar}{\overline{\mathrm{MS}}}

\DeclareCaptionJustification{justified}{\leftskip=0pt \rightskip=0pt \parfillskip=0pt plus 1fil}
\makeatletter
\g@addto@macro\bfseries{\boldmath}
\makeatother
\begin{document}

\title{Nucleon axial, scalar, and tensor charges using lattice QCD at the physical pion mass}

\author{Nesreen~Hasan}
\email{n.hasan@fz-juelich.de}
  \affiliation{Bergische Universität Wuppertal, 42119 Wuppertal, Germany}
  \affiliation{IAS, Jülich Supercomputing Centre, Forschungszentrum Jülich, 52425 Jülich, Germany}
  
\author{Jeremy~Green}
\email{jeremy.green@desy.de}
  \affiliation{NIC, Deutsches Elektronen-Synchroton, 15738 Zeuthen, Germany}

\author{Stefan~Meinel}
  \affiliation{Department of Physics, University of Arizona, Tucson, AZ 85721, USA}
  \affiliation{RIKEN BNL Research Center, Brookhaven National Laboratory, Upton, NY 11973, USA}

\author{Michael~Engelhardt}
  \affiliation{Department of Physics, New Mexico State University, Las Cruces, NM 88003-8001, USA}

\author{Stefan~Krieg}
  \affiliation{Bergische Universität Wuppertal, 42119 Wuppertal, Germany}
  \affiliation{IAS, Jülich Supercomputing Centre, Forschungszentrum Jülich, 52425 Jülich, Germany}

\author{John~Negele}
  \affiliation{Center for Theoretical Physics, Massachusetts Institute of Technology, Cambridge, MA 02139, USA}

\author{Andrew~Pochinsky}
  \affiliation{Center for Theoretical Physics, Massachusetts Institute of Technology, Cambridge, MA 02139, USA}

\author{Sergey~Syritsyn} 
  \affiliation{RIKEN BNL Research Center, Brookhaven National Laboratory, Upton, NY 11973, USA}
   \affiliation{Department of Physics and Astronomy, Stony Brook University, Stony Brook, NY 11794, USA}

\date{\today}
\begin{abstract}
We report on lattice QCD calculations of the nucleon isovector axial, scalar, and tensor charges. 
Our calculations are performed on two $2+1$-flavor ensembles generated using a 2-HEX-smeared Wilson-clover action at the physical pion mass and lattice spacings $a\approx 0.116$ and $0.093$ fm. 
We use a wide range of source-sink separations --- eight values ranging from roughly 0.4 to 1.4 fm on the coarse 
ensemble and three values from 0.9 to 1.5 fm on the fine ensemble --- which allows us to perform an extensive study of excited-state effects using different analysis and fit strategies.
To determine the renormalization factors, we use the nonperturbative Rome-Southampton approach and compare RI$'$-MOM and RI-SMOM intermediate schemes to estimate the systematic uncertainties.
Our final results are computed in the $\MSbar$ scheme at scale $2$ GeV. The tensor and axial charges have uncertainties of roughly 4\%, $g_T=0.972(41)$ and $g_A=1.265(49)$. The resulting scalar charge, $g_S=0.927(303)$, has a much larger uncertainty due to a stronger dependence on the choice of intermediate renormalization scheme and on the lattice spacing.
\end{abstract}

\maketitle

\section{Introduction}
Nucleon charges quantify the coupling of nucleons to quark-level 
interactions and play an important role in the analysis of the Standard Model and Beyond the Standard Model (BSM) physics.
The isovector charges, $g_X$, are associated with the $\beta$-decay of the neutron into a proton and are defined via the transition matrix elements,
\begin{equation}
\langle p(P,s) | \bar u \Gamma_X d | n(P,s) \rangle = g_X \bar u_p(P,s) \Gamma_X u_n(P,s)
\end{equation}
where the Dirac matrix $\Gamma_X$ is $1, \gamma_\mu\gamma_5$ and $\sigma_{\mu\nu}$ for the scalar (S), the axial (A) and the tensor (T) operators, respectively. They are straightforward to calculate in lattice QCD since they receive only connected contributions arising from the coupling of the operator to the valence quarks, i.e.\ there are no contributions from disconnected diagrams. Lattice calculations of these charges were recently reviewed by FLAG~\cite{Aoki:2019cca}, and we note some calculations of them in the last few years in Refs.~\cite{Green:2012ej, Green:2012ud, Bali:2014nma, Bhattacharya:2016zcn, Alexandrou:2017qyt, Alexandrou:2017hac, Capitani:2017qpc, Yamanaka:2018uud, Chang:2018uxx, Liang:2018pis, Gupta:2018qil, Ishikawa:2018rew, Shintani:2018ozy, Harris:2019bih}.

The nucleon axial charge is experimentally well determined; the latest PDG value is $g_A = 1.2724(23)$~\cite{PhysRevD.98.030001}. In addition to its role in beta decay, the axial charge gives the intrinsic quark spin in the nucleon, and its deviation from unity is a sign of chiral symmetry breaking. Since the axial charge is so well measured, it is considered to be a benchmark quantity for lattice calculations, and it is essential for lattice QCD to reproduce its experimental value.

Unlike the axial charge, the nucleon scalar and tensor charges are difficult to directly
measure in experiments. 
Thus, computations of those observables within lattice QCD will provide useful input for ongoing experimental searches for BSM physics.
The generic BSM contributions to neutron beta decay were studied in Ref.~\cite{Bhattacharya:2011qm}, where it was shown that the leading effects are proportional to these two couplings; thus, calculations of $g_S$ and $g_T$ are required in order to find constraints
on BSM physics from beta-decay experiments. 
The tensor charge is also equal to the isovector first moment of the proton's transversity parton distribution function (PDF), $\langle 1\rangle_{\delta u - \delta d}$. Constraining the experimental data with lattice estimates of the tensor charge reduces the uncertainty of the transversity PDF significantly~\cite{Lin:2017stx}. In experiment, there are multiple observables that could be used to constrain $g_T$~\cite{Courtoy:2015haa,Radici:2018iag}, and the overall precision will be greatly improved by the SoLID experiment at Jefferson Lab~\cite{Ye:2016prn}, providing a test of predictions from lattice QCD. In addition, the tensor charge controls the contribution of the quark electric dipole moments (EDM) to the neutron EDM, which is an important observable in the search for new sources of CP violation.
The scalar charge is related via the conserved vector current relation
to the contribution from the difference in $u$ and $d$ quark masses to the neutron-proton mass splitting in the absence of electromagnetism, $g_S = \delta M_N^{QCD}/\delta m_q$~\cite{Gonzalez-Alonso:2013ura}.

In this paper, we present 
a lattice QCD calculation of the isovector axial, scalar, and tensor 
charges of the nucleon using two ensembles at the physical pion mass 
with different lattice spacings.
This paper is organized as follows. In Sec.~\ref{sec:lat_setup}, we describe the parameters of the gauge ensembles analyzed, the lattice methodology, and the fits to the two-point functions used to extract the energy gaps to the first excited state on each ensemble. We discuss different analysis methods for estimating the three bare charges and eliminating the excited-state contaminations and present a procedure for combining the multiple results in Sec.~\ref{sec:estimation_of_bare_charges}. The procedure we follow for determining the renormalization factors for the different observables using both RI$'$-MOM and RI-SMOM schemes is described in Sec.~\ref{sec:NPR}.
In Sec.~\ref{sec:renormalized_charges}, we give the final estimates of the renormalized charges and discuss the continuum and infinite volume effects. Finally, we give our conclusions in Sec.~\ref{sec:summary}. In Appendix~\ref{app:Mstate}, we show analysis results of the bare charges using the many-state fit, which is an alternative model for excited-state contributions based on the contributions of noninteracting $N\pi$ states with relative momentum $(\vec p)^2<(\vec p_\mathrm{max})^2$. In Appendix~\ref{app:table}, we list the bare charges determined on the two ensembles studied in this work, along with data used in previous publications~\cite{Green:2012ej,Green:2012ud}.

\section{Lattice setup}
\label{sec:lat_setup}

\subsection{Correlation functions}

To determine the nucleon matrix elements in lattice QCD, we compute the nucleon two-point and three-point functions at zero momentum,
 \begin{equation}\label{defc2}
 C_{2}(t) = \sum_{\vec x}  (\Gamma_{\rm pol})_{\alpha\beta} \left\langle \chi_\beta(\vec x,t) \bar{\chi}_\alpha(0)\right\rangle,
 \end{equation}
 \begin{equation} \label{defc3}
 C_{3}^X(\tau, T) = \sum_{\vec x, \vec y} (\Gamma_{\rm pol})_{\alpha\beta} \left\langle \chi_{\beta}(\vec x, T) \mathcal{O}_X (\vec y, \tau) \bar \chi_{\alpha}(0)\right\rangle.
 \end{equation}
Here, we place the source at timeslice $0$, the sink at timeslice $T$, and insert the operator $\mathcal O_X$ at the intermediate timeslice $\tau$. The latter is the isovector current $\mathcal O_X = \bar q \Gamma_{X} \frac{\tau^3}{2} q$, where $q$ is the quark doublet $q=(u,d)^T$, and $\chi=\epsilon^{abc} (\tilde{u}_a^T C \gamma_5 \frac{1+\gamma_4}{2} \tilde{d}_b) \tilde{u}_c$ is a proton interpolating operator constructed using smeared quark fields $\tilde q$. We use Wuppertal smearing~\cite{Gusken:1989qx}, $\tilde q\propto (1+\alpha H)^Nq$, where $H$ is the nearest-neighbor gauge-covariant hopping matrix constructed using the same smeared links used in the fermion action; the parameters are chosen to be $\alpha=3.0$ on both ensembles, $N=60$ on the coarse ensemble, and $N=100$ on the fine ensemble. The spin and parity projection matrices are defined\footnote{In this paper we use Euclidean conventions, $\{\gamma_\mu,\gamma_\nu\}=2\delta_{\mu\nu}$.} as $\Gamma_{\rm pol}=\frac{1}{2}(1+\gamma_4)(1-i\gamma_3\gamma_5)$.

 In order to compute $C_{3}^X$, we use sequential propagators through the sink~\cite{Martinelli:1988rr}. This has the advantage of allowing for any operator to be inserted at any time using a fixed set of quark propagators, but new backward  propagators must be computed for each source-sink separation $T$. The three-point correlators have contributions from both connected and disconnected quark contractions, but we compute only the connected part since  for the isovector flavor combination the disconnected contributions cancel out.
\subsection{Simulation details}
We perform our lattice QCD calculations using a tree-level Symanzik-improved gauge action and $2 + 1$ flavors of tree-level improved Wilson-clover quarks, which couple
to the gauge links via two levels of HEX smearing~\cite{Durr:2010aw}. We use two ensembles at the physical pion mass: one with size $48^4$ and lattice spacing $a\approx 0.116$ fm which we call coarse, and another with $64^4$ and $a\approx 0.093$ fm which we call fine. Both volumes satisfy $m_\pi L\approx 4$. On the coarse ensemble, we perform measurements on $212$ gauge configurations using source-sink separations $T/a \in \{3,4,5,6,7,8,10,12\}$ ranging roughly from $0.4$ to $1.4$ fm. In addition, we make
use of all-mode-averaging (AMA)~\cite{Blum:2012uh,Shintani:2014vja} to reduce the computational cost through inexpensive approximate quark propagators. For $T/a \in \{3,4,5\}$, we use approximate samples from $96$ source positions per gauge configurations and high-precision samples from one source position for bias correction, and for $T/a \in \{6,7,8,10,12\}$ we use double those numbers.
On the fine ensemble, we perform the calculations on $442$ gauge configurations using source-sink separations $T/a \in \{10,13,16\}$ ranging roughly from $0.9$ to $1.5$ fm. AMA is applied with $64$ sources with approximate propagators and one source for bias correction per gauge configuration.
Table~\ref{tab:lat_setup} summarizes the parameters and the number of measurements performed on each of the ensembles.

On each gauge configuration, a random initial source position is
chosen and the others are evenly separated and distributed throughout
the volume. We always bin all of the samples on each configuration to
account for any spatial correlations. In addition, we have tested for
autocorrelations by binning the configurations in groups of 2, 4, 8,
and 16: no significant trends were identified in the estimated
statistical uncertainty and therefore we elected not to bin samples
from different configurations.

\begin{table}[!htbp]
\centering
\begin{tabular}{c | c c c c | c c c c | c c c c}
Ensemble ID  & Size & $\beta$ & $am_{ud}$ & $am_s$
& $a$ [fm] & $am_\pi$ & $m_\pi$ [MeV] & $m_\pi L$
& $N_\mathrm{conf}$ & $T/a$ & $N_\mathrm{meas}^\mathrm{AMA}$ & $N_\mathrm{meas}^\mathrm{HP}$\\
\hline
\hline
coarse & $48^4$ & 3.31 & $-0.09933$ & $-0.04$ & 0.1163(4) & 0.0807(12) & 137(2) & 3.9 & 212&
\begin{tabular}{c}
$\{3,4,5\}$\\
$\{6,7,8,10,12\}$
\end{tabular}
&
\begin{tabular}{c}
 40704\\
 81408
 \end{tabular} & 
\begin{tabular}{c}
 424\\
 848
 \end{tabular}\\
 \hline
fine & $64^4$ & 3.5 & $-0.05294$ & $-0.006$ & 0.0926(6) & 0.0626(3) & 133(1) & 4.0 & 442 & $\{10,13,16\}$ & 56576 &884
\end{tabular}
\caption{Parameters of the ensembles and measurements used in this work. The lattice spacing is taken from Ref.~\cite{Durr:2010aw}, where it is set using the mass of the $\Omega$ baryon at the physical point. $N_\mathrm{conf}$ refers to the number of gauge configurations analyzed and $N_\mathrm{meas}^\mathrm{AMA} = 2\times N_\mathrm{conf} \times N_\mathrm{src}^\mathrm{AMA}$ is the number of measurements performed using the AMA method with $N_\mathrm{src}^\mathrm{AMA}$ being the number of source positions used on each gauge configuration. The factor of 2 in $N_\mathrm{meas}$ accounts for the use of forward- and
backward-propagating states. Finally, $N_\mathrm{meas}^\mathrm{HP}$ refers to the number of measurements made with high-precision.}
\label{tab:lat_setup}
\end{table}
\subsection{Fitting two-point functions}\label{sec:2stateC2}

Inserting a complete set of states $I=\sum_{n} | n \rangle \langle n|$
into Eq.~\eqref{defc2} yields the spectral decomposition
\begin{equation}\label{eq:C2spectral}
C_2(t) = \sum_n e^{-E_n t} (\Gamma_{\rm pol})_{\alpha\beta} \langle \Omega | {\chi}_\beta |n \rangle \langle n|{\bar{ \chi}}_\alpha| \Omega \rangle,
\end{equation}
where we use the shorthand $\chi_\beta = \chi_\beta(0)$.  Truncating
this to the ground state and a single excited state, on each ensemble
we perform two-state fits to the two-point correlation functions at
zero momentum:
\begin{equation}\label{eq:2stateC2}
C_2(t) = a_0 e^{-E_0 t} + a_1 e^{-E_1 t},
\end{equation}
where $a_i$ and $E_i$ denote the amplitudes and the energies of the two states. For comparison, we also perform one-state fits with $C_2(t) = a_0 e^{-E_0 t} $ only.

\begin{figure}
\begin{center}
\includegraphics[width=0.45\textwidth]{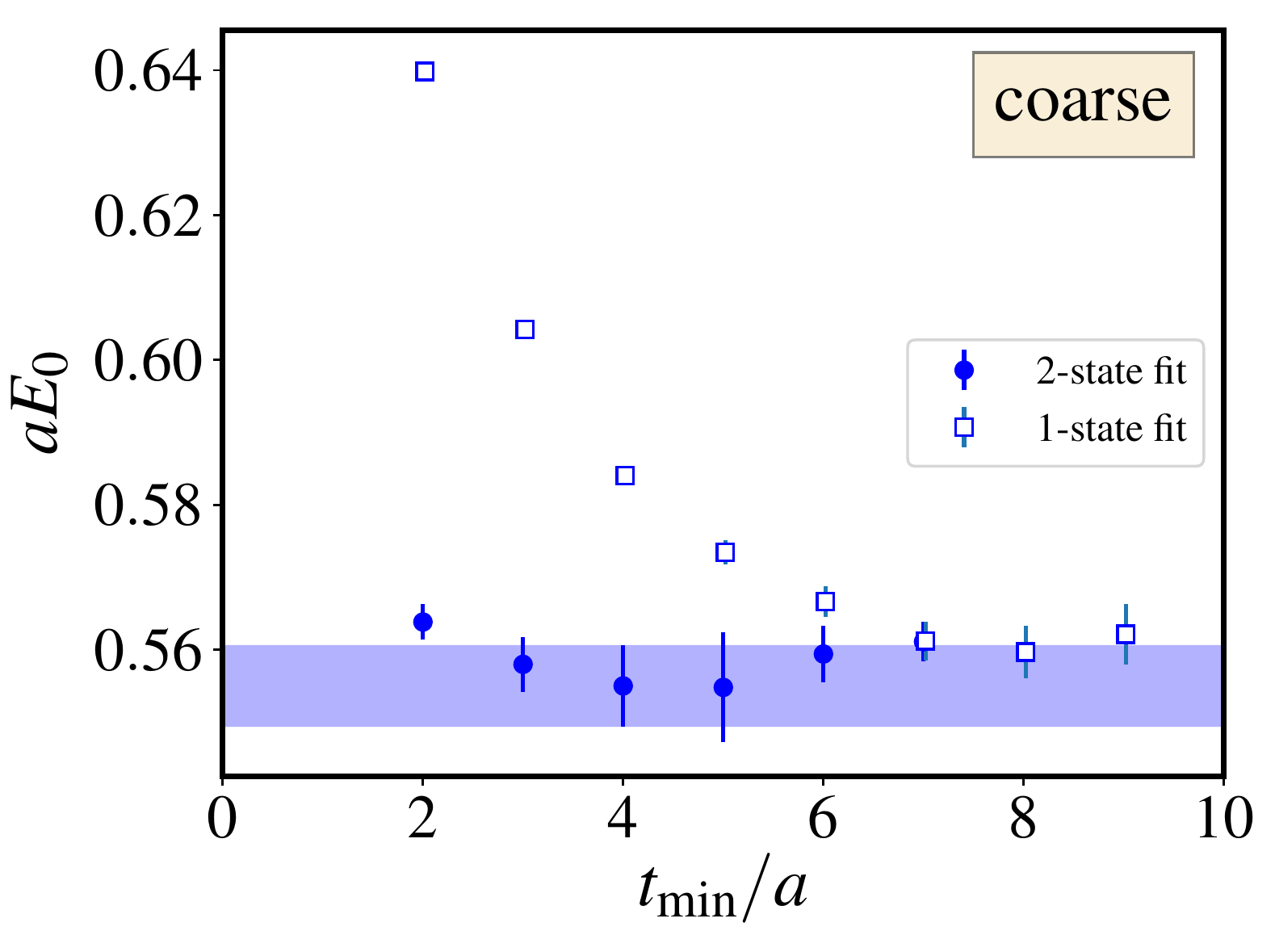}
\hspace{0.01\textwidth}
\includegraphics[width=0.45\textwidth]{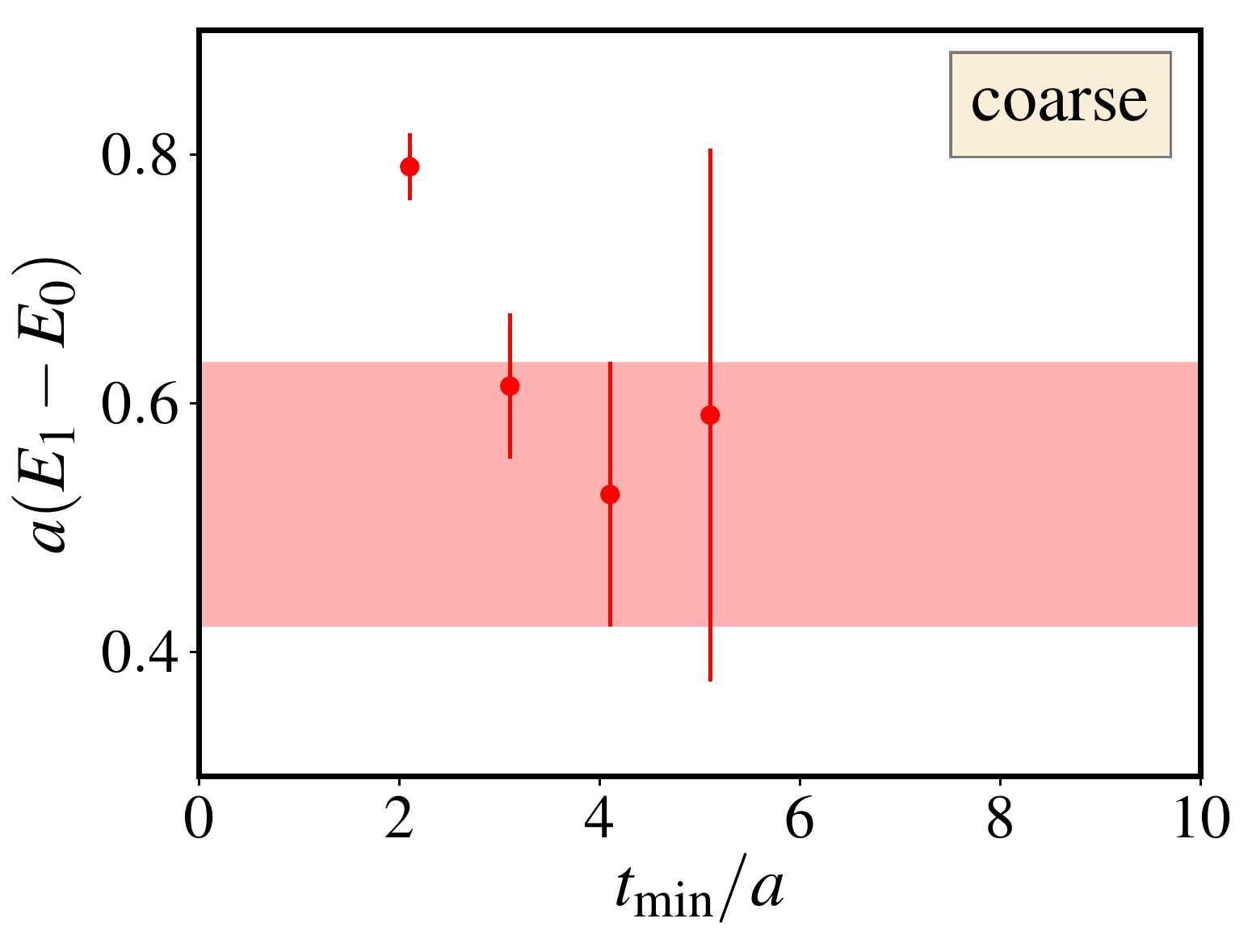}\\
\includegraphics[width=0.45\textwidth]{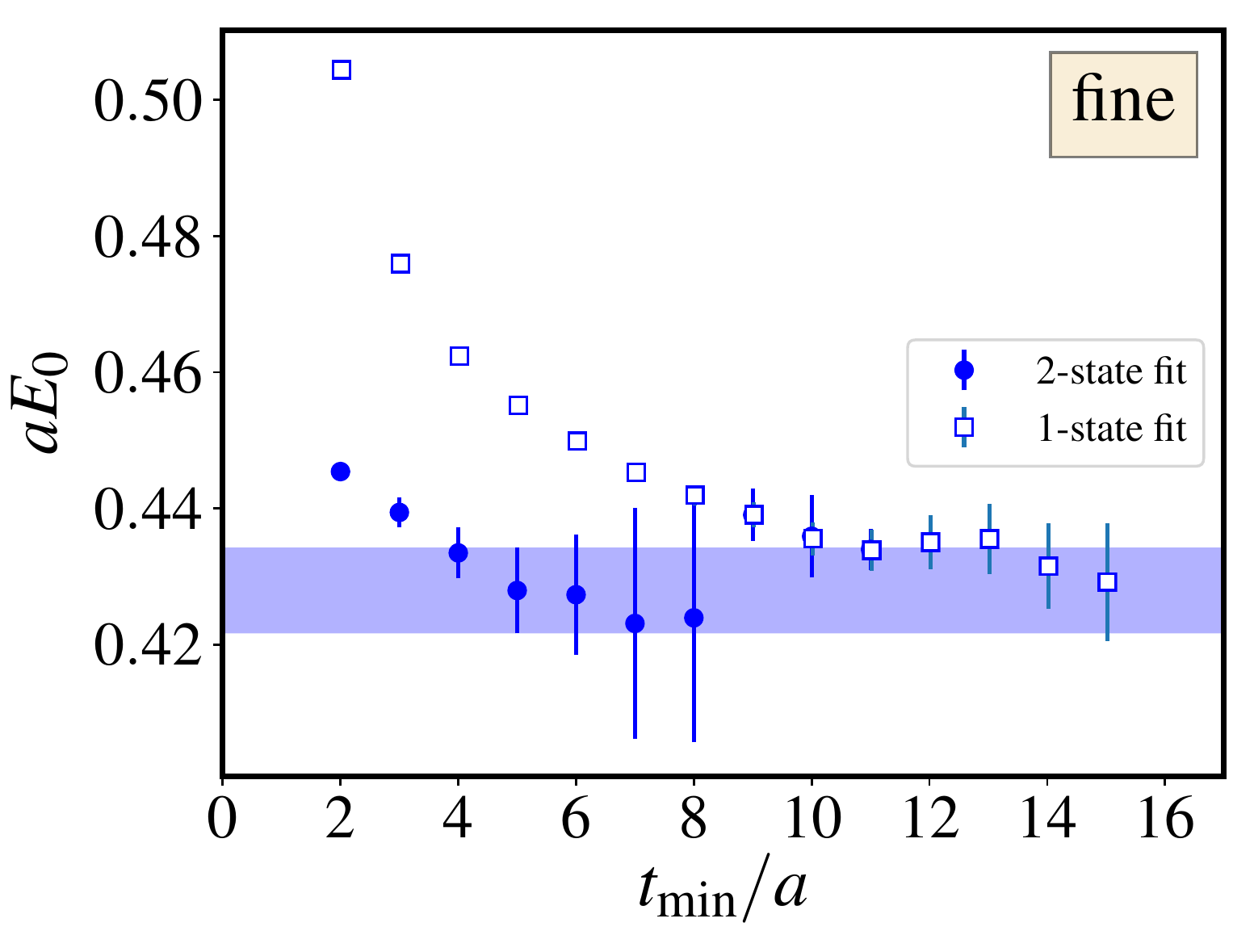}
\hspace{0.01\textwidth}
\includegraphics[width=0.45\textwidth]{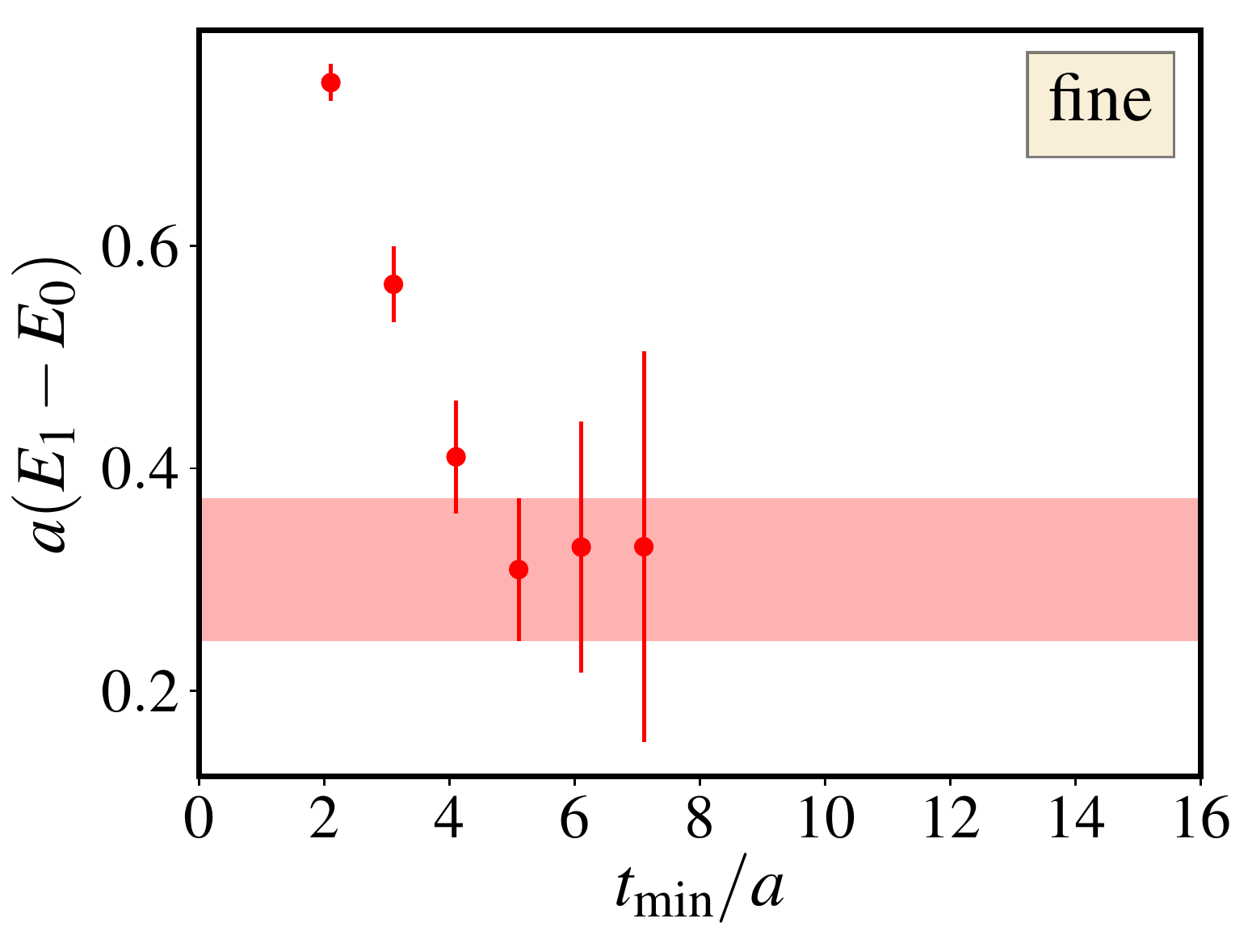}
\end{center}
  \caption{Left column: Dependence of $E_0$ on $t_\mathrm{min}$ estimated using one-state fits to the two-point function with $t_\mathrm{max}=16$ and $t_\mathrm{max}=20$ for the coarse and fine ensemble, respectively. 
Moreover, we plot $E_0$ extracted from two-state fits to the two-point function with $t_\mathrm{max}=12$ and $t_\mathrm{max}=16$ for the coarse and fine ensemble, respectively.
The blue shaded bands correspond to our preferred estimates of the ground-state masses.
Right column: Dependence of the energy gap, $E_1-E_0$, (red circles) on $t_\mathrm{min}$ using the previous two-state fits where the red shaded bands refer to our preferred estimates.}
  \label{fig:m0m1}
\end{figure}

\begin{figure}
\begin{center}
\includegraphics[width=0.45\textwidth]{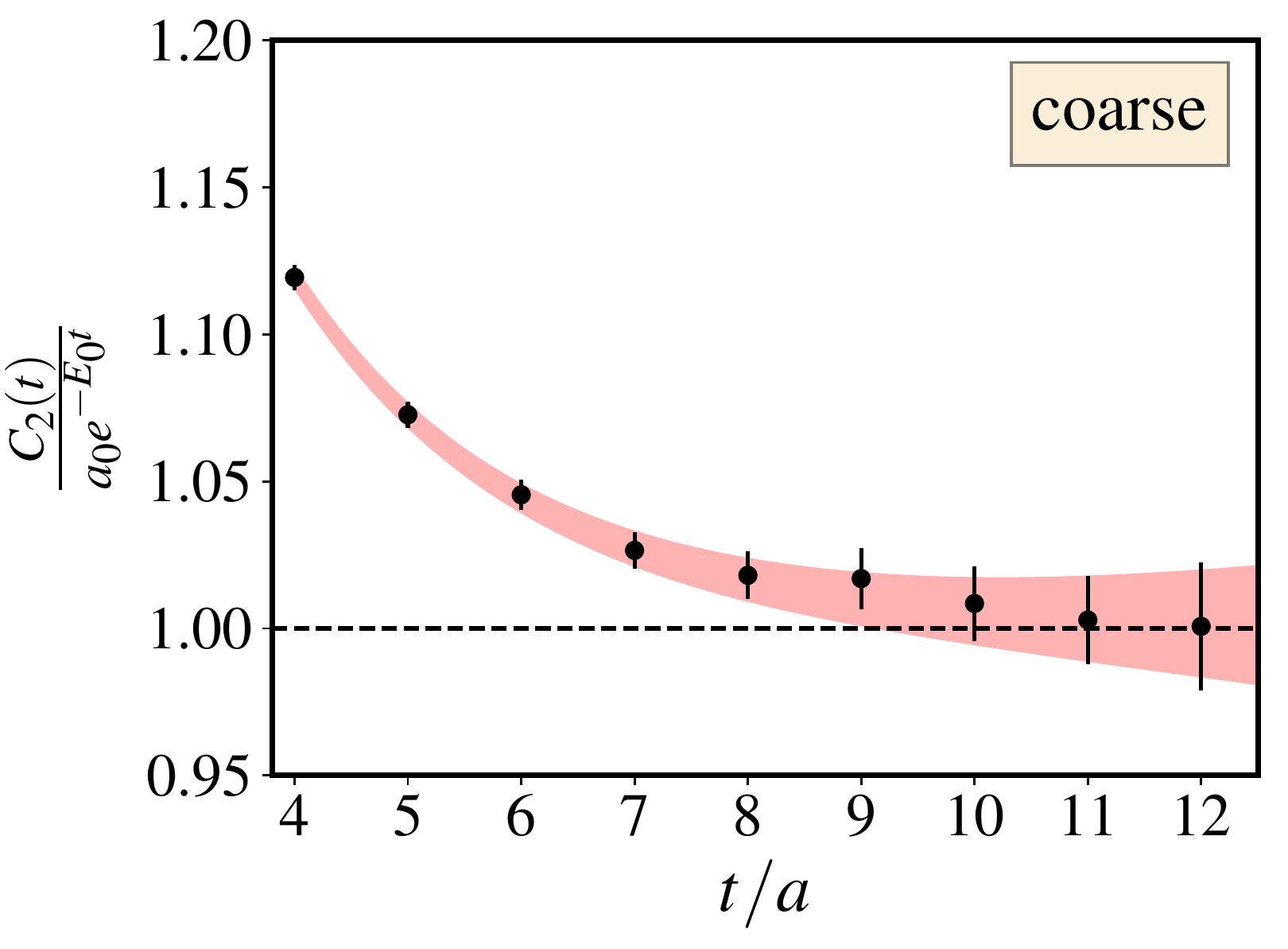}
\hspace{0.01\textwidth}
\includegraphics[width=0.45\textwidth]{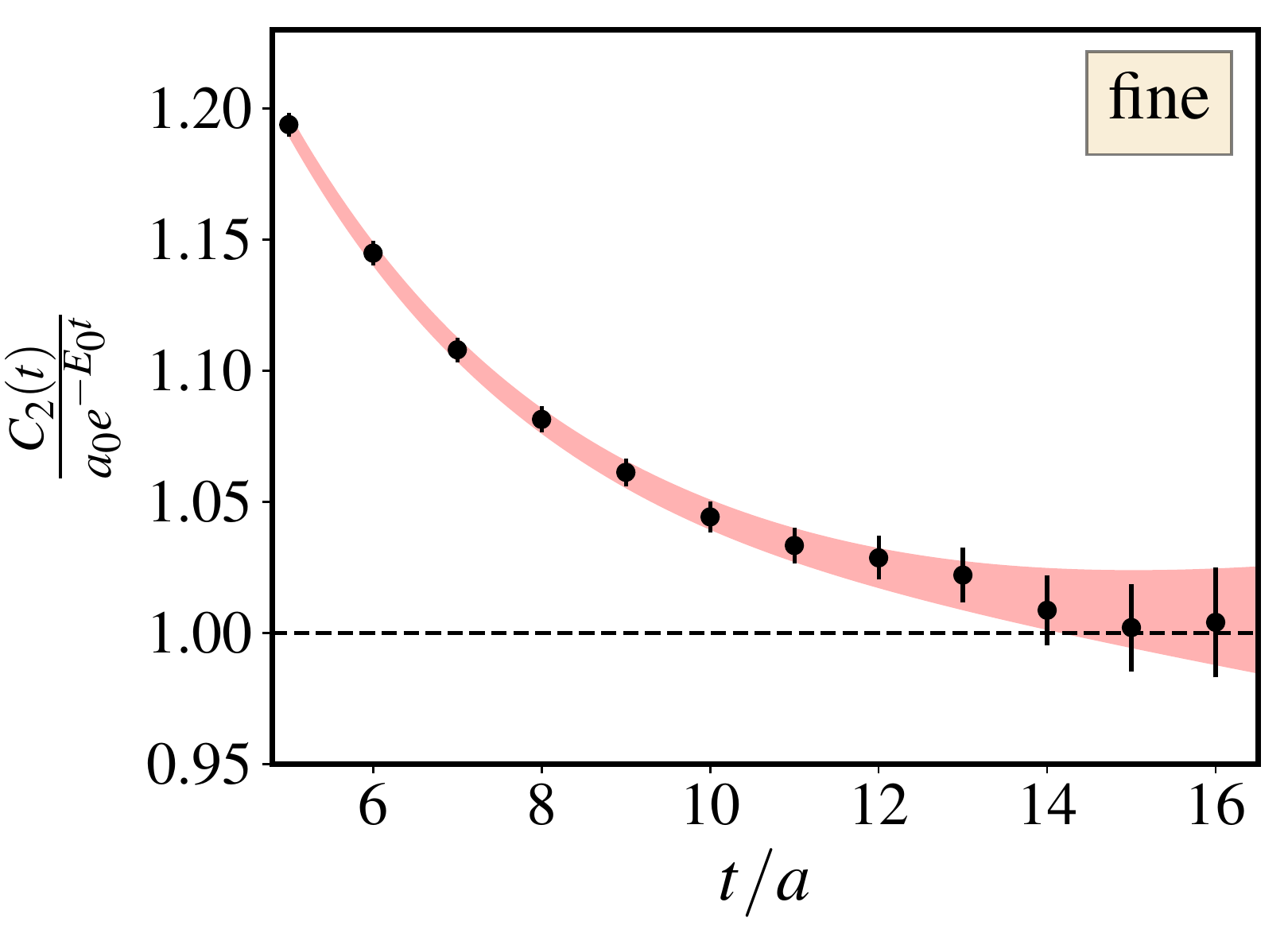}
\end{center}
  \caption{Plots of the two-point function divided by the ground-state contribution, for both the coarse (left) and the fine (right) ensembles.}
  \label{fig:c2_4864}
\end{figure}

The blue and red points in Fig.~\ref{fig:m0m1} show the dependence of $a E_0$ and $a\Delta E_{1}=a(E_1-E_0)$ on the start time slice $t_\mathrm{min}/a$ for the coarse (left) and fine (right) ensembles. The values for $aE_0$ were obtained using both the one- and two-state fits.
The shaded blue and red bands indicate our preferred estimates of $a E_0$ and $a \Delta E_{1}$, respectively. Those correspond to the two-state fits with $t_\mathrm{min}/a=4$ and $t_\mathrm{max}/a=12$ for the coarse ensemble and $t_\mathrm{min}/a=5$ and $t_\mathrm{max}/a=16$ for the fine ensemble. 
The quality of the resulting fits is shown in Fig.~\ref{fig:c2_4864} by plotting the
two-point function divided by its fitted ground-state contribution
\begin{equation}
\frac{C_2(t)}{a_0 \mathrm{exp}(-E_0 t)}.
\end{equation}
Table~\ref{tab:summary_c2} gives a summary of the estimated fit parameters on both the coarse and fine ensembles.


\begin{table}
\begin{center}
\begin{tabular}{l | c | c | c | c |}
Ensemble & $aE_0$ & $aE_1$ & $a_1/a_0$ & $\chi^2/\mathrm{dof}$\\
\hline \hline
coarse & 0.5550(56) & 1.08(11) & 0.97(20) & 0.45\\ 
fine & 0.4279(36)& 0.737(70) & 0.89(12) & 0.33\\
\end{tabular}
\end{center}
\caption{Estimated parameters of two-state fit to two-point correlation functions.}
\label{tab:summary_c2}
\end{table}
\section{Estimation of bare charges}
\label{sec:estimation_of_bare_charges}

The spectral decomposition of the three-point function is
\begin{equation}\label{eq:C3spectral}
C_3^X(\tau,T) = \sum_{n,n'} e^{-E_n(T-\tau)} e^{-E_n' \tau} (\Gamma_{\rm pol})_{\alpha\beta} \langle \Omega | {\chi}_\beta | n \rangle \langle n|\mathcal{O}_X|n'\rangle \langle n'| {\bar{ \chi}}_\alpha| \Omega \rangle,  
\end{equation}
where $\mathcal{O}_X=\mathcal{O}_X(0)$. This decomposition, along with
Eq.~\eqref{eq:C2spectral}, formally only holds in the continuum limit
since the lattice action includes a clover term and smearing that
extend in the time direction. In practice, this means that the
shortest time separations may not be trustworthy.
When the time separations $\tau$ and $T-\tau$ are 
large, excited states are exponentially suppressed and the ground-state denoted by $n,n'=0$ dominates. In 
this limit, the ratio of $C_3^X(\tau,T)$ and $C_2(T)$ yields the bare charge:
\begin{align}\label{eq:ggtimes_ratio}
R^X(\tau,T) \equiv \frac{C_3^X(\tau,T)}{C_2(T)} \xrightarrow{\text{large }\tau,(T-\tau)} g_X^{\mathrm{bare}}  + \sum_n \left[ b_n \left( e^{-\Delta E_n(T-\tau)} + e^{-\Delta E_n\tau} \right) + b'_n e^{-\Delta E_n T}  + \ldots \right],
\end{align}
where $\Delta E_n \equiv E_n - E_0$ is the energy gap between $n$th excited state and  ground state. Increasing $T$ suppresses excited-state contamination, but it also increases the noise; the signal-to-noise ratio is expected to decay asymptotically as $e^{-(E -\frac{3}{2} m_\pi) T}$~\cite{Lepage:1989hd}. 
The ratio $R^X(\tau,T)$ produces at large $T$ a plateau with ``tails'' at both ends caused by excited states. In practice, for each fixed $T$, we average over the central two or three points near $\tau=T/2$, which allows for matrix elements to be computed with errors that decay asymptotically as $e^{-\Delta E_1T/2}$. 

Excited-state contamination is a source of significant systematic uncertainties in the calculation of nucleon structure observables. 
These contributions to different nucleon structure observables have been studied recently using baryon chiral perturbation theory (ChPT) ~\cite{Tiburzi:2015tta,Bar:2016uoj,Hansen:2016qoz,Bar:2017kxh}. 
Contamination from two-particle $N\pi$ states in the plateau estimates of various nucleon charges, which becomes more pronounced in physical-point simulations, has been studied in Refs.~\cite{Bar:2016uoj,Bar:2017kxh}.
It was found that this particular contamination leads to an overestimation at the 5--$10\%$ level for source-sink separations of about 2 fm. This suggests that the source-sink separations of $\sim 1.5$ fm reached in present-day calculations may not be sufficient to isolate the contribution of the ground-state matrix element with the desired accuracy.
On the other hand, in Ref.~\cite{Hansen:2016qoz} a model was used to study corrections to the LO ChPT result for the axial charge; it was found that high-momentum $N\pi$ states with energies larger than about $1.5 M_N$ can be the cause for the underestimating of the axial charge observed in Lattice QCD calculations. These contributions, however, cannot be estimated in chiral perturbation theory. 
Refs.~\cite{Tiburzi:2015tta,Bar:2016uoj,Hansen:2016qoz,Bar:2017kxh} find that multiple low-lying nucleon-pion states give important contributions to $R^X(\tau,T)$, which is in stark contrast to the commonly-used fit model based on a single excited state.

In the remainder of this section, we discuss the analysis methods we employ to study and suppress excited-state contributions to the axial, scalar, and tensor charges.
We start with estimating the bare charges using the standard `ratio method' in Sec.~\ref{sec:Ratio}. In Sec.~\ref{sec:summ_2stateS}, we discuss the use of the summation method in addition to presenting a two-state fit model to the summations, which was inspired by the calculation in Ref.~\cite{Chang:2018uxx} that quotes a percent-level uncertainty for $g_A$.
Furthermore, we employ a two-state fit to the ratios $R^X(\tau,T)$, which is presented in Sec.~\ref{sec:2stateR}.
Finally, in Sec.~\ref{sec:final_analyis}, we explain the procedure we follow to combine the estimates from the different fit strategies and extract final values for the bare charges.

\subsection{Ratio method}\label{sec:Ratio}
\begin{figure}
\begin{center}
 \includegraphics[width=0.49\textwidth]{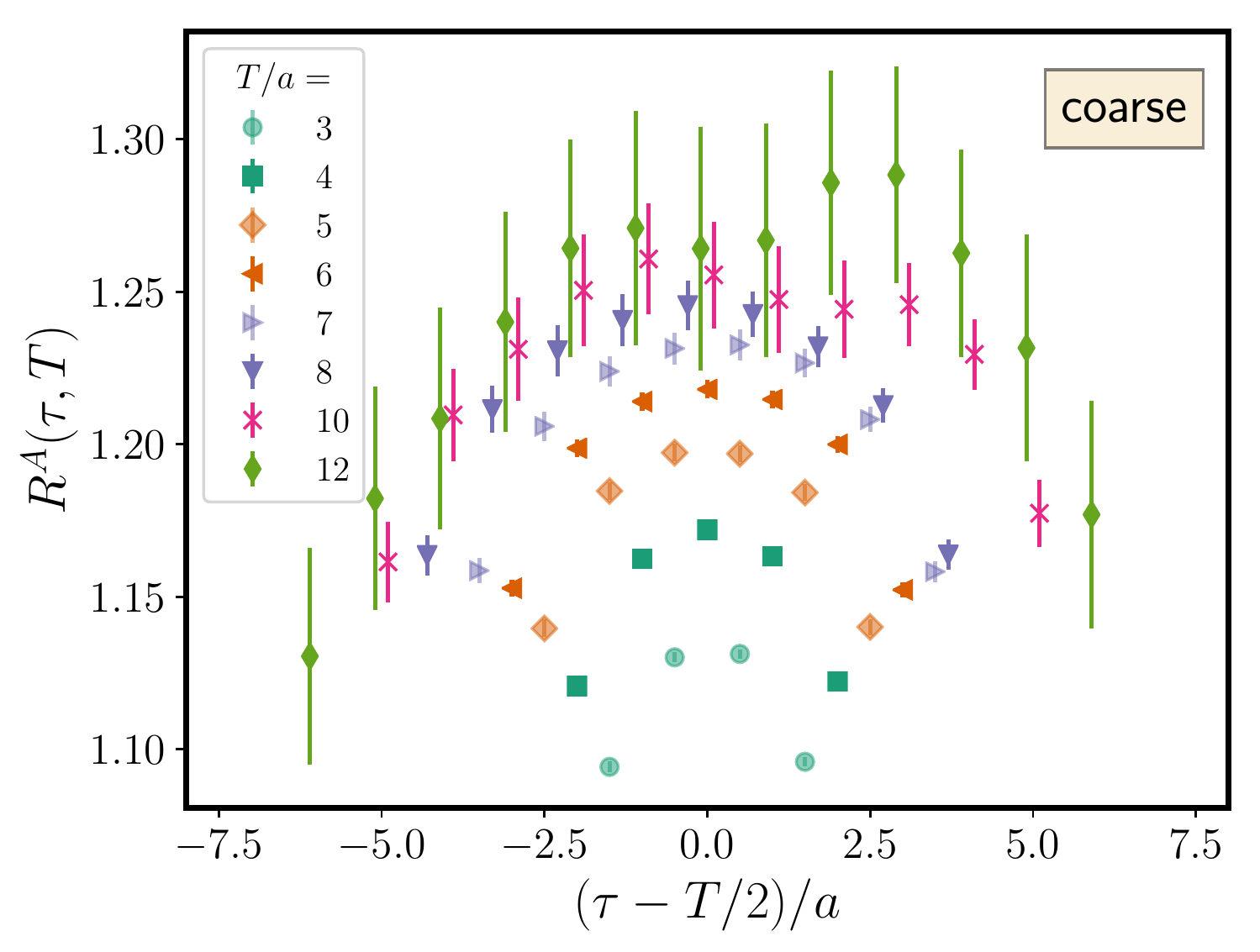}
  \hspace{0.001\textwidth}
   \includegraphics[width=0.49\textwidth]{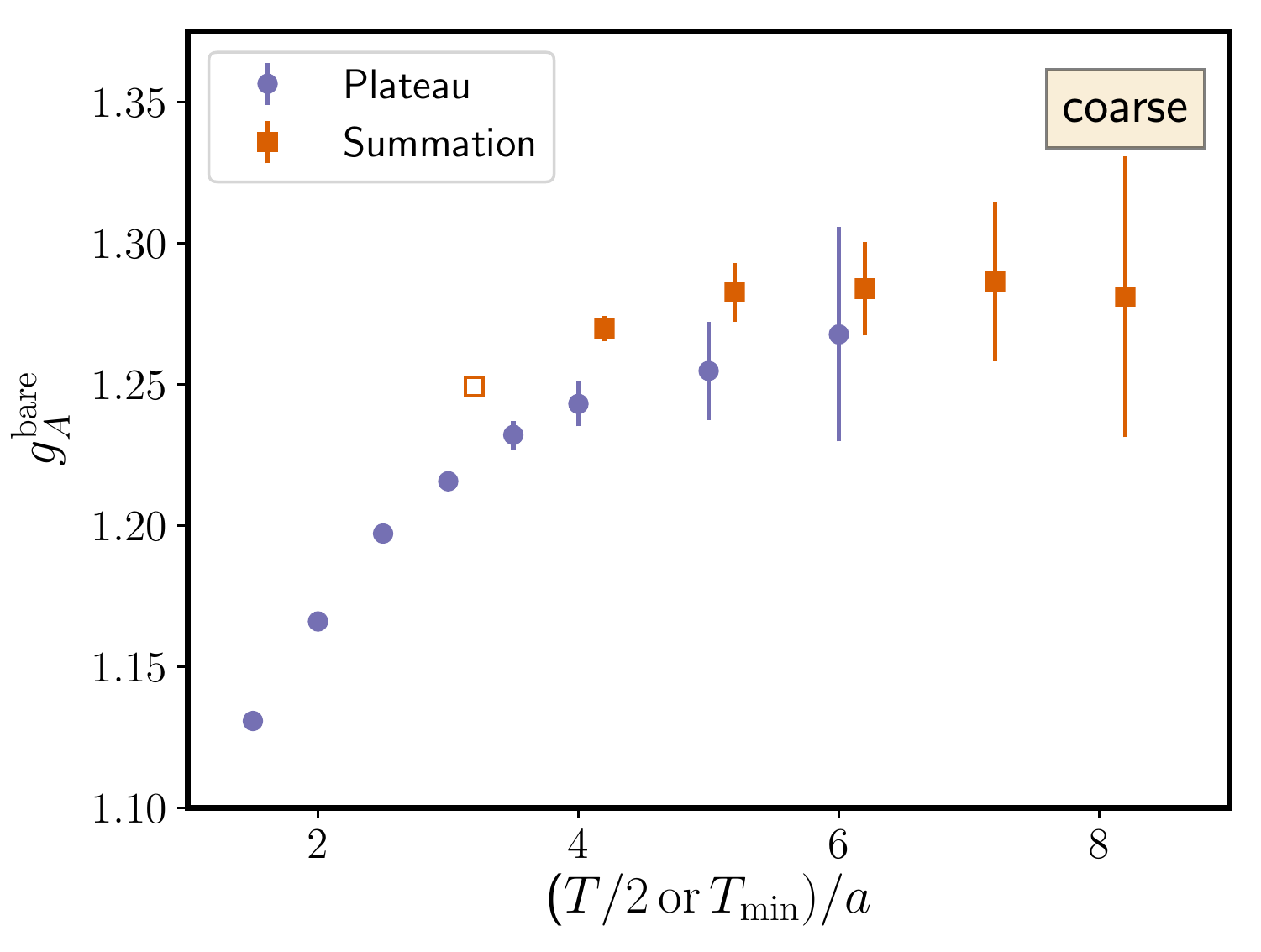}
   \\
     \includegraphics[width=0.49\textwidth]{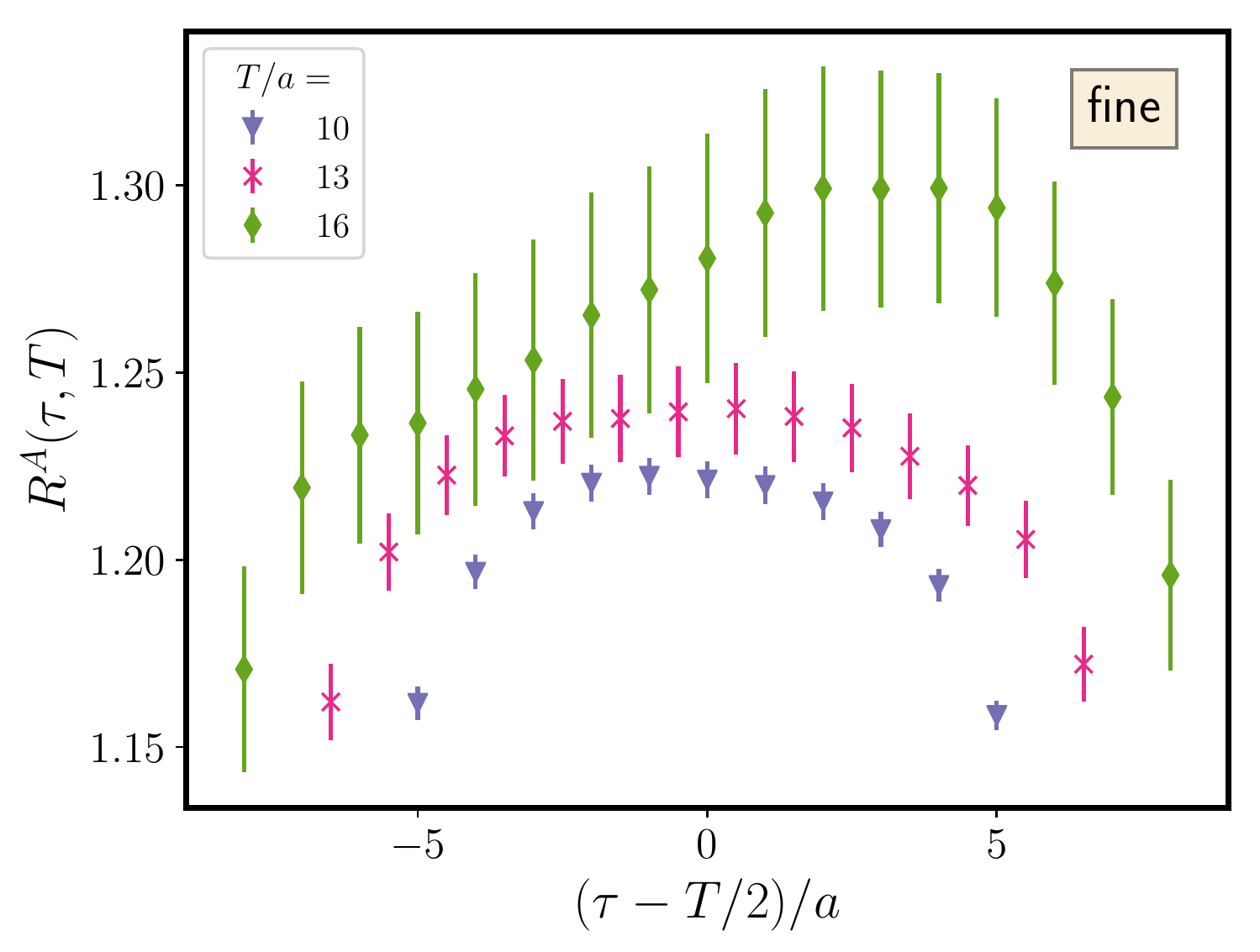}
 \hspace{0.001\textwidth}
   \includegraphics[width=0.49\textwidth]{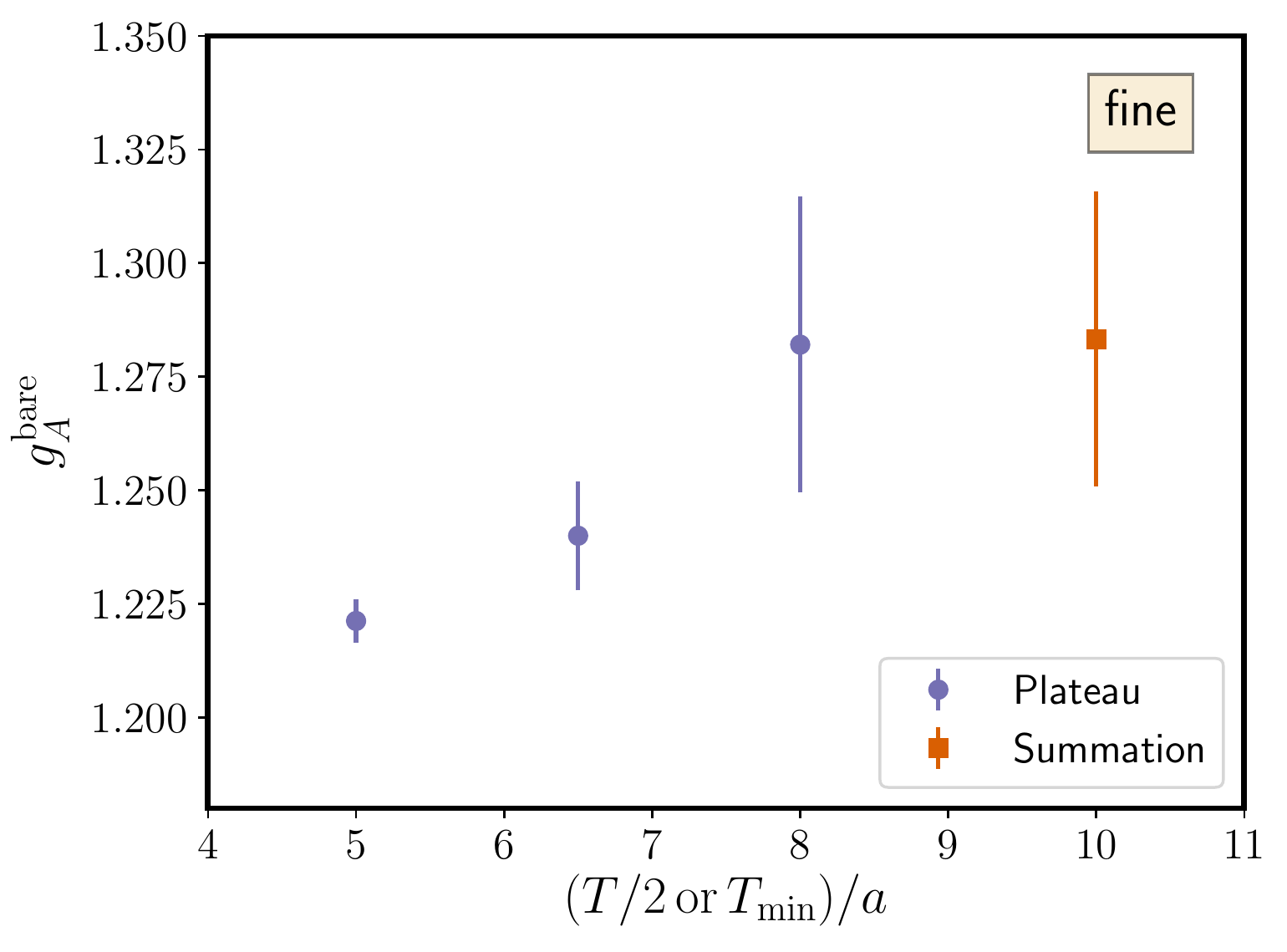}
   \end{center}
  \caption{Results for the isovector axial charges on the coarse (top row) and fine (bottom row) ensembles using the ratio and summation methods. The first column shows 
the dependence of the ratios on the operator insertion time $\tau$ and the source-sink separation $T$.
Different source-sink separations are displayed in different colors. The blue circles in the second column show the values of the charges estimated by averaging the two or three central points of $R^A(\tau,T)$ near $\tau=T/2$ and their dependences on $T/2$. The red squares in the second column show the resulting bare isovector axial charges using the summation method. Here, we show the dependences of the obtained axial charge on the minimal source-sink separations included in the fit $T_\mathrm{min}$. The open symbol
indicates a poor fit with $p$-value less than $0.02$.}
  \label{fig:gA_plat_summary}
\end{figure}

\begin{figure}
\begin{center}
 \includegraphics[width=0.49\textwidth]{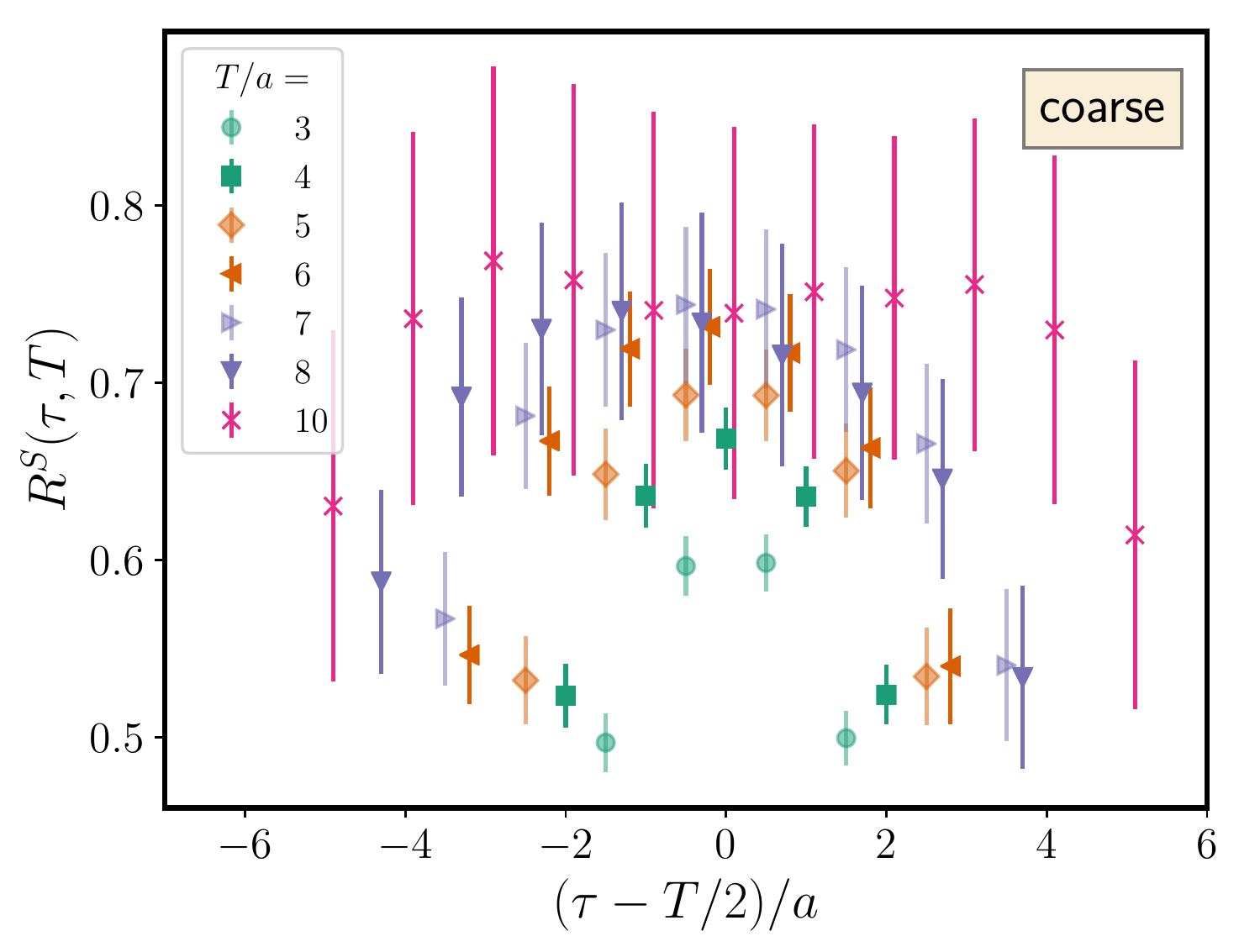}
  \hspace{0.001\textwidth}
   \includegraphics[width=0.49\textwidth]{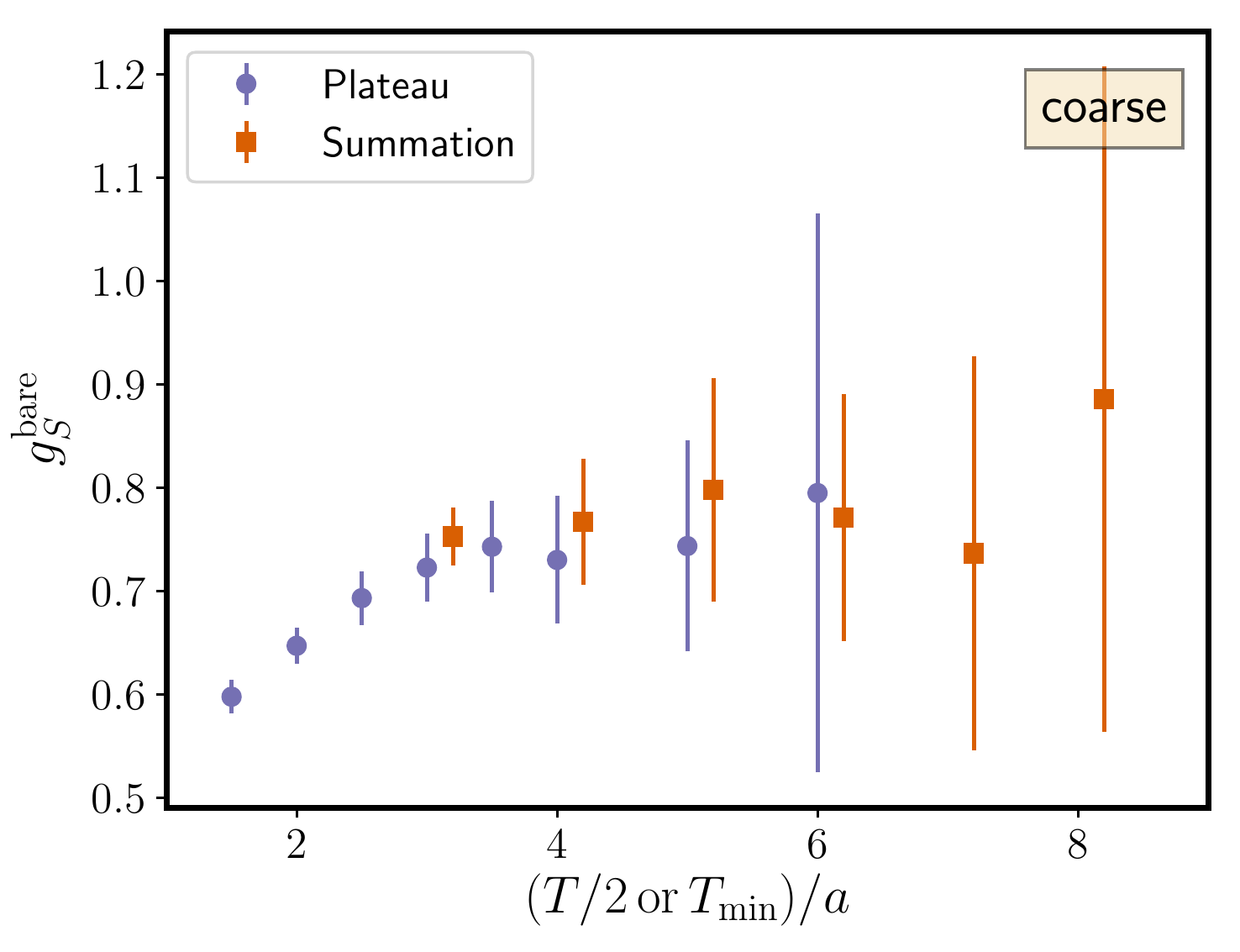}
   \\
     \includegraphics[width=0.49\textwidth]{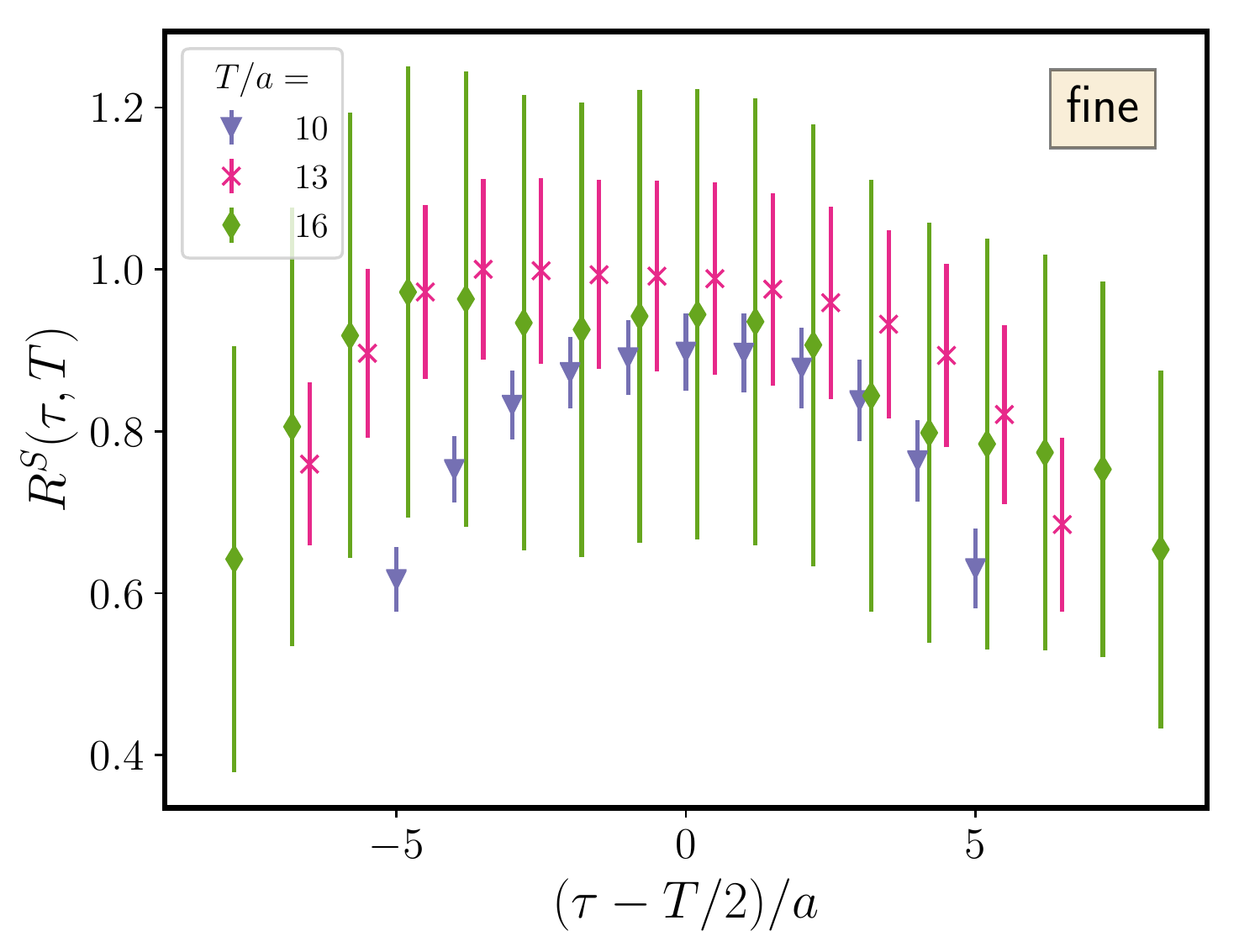}
 \hspace{0.001\textwidth}
   \includegraphics[width=0.49\textwidth]{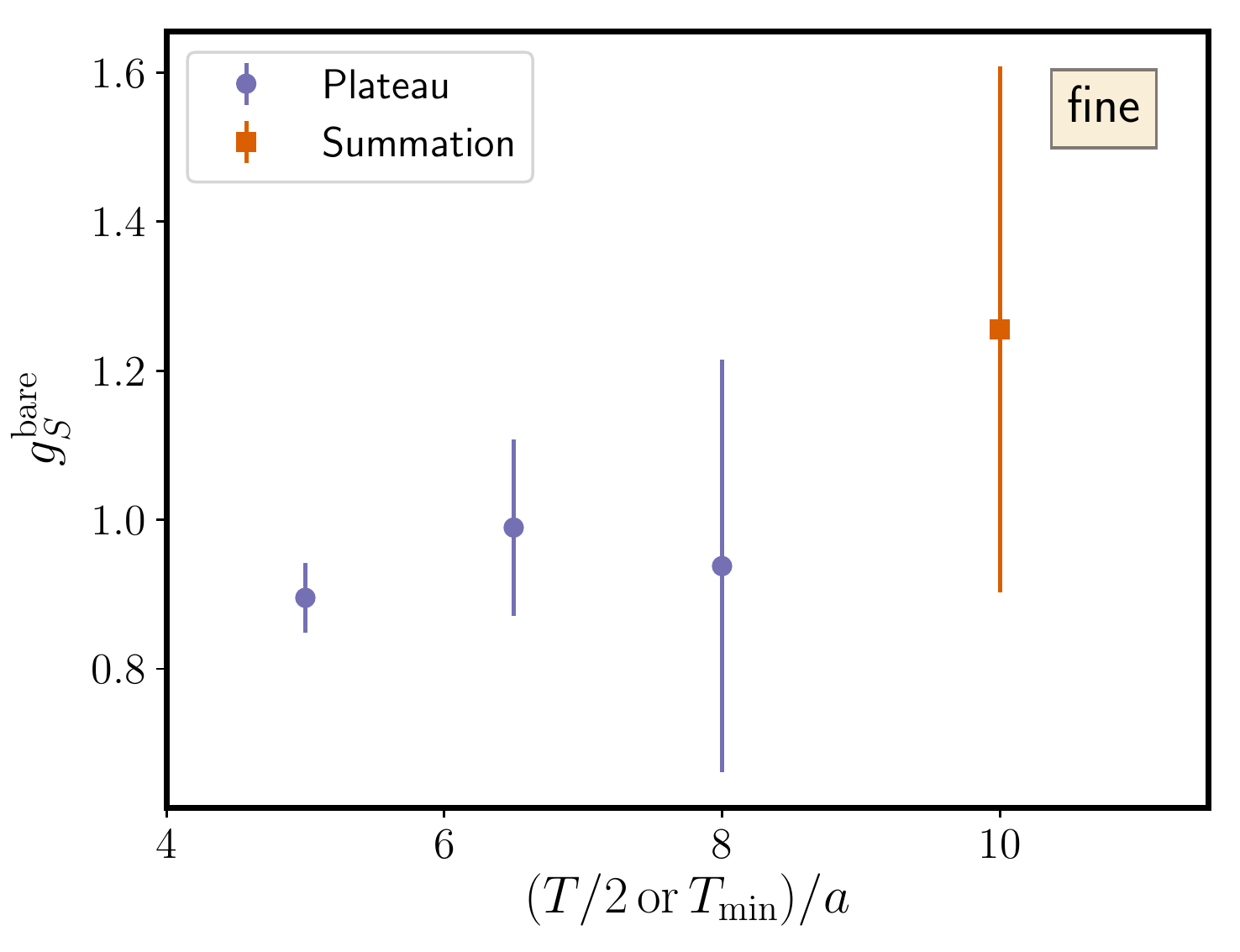}
   \end{center}
  \caption{Results for the isovector scalar charge using the ratio and summation methods. See the caption of Fig.~\ref{fig:gA_plat_summary} for explanations.}
  \label{fig:gS_plat_summary}
\end{figure}

\begin{figure}
\begin{center}
 \includegraphics[width=0.49\textwidth]{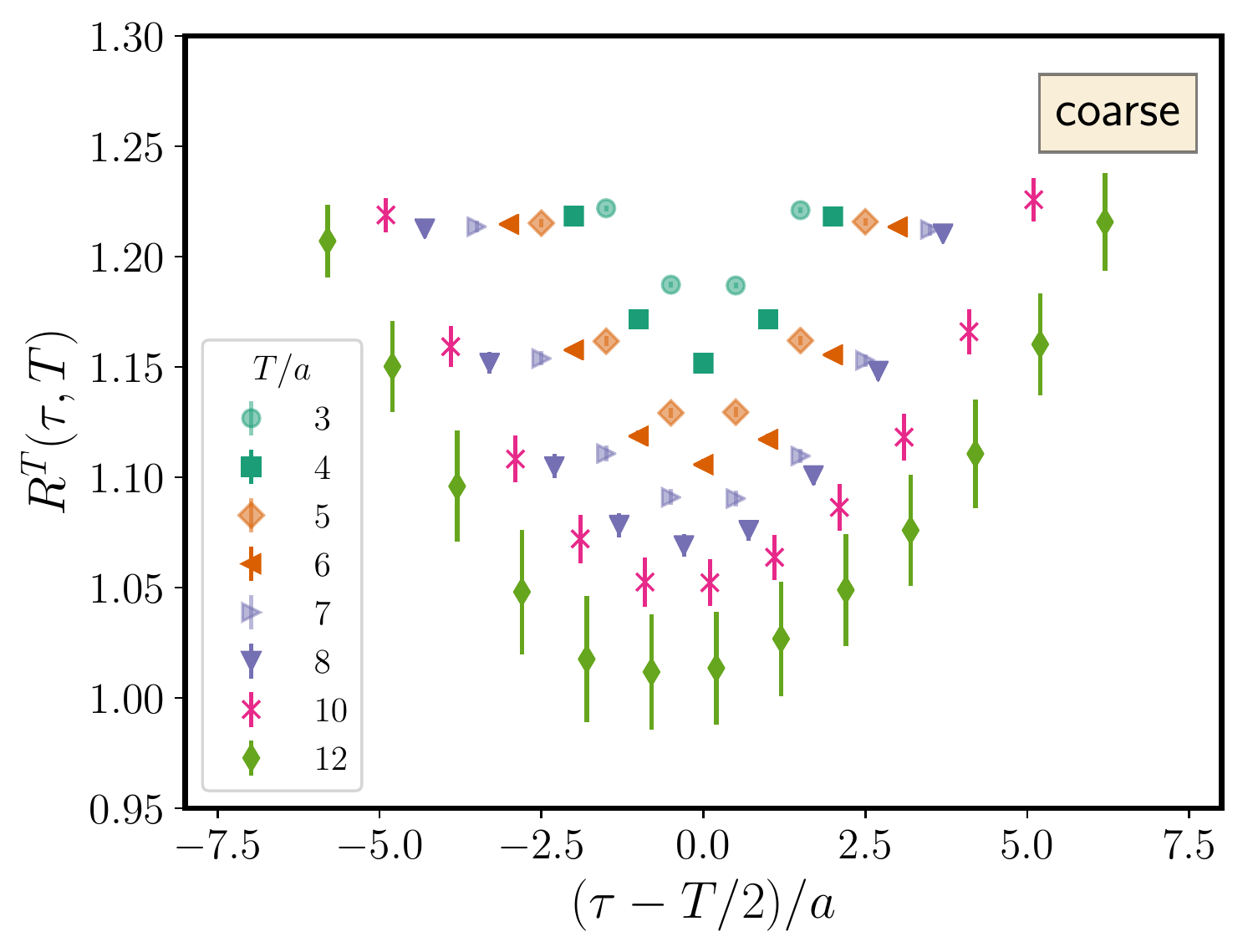}
  \hspace{0.001\textwidth}
   \includegraphics[width=0.49\textwidth]{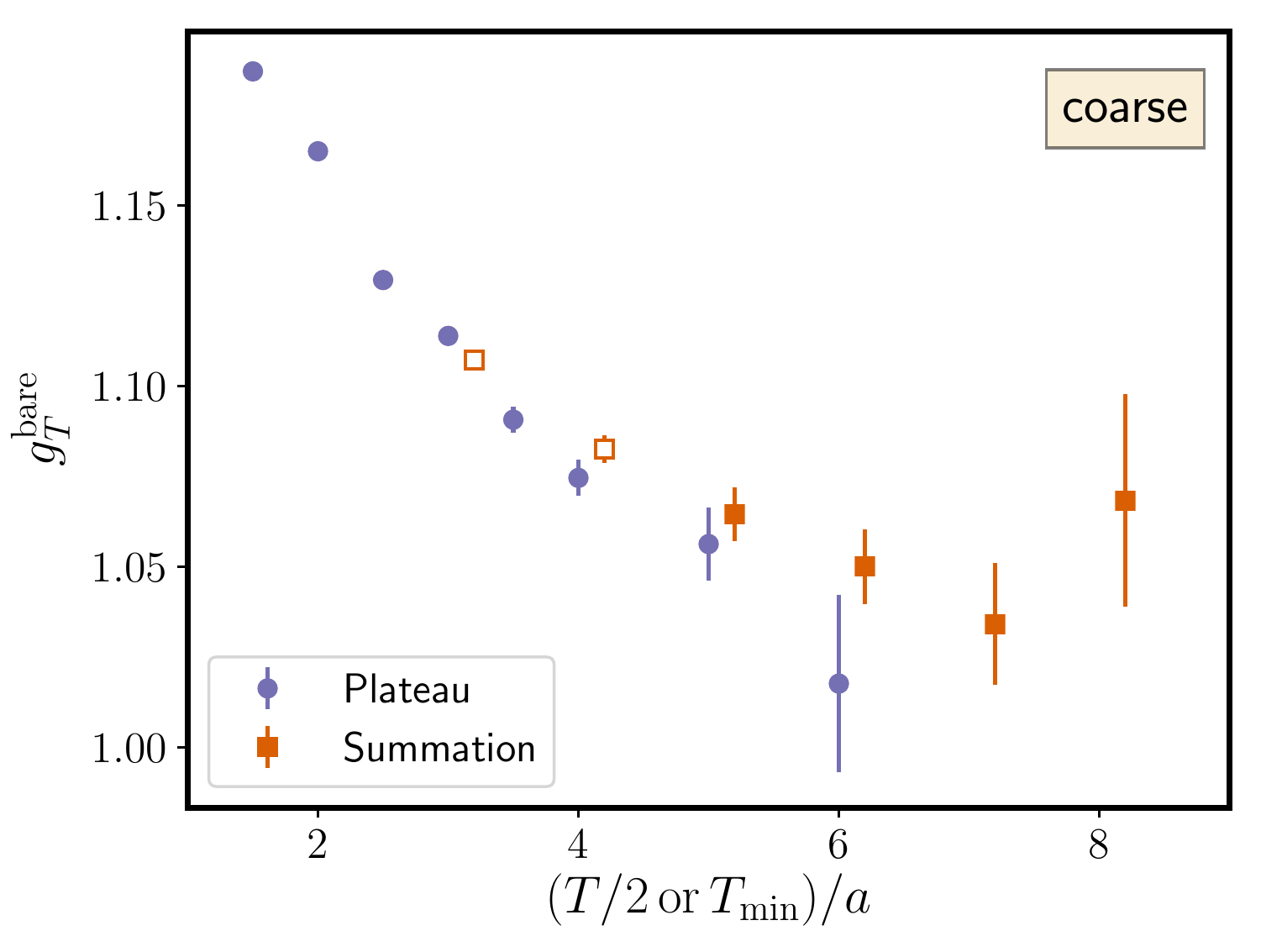}
   \\
     \includegraphics[width=0.49\textwidth]{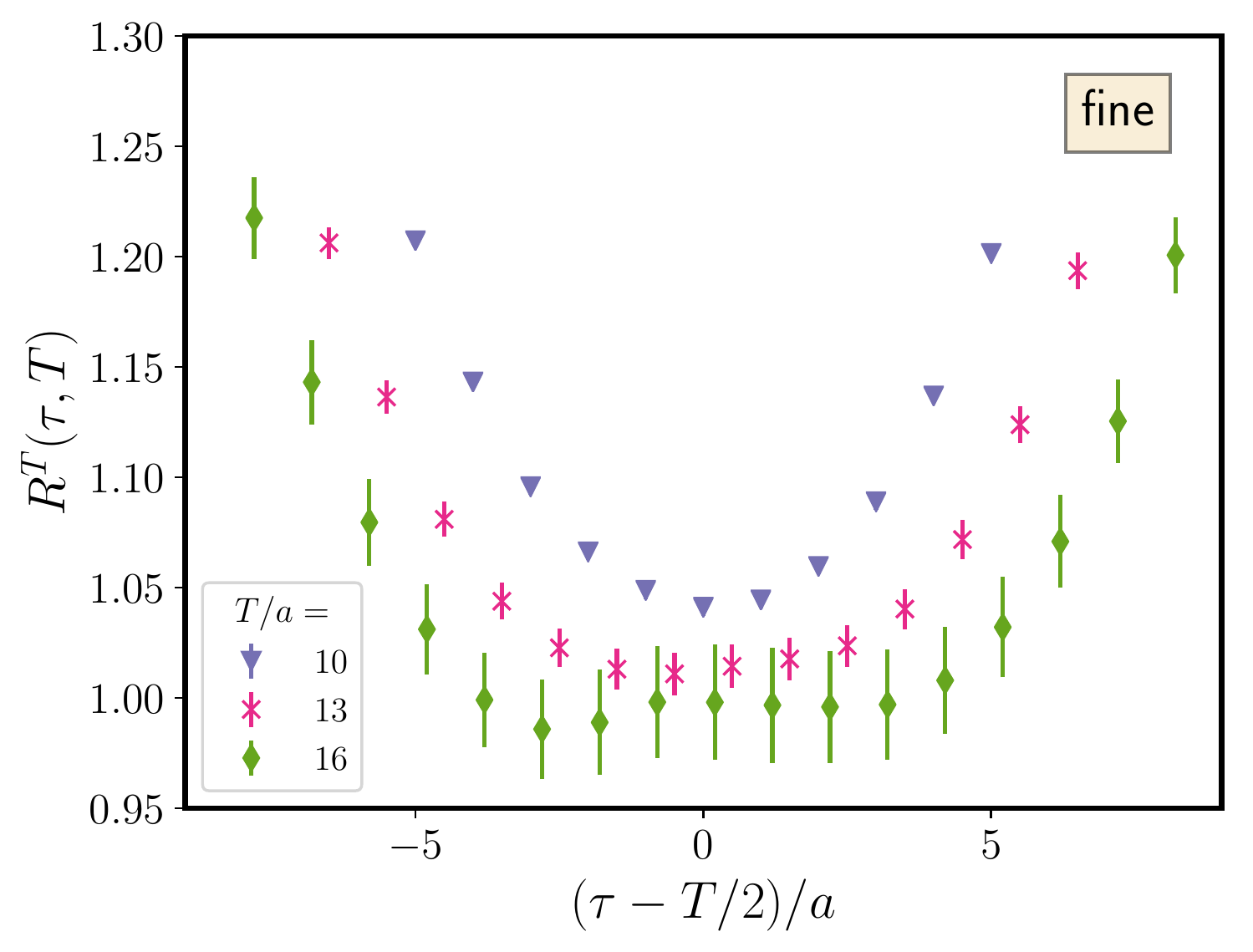}
 \hspace{0.001\textwidth}
   \includegraphics[width=0.49\textwidth]{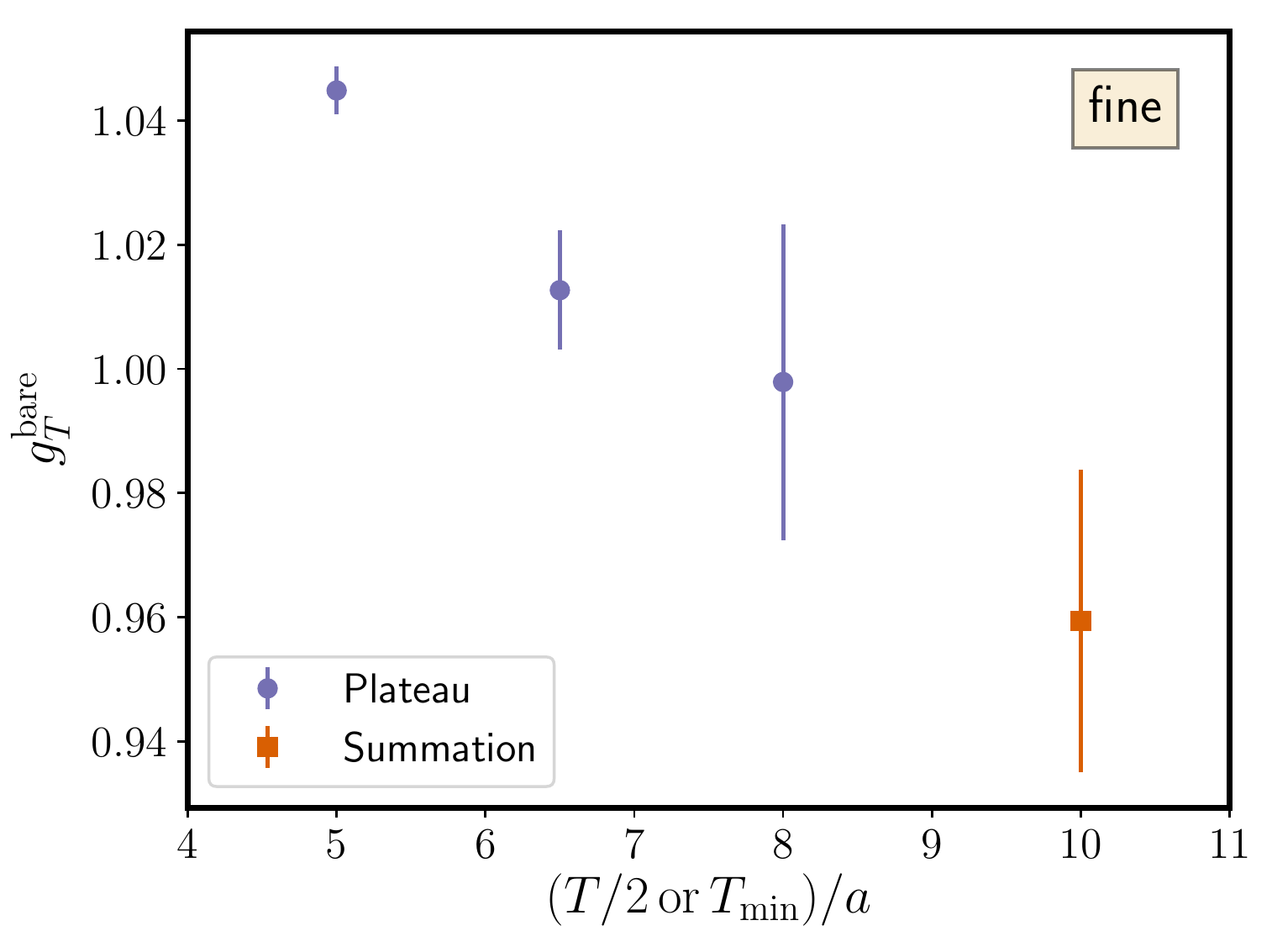}
   \end{center}
  \caption{Results for the isovector tensor charge using the ratio and summation methods. See the caption of Fig.~\ref{fig:gA_plat_summary} for explanations.}
  \label{fig:gT_plat_summary}
\end{figure}

The ratio method is a simple approach that allows for excited-state effects to be clearly seen.
\Cref{fig:gA_plat_summary,fig:gS_plat_summary,fig:gT_plat_summary} show our results for the isovector axial, scalar, and tensor charges on the coarse (top rows) and fine (bottom rows) ensembles.
The first columns of those figures show the ratios yielding the different charges as functions of the insertion time $\tau/a$ shifted by half the source-sink separation, i.e.\ $(\tau-T/2)/a$.
The different colors correspond to the ratios obtained using different source-sink separations.
For the axial and scalar charges (particularly on the coarse
ensemble), there appears to be a jump in the data from $\tau=0$ to
$\tau/a=1$, which is expected because even in the continuum limit the
spectral decomposition in Eq.~\eqref{eq:C3spectral} assumes a nonzero
separation between the interpolating operator and the current. On the
other hand, the data appear to vary smoothly between $\tau=a$ and
$\tau=T-a$, suggesting that the effect of the lack of a transfer
matrix is mild.

As explained in Sec.~\ref{sec:lat_setup}, when the times $\tau$ and $T-\tau_0$ are large, the ratios become time-independent. One observes increasing (for $g_A^\mathrm{bare}$ and $g_S^\mathrm{bare}$) or decreasing (for $g_T^\mathrm{bare}$) trends for the plateau values  as the time separations are increased and clear curvatures indicating the significant contributions from excited states.
We estimate the different charges by averaging the central two or three points near $\tau=T/2$. The blue circles in the second columns of \cref{fig:gA_plat_summary,fig:gS_plat_summary,fig:gT_plat_summary} are the estimated charges from the plateaus plotted against $T/2$. We know that the excited-state contributions to $R^X(\tau,T)$ decay as $e^{-\Delta E_1 T/2}$ which results eventually in a plateau when the source-sink separation is large enough.
We observe on both the coarse and fine ensembles that the scalar
charge reaches a plateau as expected with increasing $T/2$.  This does
not happen in the case of the tensor charge, indicating that this
method fails to reliably control excited states for $g_T$. For the
axial charge, a plateau is possibly reached at the largest values of
$T/2$, although this coincides with the presence of particularly large
statistical uncertainties.

\subsection{Summation method}\label{sec:summ_2stateS}
For studying the excited-state contributions, we use in addition to the aforementioned ratio method, the \emph{summation method}~\cite{Capitani:2010sg,Bulava:2010ej}. The summation method allows improving the asymptotic behavior of excited-state contributions through summing ratios at each source-sink separation $T$. The summed ratios can be shown to be asymptotically linear in the source-sink separation,
 \begin{equation}\label{eq:summfit}
 S^X(T) \equiv \sum_{\tau=\tau_0}^{T-\tau_0} R^X(\tau,T) = c_0+T g_X^{\mathrm{bare}} + O(Te^{-\Delta E_1 T}) + O(e^{-\Delta E_1 T}).
 \end{equation}
We choose $\tau_0/a=1$. The matrix element can then be extracted from the slope of a linear fit to $S^X(T)$ at several values of $T$. The leading excited-state contaminations decay as $Te^{-\Delta E_1 T}$.

For performing the fits of the summation method on the coarse ensemble, we vary the fit range by fixing the maximum source-sink separation included in the fit to $T_\mathrm{max}/a=12$ and changing the minimal source-sink separation, $T_\mathrm{min}/a$. The obtained results for the three charges on the coarse ensemble are displayed as red squares in the upper right panels
of~\cref{fig:gA_plat_summary,fig:gS_plat_summary,fig:gT_plat_summary} which demonstrate the dependences of $g_X^\mathrm{bare}$ on $T_\mathrm{min}/a$. 
Here, we see that the obtained $g_A^\mathrm{bare}$ shows a slight increase when increasing from the shortest 
$T_\mathrm{min}$ and $g_T^\mathrm{bare}$ shows a somewhat larger decrease, whereas $g_S^\mathrm{bare}$ is flat. We eventually reach a plateau in all cases. The fit quality is measured by computing the $p$-value and  
the open symbols refer to fits with $p$-value $<0.02$. 
The red squares in the lower right panels
of~\cref{fig:gA_plat_summary,fig:gS_plat_summary,fig:gT_plat_summary} show the results for the summation method on the fine ensemble including all three available source-sink separations, which leads to one summation point at $T_\mathrm{min}/a=10$.

\begin{figure}
\begin{center}
\includegraphics[width=0.49\textwidth]{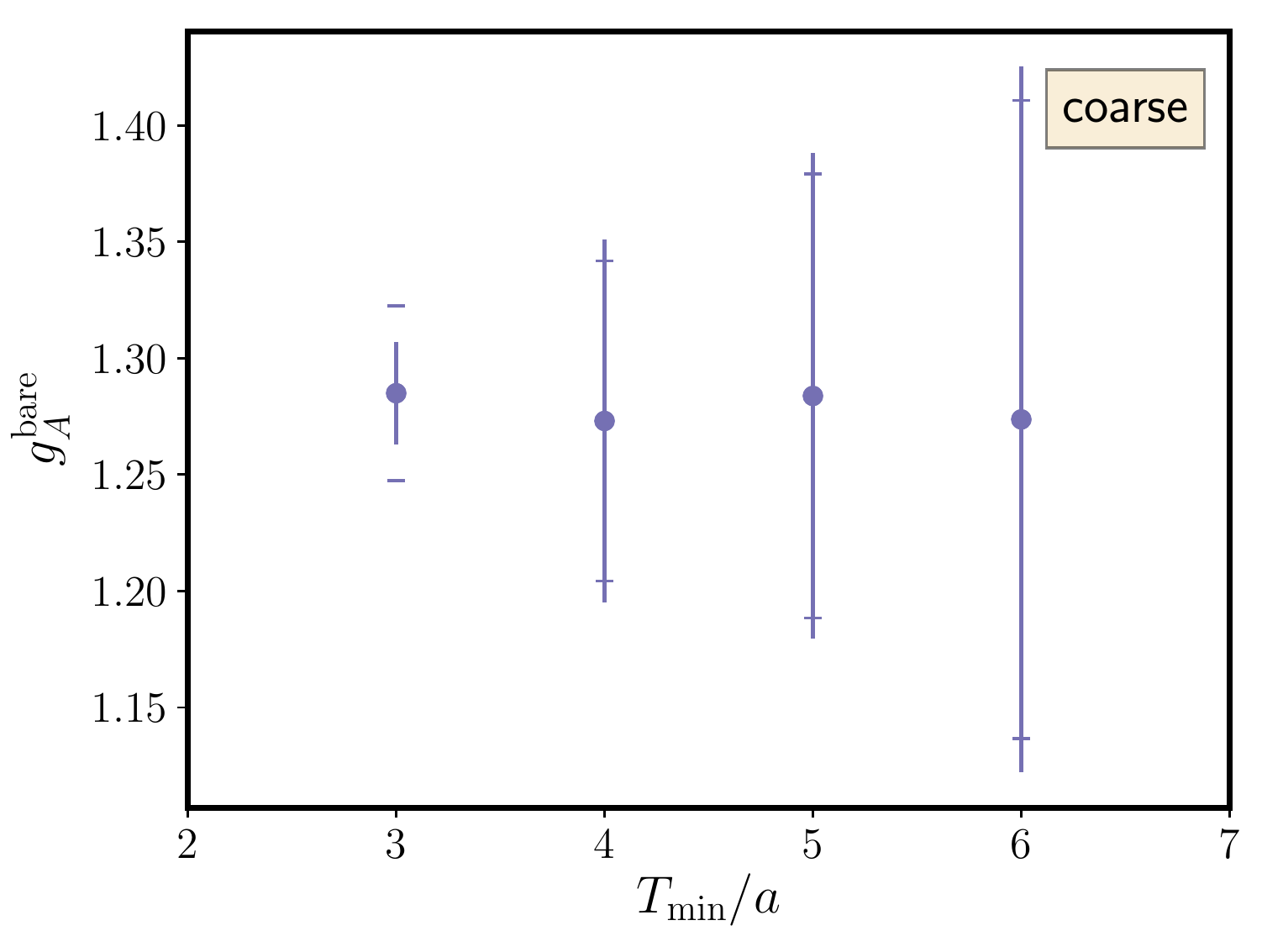}
\hspace{0.001\textwidth}
\includegraphics[width=0.49\textwidth]{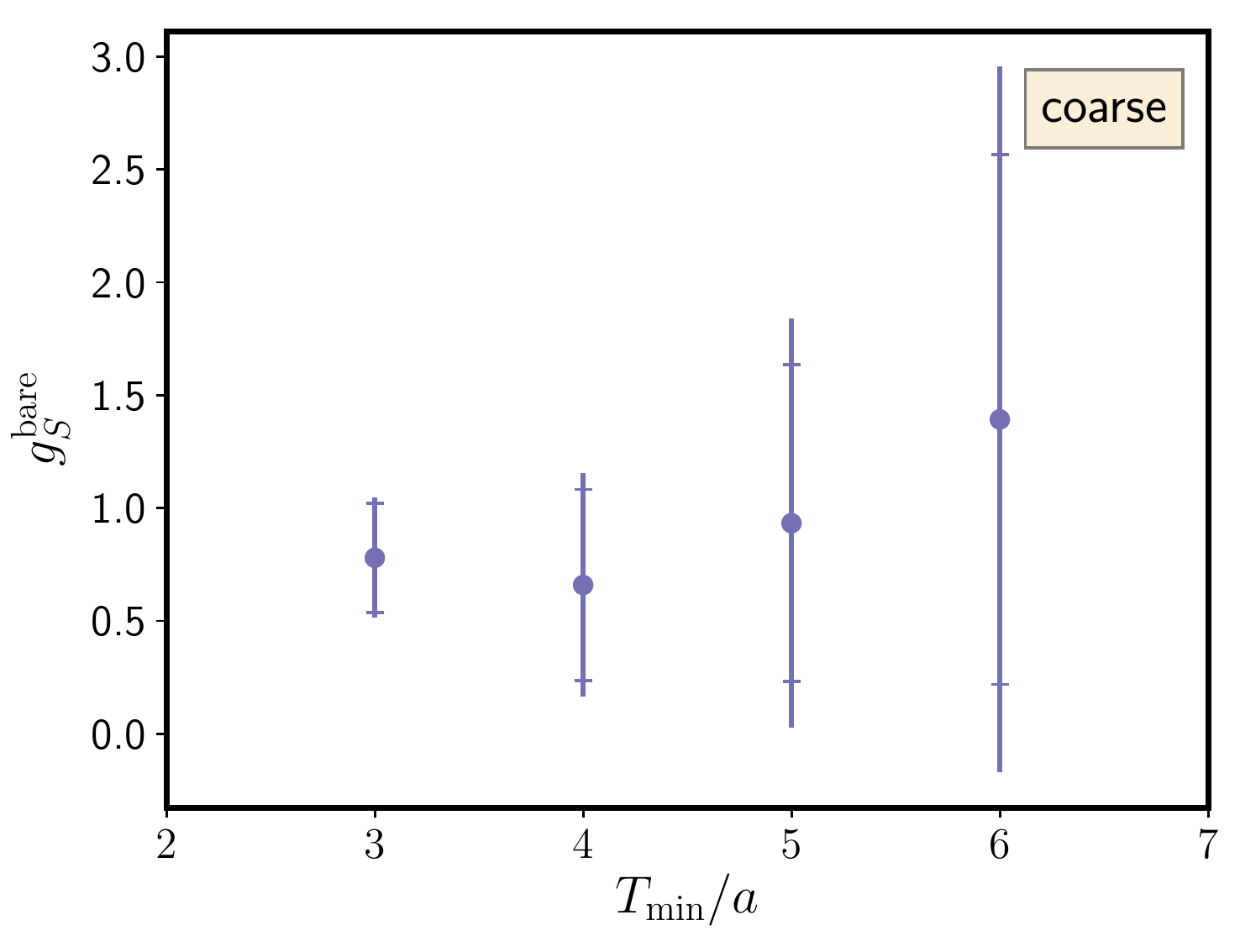}
\hspace{0.001\textwidth}
\includegraphics[width=0.49\textwidth]{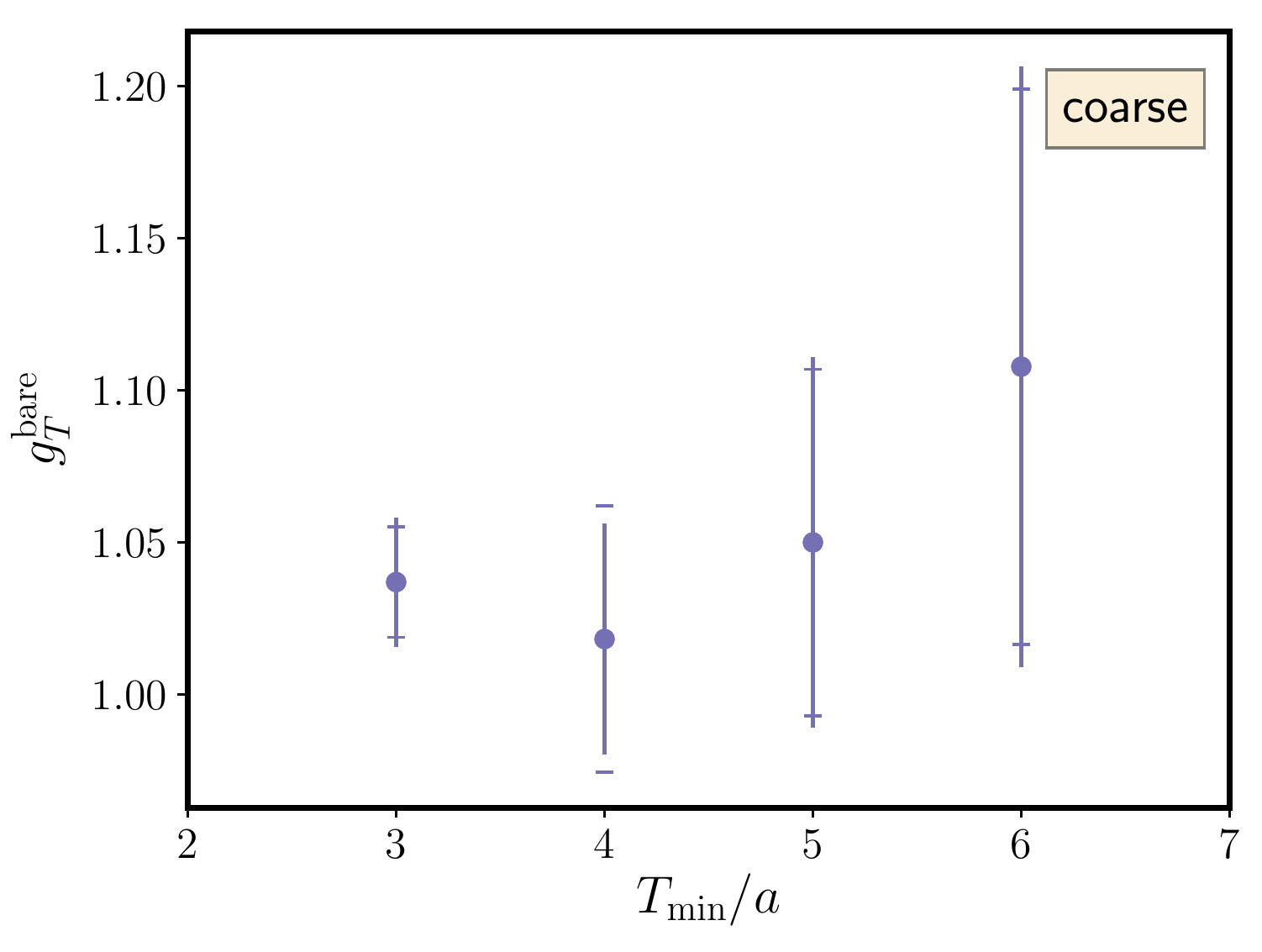}
\end{center}
  \caption{The obtained unrenormalized axial, scalar, and tensor charges on the coarse ensemble from fits to the summations $S^X(T)$ including contributions from a single 
excited state.
The endcaps of the error bars refer to the resulting statistical uncertainties when fixing $\Delta E_1$ in Eq.~\eqref{eq:summST} to its central value, whereas the vertical lines of the error bars result from taking the uncertainties in $\Delta E_1$ into consideration when evaluating the fit in Eq.~\eqref{eq:summST}.}
  \label{fig:2stateS_summary}
\end{figure}

The numerous source-sink separations used for calculations on the coarse ensemble allow us to perform the fit to the summations in Eq.~\eqref{eq:summfit} including contributions from the first excited state. This leads to two additional fit parameters $c_1$ and $c_2$ where $\Delta E_1$ is estimated from two-state fit to the two-point correlation function. The fit function becomes
\begin{equation}\label{eq:summST}
S^X(T) = c_0 + g_X^\mathrm{bare} T + c_1 T e^{-\Delta E_1 T} + c_2 e^{-\Delta E_1 T}.
\end{equation}
In Fig.~\ref{fig:2stateS_summary}, we show the results of such a fit for all three charges.
As before, we fix $T_\mathrm{max}/a=12$ and vary $T_\mathrm{min}$. The results in Fig.~\ref{fig:2stateS_summary} show that the fits for the different charges are stable although with relatively large statistical errors. The endcaps of the error bars refer to the resulting statistical uncertainties when fixing $\Delta E_1$ in Eq.~\eqref{eq:summST} to its central value whereas the vertical lines of the error bars result from taking the uncertainties in $\Delta E_1$ into consideration when evaluating the fit in Eq.~\eqref{eq:summST}. We see that fixing $\Delta E_1$ to its central value has little to no effect on the final results.

\subsection{Two-state fit of the ratio}\label{sec:2stateR}

In this section, we study including the contribution from a single excited state when fitting the ratio, $R^X(\tau,T)$. This is performed using the fit function
\begin{equation}\label{eq:2stateR}
R^X(\tau,T) = g_X^{\mathrm{bare}} + b_X \left(e^{-\Delta E_1\tau} + e^{-\Delta E_1(T-\tau)}\right) + b'_X e^{-\Delta E_1 T}.
\end{equation}
Here, $\Delta E_1$ is estimated from two-state fit to the two point function.
We perform the stability analysis for this method by 
fitting to all points with $\tau \in [\tau_0,T-\tau_0]$ and varying $\tau_0$.
As previously noted, the ratios appear smooth starting from $\tau/a=1$
and therefore we choose to start from $\tau_0/a=1$ in order to judge
the approach to a plateau. However, for our final selection of results
in the next subsection, we will not use $\tau_0/a$ smaller than 3.

\begin{figure}
\begin{center}
\includegraphics[width=0.49\textwidth]{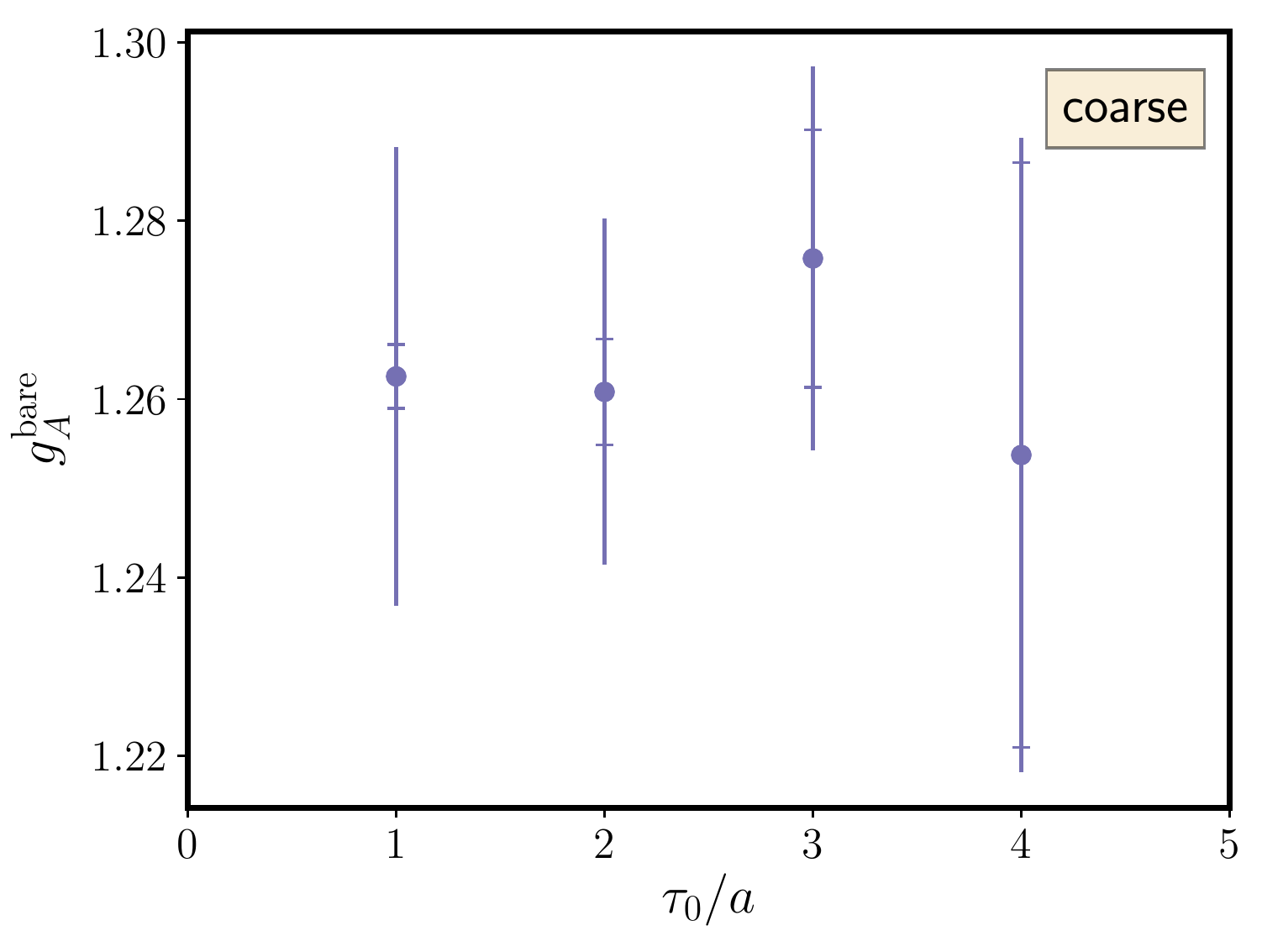}
\hspace{0.001\textwidth}
\includegraphics[width=0.49\textwidth]{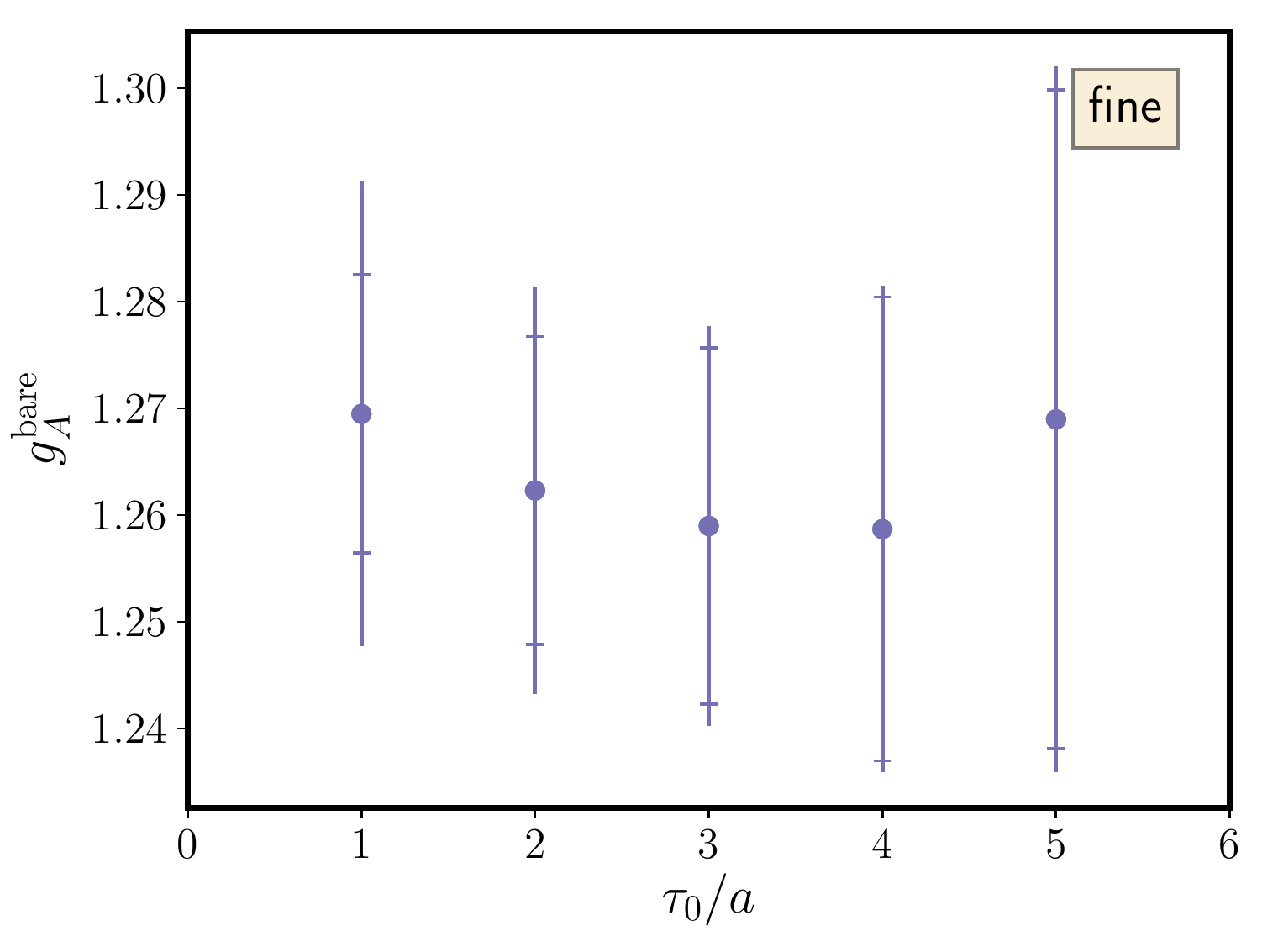}
\\
\includegraphics[width=0.49\textwidth]
{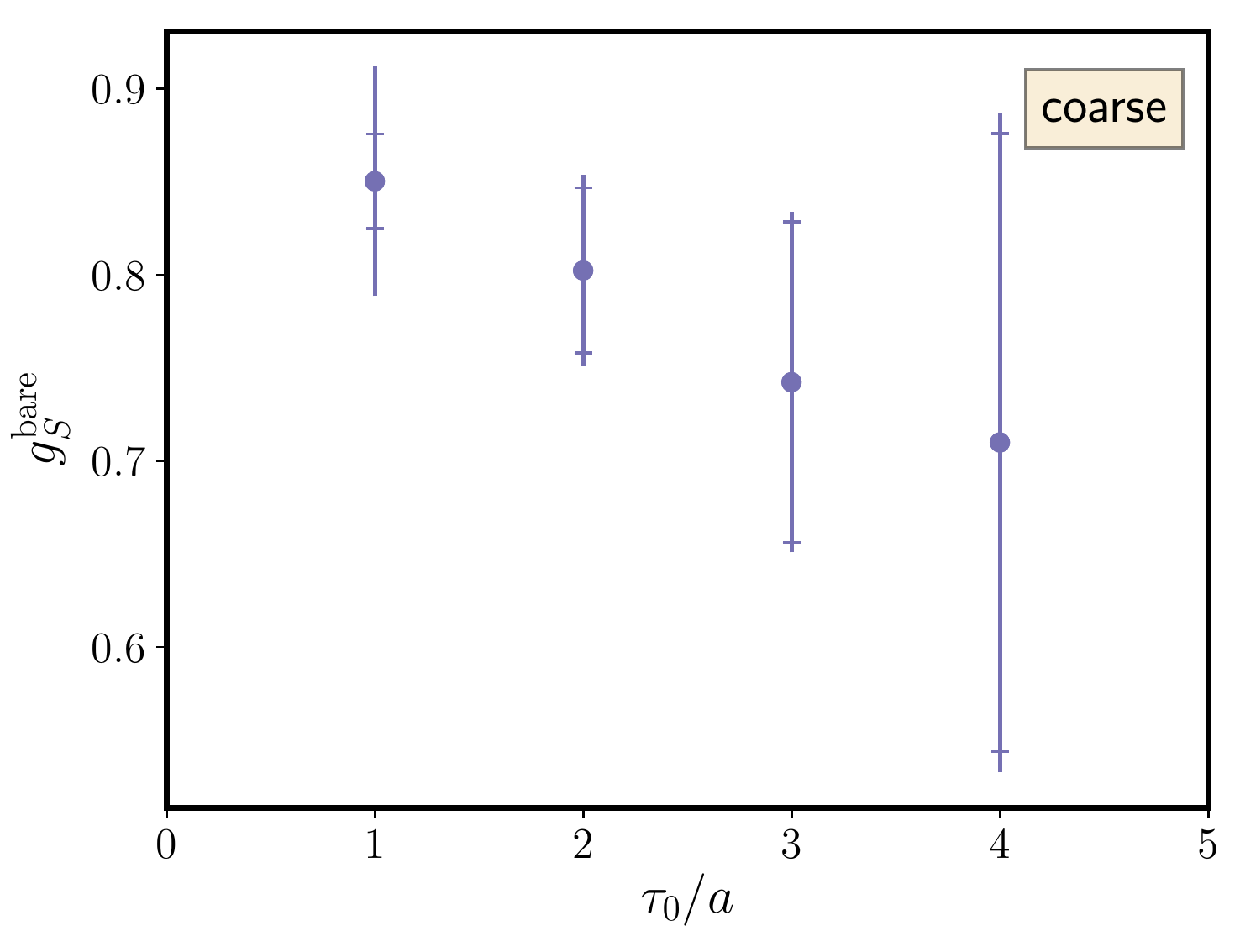}
\hspace{0.001\textwidth}
\includegraphics[width=0.49\textwidth]
{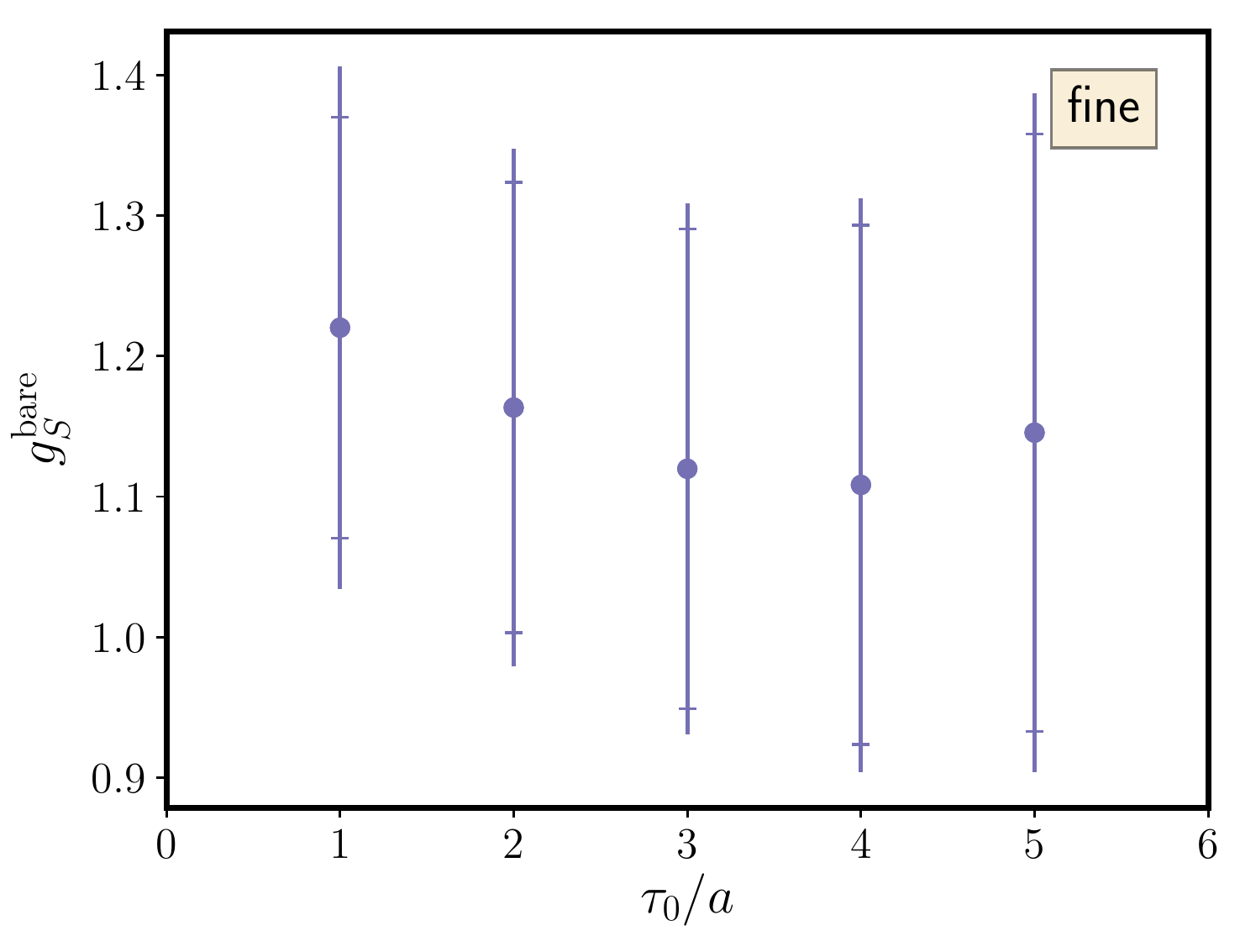}
\\
\includegraphics[width=0.49\textwidth]
{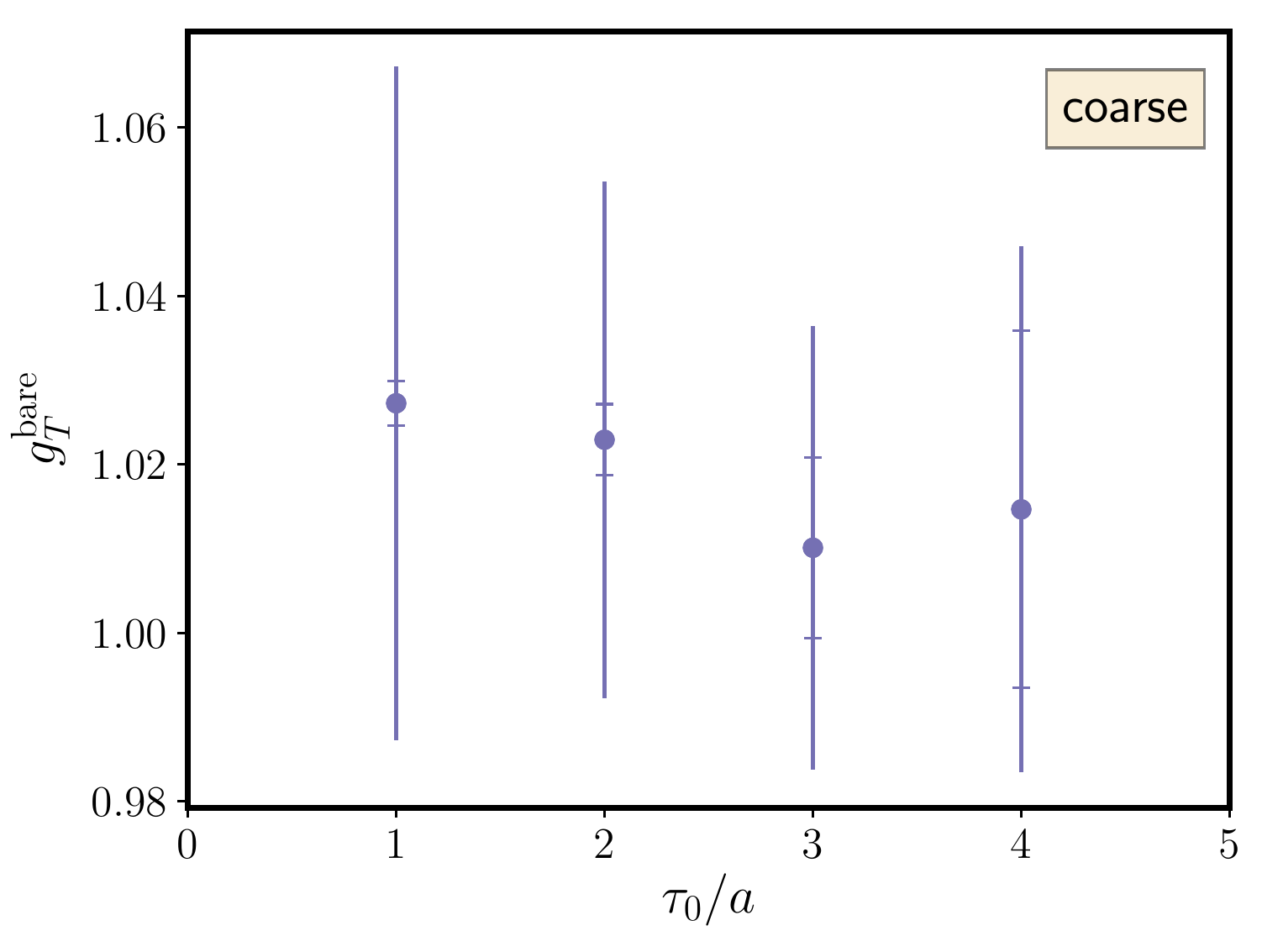}
\hspace{0.001\textwidth}
\includegraphics[width=0.49\textwidth]
{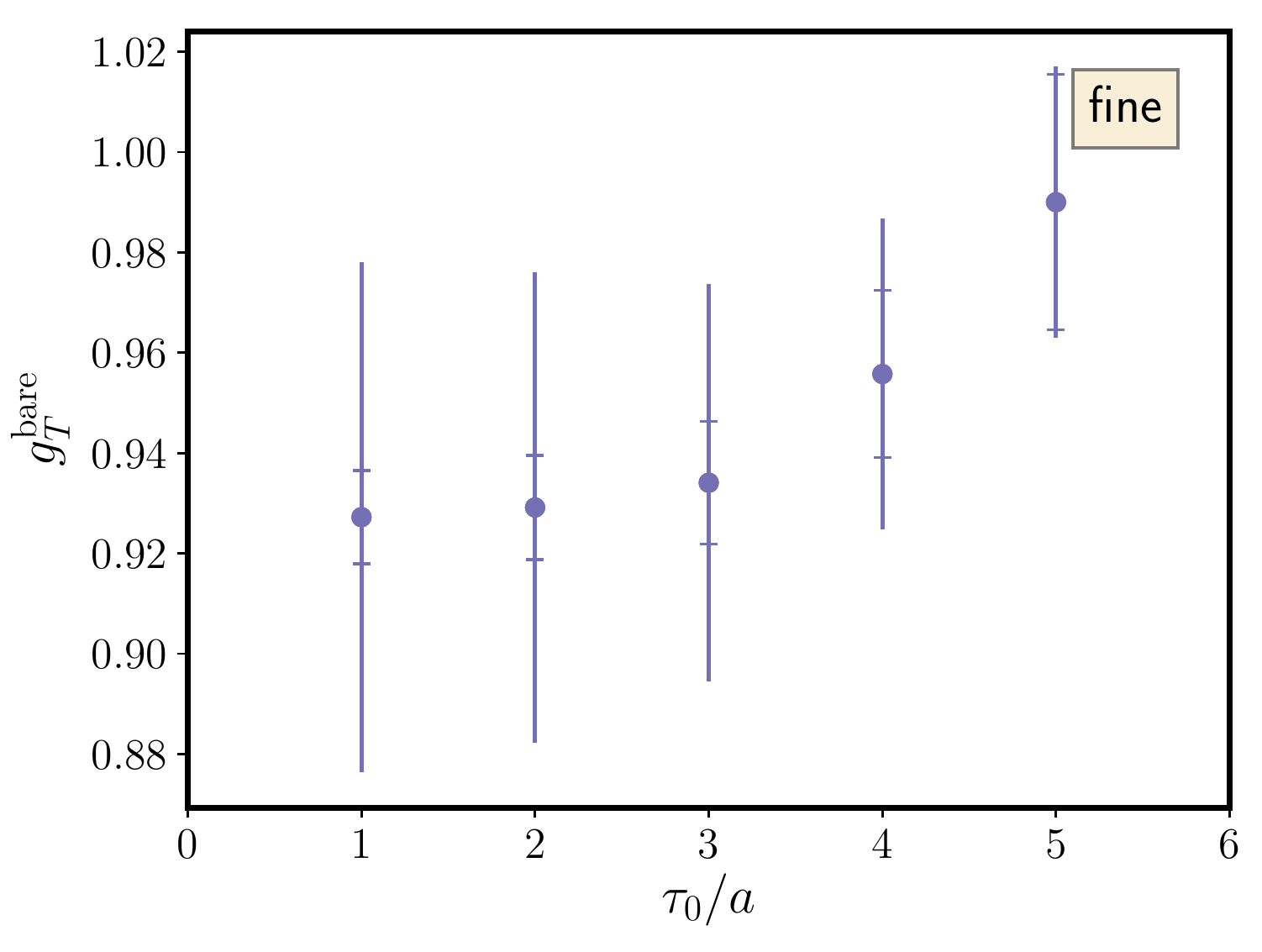}
\end{center}
  \caption{Estimates of the unrenormalized isovector axial, scalar, and tensor charges from the two-state fit to $R^X(\tau,T)$ as functions of $\tau_0$ for the coarse and fine ensembles. The inner error bars (endcaps) refer to the resulting statistical uncertainties when fixing $\Delta E_1$ in Eq.~\eqref{eq:2stateR} to its central value whereas the outer error bars (vertical lines) result from taking the uncertainties in $\Delta E_1$ into consideration when evaluating the fit in Eq.~\eqref{eq:2stateR}.}
  \label{fig:2stateR_summary}
\end{figure}

The circles with the outer statistical uncertainties in the plots of Fig.~\ref{fig:2stateR_summary} show the resulting unrenormalized isovector charges as we vary $\tau_0$ for the coarse (left column) and fine (right column) ensembles. 
The fit range includes source-sink separations satisfying $T\geq 2\tau_0$; this means that for the coarse ensemble, as $\tau_0$ is increased the shorter source-sink separations (which have the most precise data) will be excluded from the fit.
We notice that for $g_A^\mathrm{bare}$, there is no 
significant dependence on $\tau_0$. 
The estimates for $g_S^\mathrm{bare}$ show a noisier signal on the fine ensemble. The signal for $g_T^\mathrm{bare}$ on the fine ensemble shows an upward trend in the central value for increased $\tau_0$ while the statistical uncertainties are decreasing; this is normally not expected, whereas the signal on the coarse ensemble shows no to little dependence on $\tau_0$. 
The inner error bars of Fig.~\ref{fig:2stateR_summary} show the uncertainties when $\Delta E_1$ is fixed to its central value.
The difference between the inner and outer statistical uncertainties for the axial charge shows that the uncertainty on the energy gap makes a large contribution to the uncertainties of the final results, particularly when including small time separations in the fit. This may be because the small time separations are more sensitive to the model parameters used to remove excited-state contributions.
This observation applies also to the tensor charge but less to the scalar charge.

\subsection{Combining different analyses}\label{sec:final_analyis}
We have so far applied four methods for analyzing the excited-state contributions to the different observables on each ensemble, namely
\begin{enumerate}
\item Ratio method
\item Summation method
\item Two-state fit to $R^X(\tau,T)$
\item Two-state fit to $S^X(T)$ (on the coarse ensemble)
\end{enumerate}
For each method, we have plotted the estimated charges as functions of a Euclidean time separation $T/2$, $T_\mathrm{min}$, or $\tau_0$, which we will generically call $\delta t$.
For each bare charge, we want to choose a preferred $\delta t$ for each method and then combine the results from all methods to obtain a final result. In order to reduce the number 
of case-by-case decisions, we have devised a procedure that we will follow to accomplish this.
Our procedure is designed to fulfill the following requirements ordered in decreasing importance:
\begin{itemize}
\item Fit ranges with poor fit quality are excluded, since that indicates the data are not compatible with the fit model and therefore the fit result is not trustworthy.
\item Estimations should be taken from the asymptotic plateau regime, where there is no significant dependence on $\delta t$.
\item Smaller statistical uncertainties are preferred.
\item Larger time separations are preferred so that we reduce the residual excited-state contamination.
\end{itemize}

In the following, we outline the first part of the procedure which aims to find a preferred $\delta t$ from each analysis method. 
\begin{enumerate}
\item If data are obtained from fits (all methods except the ratio method), we start from the smallest $\delta t$, $\delta t_\mathrm{min}$, and increase it until the fit quality is good. The criterion is for the fit to have a $p$-value greater than $0.02$. We call the smallest $\delta t$ that fulfills this criterion $\delta t_0$.
\item We fit the data starting from $\delta t_0$ with a constant and test if the $p$-value of that fit is greater than $0.05$.
We increase $\delta t$ until this is the case. We name the smallest $\delta t$ that fulfills this requirement $\delta t_1$.
\item In order to make sure that we are well inside a plateau region, we take $\delta t_2 = \delta t_1 + 0.2$~fm. Rounded to the nearest lattice spacing, this corresponds to the addition of $2a$ on each ensemble.
\item We find the data point with $\delta t \geq \delta t_2$ that has the smallest statistical uncertainty. We denote this point as $\delta t_3$.
\item Starting from the largest available $\delta t$, we decrease $\delta t$ until we find a data point with uncertainty no more than $20\%$ larger than the uncertainty at $\delta t_3$. We consider this data point to be the final estimation for the analysis method under consideration. We name the time separation at this point $\delta t_f$. The motivation here is that for 
points of similar statistical uncertainty, larger $\delta t$ is preferred 
because of the reduced residual excited-state contamination.

\end{enumerate}
\begin{table}[!htbp]
\centering
\begin{tabular}{l | c | c c c c | c c c c |c c c c }
Method   & $\delta t_\mathrm{min}$ & $\delta t_0$ &  $\delta t_1$& $\delta t_f$  & $g_A^{\mathrm{bare}}$  & $\delta t_0$ &  $\delta t_1$&   $\delta t_f$   & $g_S^{\mathrm{bare}}$   & $\delta t_0$ &  $\delta t_1$& $\delta t_f$ & $g_T^{\mathrm{bare}}$  \\
\hline
\hline
Ratio       & $1.5 a$ &  & $3.5a$ & $6a$  & 1.268(38) & & $2a$ & $4a$ &0.730(62) &   & $5a$& - &- \\
Summation   & $3a$ & $4a$ & $4a$ & $6a$  & 1.284(17) & $3a$ & $3a$ & $6a$ & 0.77(12)  & $5a$ & $5a$& $7a$ & 1.034(17) \\
Two-state fit to $R^X(\tau,T)$ & $1a$ & $1a$ & $1a$ & $3a$ & 1.276(22) &  $1a$ & $1a$ & $3a$ & 0.742(91) & $1a$ & $1a$ & $4a$ & 1.015(31)\\
Two-state fit to $S^X(T)$ & $3a$ &$3a$ & $3a$ &$5a$&1.28(10) & $3a$ & $3a$& $5a$  & 0.93(91) & $3a$& $3a$&$5a$ & 1.050(61) \\
\end{tabular}
\caption{The final estimates for each method and observable on the coarse ensemble.}
\label{tab:summary_coarse}
\end{table}
\begin{table}[!htbp]
\centering
\begin{tabular}{l | c | c c c c | c c c c | c c c c}
Method   & $\delta t_\mathrm{min}$ & $\delta t_0$& $\delta t_1$ &$\delta t_f$  & $g_A^{\mathrm{bare}}$  & $\delta t_0$& $\delta t_1$ & $\delta t_f$ & $g_S^{\mathrm{bare}}$  & $\delta t_0$& $\delta t_1$ & $\delta t_f$ & $g_T^{\mathrm{bare}}$ \\
\hline
\hline
Ratio       & $5a$ & & $5a$& $8a$ & 1.282(33) &&$5a$& $5a$ &0.895(47) & &$6.5a$& - & - \\
Summation   & $10a$ &$10a$&$10a$& $10a$ &1.283(32)  & $10a$ &$10a$& $10a$ &1.25(35) & $10a$ &$10a$& $10a$ &  0.959(24)\\
Two-state fit to $R^X(\tau,T)$ & $1a$ & $1a$ & $1a$ & $4a$  &1.259(23) & $1a$ & $1a$ & $4a$ &1.11(20) & $1a$ & $1a$ & $5a$ & 0.990(27)\\
\end{tabular}
\caption{The final estimates of the charges for each method on the fine ensemble.}
\label{tab:summary_fine}
\end{table}
On the fine ensemble, we do not have small values of $\delta t$ for the ratio and summation methods. When $\delta t_1=\delta t_\mathrm{min}$, this suggests that the plateau could start earlier than our available data. In this case, we choose to take $\delta t_2$ determined on the coarse ensemble for the same method and same charge, and use it (scaled to account for the different lattice spacings) as $\delta t_2$ on the fine ensemble.

The above procedure gives multiple estimates for each observable: at most one from each method. The obtained estimates of the charges for the coarse and fine ensembles are listed in Tab.~\ref{tab:summary_coarse} and Tab.~\ref{tab:summary_fine}, respectively. 
In those tables, we also outline for each case the obtained $\delta t_\mathrm{min}$ which is the smallest available $\delta t$, $\delta t_0$ resulting from the first step in the above procedure and $\delta t_1$ from the second step.
For cases where $\delta t_\mathrm{min}=\delta t_1$, this indicates that there is no significant residual excited-state contamination. This is always the case for two-state fits, indicating that the data are compatible with the single-excited-state model.
There are cases in the two tables where we have no remaining data after the second or the third step of the above procedure to define a $\delta t_f$ and therefore we leave those fields empty, as no reliable result could be obtained.
We notice that we obtain similar $\delta t_1$ for the ratio and summation methods which indicates that it is appropriate to compare ratios at separation $T$ with summation points at $T_\mathrm{min}=T/2$. In this case (and it can be seen in Figs.~\ref{fig:gA_plat_summary}--\ref{fig:gT_plat_summary}), the summation method provides more precise results than the ratios; this is in contrast to the usual comparison of ratio at separation $T$ and summation at $T_\mathrm{min}=T$, which
finds that the summation method has larger uncertainties.
The values for the axial, scalar, and tensor charges in both tables show consistency within error bars between the different methods.
The statistical uncertainties differ between the different fit strategies; in particular we obtain relatively large error bars for the scalar charge on both ensembles.

For obtaining a final estimate of the charges, we combine the different analysis methods by performing a weighted average
to determine the central value. The statistical uncertainty is determined using bootstrap resampling.
We test the compatibility of the central value with the set of analysis methods using a correlated $\chi^2$. If the reduced $\chi^2$ is greater than one, then this indicates the different analysis methods are not in agreement, and the corresponding systematic uncertainty can be accounted for by scaling the statistical uncertainty by $\sqrt{\chi^2/\mathrm{dof}}$. We list our final estimates of the charges on both ensembles in Tab.~\ref{tab:summary_final}. In this table, the given uncertainties are obtained from bootstrap resampling and all the $\chi^2$ values are acceptable. We obtain the largest $\chi^2/\mathrm{dof}=1.04$ for $g_S^\mathrm{bare}$ from the fine ensemble. 

\begin{table}
\begin{tabular}{l | c | c | c  }
Ensemble & $g_A^{\mathrm{bare}}$  & $g_S^{\mathrm{bare}}$ &$g_T^{\mathrm{bare}}$  \\
\hline \hline
Coarse & 1.282(17) & 0.740(74) & 1.029(20)\\
Fine & 1.271(24) & 0.913(54) &  0.972(23)
\end{tabular}
\caption{Our final estimates of the charges on the coarse and fine ensembles.}
\label{tab:summary_final}
\end{table}

\section{Nonperturbative renormalization}
\label{sec:NPR}

We determine renormalization factors for isovector axial, scalar, and
tensor bilinears using the nonperturbative Rome-Southampton
approach~\cite{Martinelli:1994ty}, in both
RI$'$-MOM~\cite{Martinelli:1994ty, Gockeler:1998ye} and
RI-SMOM~\cite{Sturm:2009kb} schemes, and (for the scalar and tensor
bilinears) convert and evolve to the $\MSbar$ scheme at scale 2~GeV
using perturbation theory. Our primary data are the Landau-gauge quark
propagator
\begin{equation}
  S(p) = \int d^4x\, e^{-ip\cdot x}
  \left\langle \psi(x) \bar\psi(0) \right\rangle,
\end{equation}
where $\psi$ is a $u$ or $d$ field, and the Landau-gauge Green's
functions for operator $\mathcal{O}$,
\begin{equation}
  G_\mathcal{O}(p',p) = \int d^4x' d^4x\,e^{-ip'\cdot x'}e^{ip\cdot x}
  \left\langle \psi(x') \mathcal{O}(0) \bar\psi(x) \right\rangle.
\end{equation}
In our case, $\mathcal{O}$ is an isovector quark bilinear and there is
only one Wick contraction, which is a connected diagram. We evaluate
both of these objects using four-dimensional volume plane-wave
sources, yielding an average over all translations in the lattice
volume. From these, we construct our main objects, the amputated
Green's functions,
\begin{equation}
  \Lambda_\mathcal{O}(p',p) = S^{-1}(p') G_\mathcal{O}(p',p) S^{-1}(p).
\end{equation}
These renormalize as $\Lambda_\mathcal{O}^R =
(Z_\mathcal{O}/Z_\psi)\Lambda_\mathcal{O}$. We will not determine
$Z_\psi$ directly; instead, we will take ratios to determine
$Z_\mathcal{O}/Z_V$ and compute $Z_V$ from pion three-point functions.

\subsection{Conditions and matching}

The RI$'$-MOM scheme uses kinematics $p'=p$, whereas RI-SMOM uses
$p^2=(p')^2=q^2$, where $q=p'-p$. In both cases the scale is defined
as $\mu^2=p^2$. Note that a comparison of RI-MOM and RI-SMOM
renormalization was previously done using chiral fermions
in Ref.~\cite{Bi:2017ybi}.

For the vector current, the operator is
$V_\mu=\bar\psi\gamma_\mu\psi$. In RI$'$-MOM, the renormalization
condition is
\begin{equation}
  \frac{1}{36}\Tr\left[ \Lambda_{V_\mu}^R(p,p)\gamma_\nu P_{\mu\nu}\right] = 1,
\end{equation}
where $P_{\mu\nu}=\delta_{\mu\nu}-p_\mu p_\nu/p^2$ is the projector
transverse to $p$, and for RI-SMOM the condition is
\begin{equation}
  \frac{1}{12q^2}\Tr\left[ q_\mu \Lambda_{V_\mu}^R(p',p)\slashed{q} \right] = 1.
\end{equation}
Imposing the vector Ward identity, both of these imply that the quark
field renormalization condition must be
\begin{equation}
  \frac{-i}{12p^2}\Tr\left[ S_R^{-1}(p) \slashed{p} \right] = 1,
\end{equation}
although we do not evaluate this explicitly.

For the axial current, the operator is
$A_\mu=\bar\psi\gamma_\mu\gamma_5\psi$. In RI$'$-MOM, the condition is
\begin{equation}
  \frac{1}{36}\Tr\left[ \Lambda_{A_\mu}^R(p,p)\gamma_5\gamma_\nu P_{\mu\nu}\right] = 1,
\end{equation}
and for RI-SMOM, it is
\begin{equation}
  \frac{1}{12q^2}\Tr\left[ q_\mu \Lambda_{A_\mu}^R(p',p)\gamma_5\slashed{q} \right] = 1.
\end{equation}
Each of these is related by a chiral rotation to the corresponding
condition on the vector current. This implies that in the chiral
limit, the renormalized axial current will satisfy the axial Ward
identity, and therefore no matching to $\MSbar$ is needed.

For the scalar bilinear, the operator is $S=\bar\psi\psi$. In
RI$'$-MOM, the condition is
\begin{equation}
  \frac{1}{12}\Tr\left[ \Lambda_S^R(p,p) \right] = 1,
\end{equation}
and for RI-SMOM, it has the same form,
\begin{equation}
  \frac{1}{12}\Tr\left[ \Lambda_S^R(p',p) \right] = 1.
\end{equation}
For RI$'$-MOM, the matching to $\MSbar$ is known to three
loops~\cite{Chetyrkin:1999pq,Gracey:2003yr}, and for RI-SMOM it is
known to two loops~\cite{Gracey:2011fb}. The anomalous dimension is
obtained from the quark mass anomalous dimension via
$\gamma_S=-\gamma_m$; we use the four-loop $\MSbar$
result~\cite{Chetyrkin:1997dh,Vermaseren:1997fq}.

We write the tensor operator as
$T_{\mu\nu}=\bar\psi\tfrac{1}{2}[\gamma_\mu,\gamma_\nu]\psi$. In
RI$'$-MOM, Gracey~\cite{Gracey:2003yr} starts from the decomposition
\begin{equation}
  \Lambda_{T_{\mu\nu}}(p,p) = \Sigma^{(1)}_T(p^2) \tfrac{1}{2}[\gamma_\mu,\gamma_\nu]
  + \Sigma^{(2)}_T(p^2)\frac{\slashed{p}}{p^2}\left(\gamma_\mu p_\nu-\gamma_\nu p_\mu\right),
\end{equation}
and then imposes the condition $\Sigma^{(1),R}_T(p^2)=1$. Note that
chiral symmetry breaking allows more terms to appear, but they won't
contribute to any relevant trace. As Gracey notes, this term can be
isolated via
\begin{equation}
  \Sigma^{(1)}_T(p^2) = \frac{-1}{72}\Tr\left[\Lambda_{T_{\mu\nu}}(p,p)\left(
      \tfrac{1}{2}[\gamma_\mu,\gamma_\nu]
    + \frac{\slashed{p}}{p^2}\left(\gamma_\mu p_\nu-\gamma_\nu p_\mu\right)
    \right)\right].
\end{equation}
This can be rewritten to obtain the renormalization condition in a
simple form:
\begin{equation}
  \frac{1}{72}\Tr\left[\Lambda_{T_{\mu\nu}}^R(p,p) \tfrac{1}{2}[\gamma_\beta,\gamma_\alpha] P_{\mu\alpha} P_{\nu\beta} \right] =1.
\end{equation}
For RI-SMOM, the condition is
\begin{equation}
  \frac{1}{144}\Tr\left[ \Lambda_{T_{\mu\nu}}^R(p',p) \tfrac{1}{2}[\gamma_\nu,\gamma_\mu]\right] = 1.
\end{equation}
For RI$'$-MOM, the matching to $\MSbar$ is known to three
loops~\cite{Gracey:2003yr}, and for RI-SMOM it is known to two
loops~\cite{Gracey:2011fb}. We use the four-loop $\MSbar$ anomalous
dimension~\cite{Baikov:2006ai}\footnote{Note that the sign of the
  three-loop term proportional to $N_f^2$ disagrees between the
  proceedings of Baikov and Chetyrkin~\cite{Baikov:2006ai} and the
  first three-loop calculation, done by
  Gracey~\cite{Gracey:2000am}. However, in an appendix of a later
  publication by Chetyrkin and Maier~\cite{Chetyrkin:2010dx}, the sign
  agrees with Gracey, and therefore we use that sign.}.

\subsection{Vector current}

\begin{figure*}
  \centering
  \includegraphics[width=0.495\textwidth]{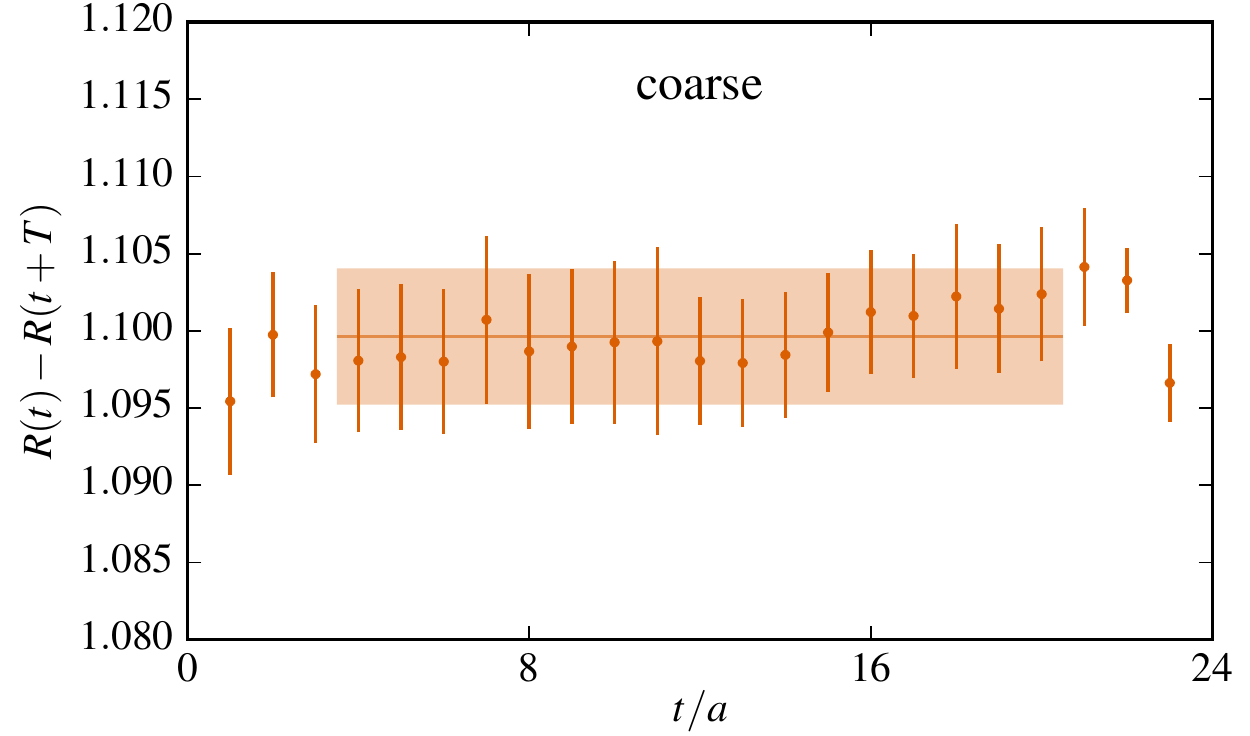}
  \includegraphics[width=0.495\textwidth]{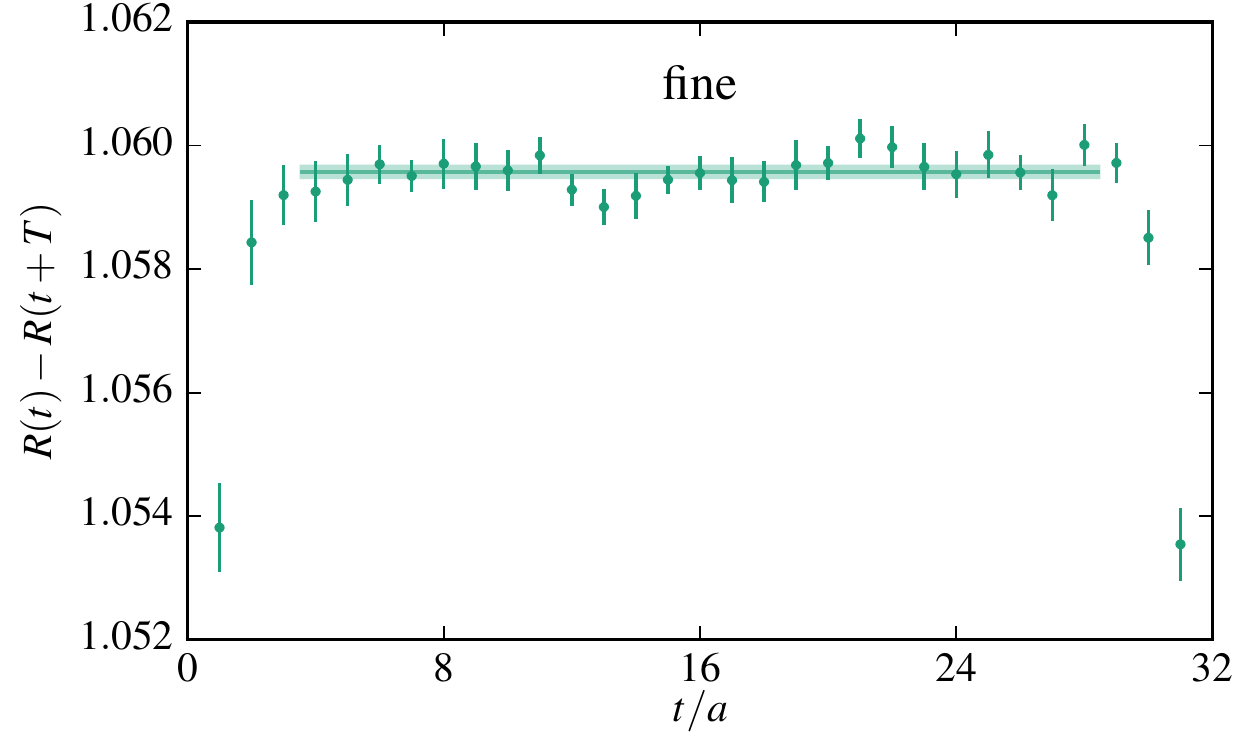}
  \caption{Determination of $Z_V$: coarse ensemble (left) and fine
    ensemble (right). This difference of ratios provides an estimate
    of $Z_V^{-1}$. Note that the vertical scale is a factor of four
    smaller for the fine ensemble.}
  \label{fig:ZV}
\end{figure*}

Following e.g.\ Refs.~\cite{Durr:2010aw,Green:2017keo}, we determine
$Z_V$ by computing the zero-momentum pion two-point function $C_2(t)$
and three-point function $C_3(t)$, where the latter has source-sink
separation $T=L_t/2$ and an operator insertion of the time component
of the local vector current at source-operator separation $t$. The
charge of the interpolating operator gives the renormalization
condition
\begin{equation}
  Z_V\left[ R(t_1) - R(t_2) \right] = 1,
\end{equation}
for $0 < t_1 < T < t_2 < L_t$, where $R(t)=C_3(t)/C_2(T)$. We choose
$t_2=t_1+T$; the difference results in a large cancellation of
correlated statistical uncertainties, so that precise results can be
obtained with relatively low statistics; see
Fig.~\ref{fig:ZV}. Results on the coarse ensemble are much noisier
than on the fine one, although the statistical errors are still below
1\%. We take the unweighted average across the plateau, excluding the
first and last three points. This yields $Z_V=0.9094(36)$ on the
coarse ensemble and $Z_V=0.94378(10)$ on the fine one.

\subsection{Axial, scalar, and tensor bilinears}

We use partially twisted boundary conditions, namely periodic in time
for the valence quarks rather than the antiperiodic condition used for
sea quarks. The plane-wave sources are given momenta
$p=\frac{2\pi}{L}(k,k,k,\pm k)$, $k=2,3,\dots,\frac{L}{4a}$. By
contracting them in different combinations, we get data for both
RI$'$-MOM kinematics, $p'-p=0$, and RI-SMOM kinematics,
$p'-p=\frac{2\pi}{L}(0,0,0,\pm 2k)$. We used 54 gauge configurations
from each ensemble. However, the modified boundary condition rendered
one configuration on the coarse ensemble exceptional and the multigrid
solver was unable to converge; therefore, we omitted this
configuration and used only 53 on the coarse ensemble. In addition, on
the coarse ensemble we also performed a cross-check using different
kinematics, $p,p'\in\{\frac{2\pi}{L}(k,k,0,0),
\frac{2\pi}{L}(k,0,k,0)\}$, which ensure that in the RI-SMOM setup the
components of $p'-p$ are not larger than those of $p$ and $p'$. Since
the primary kinematics have $p$ and $p'$ oriented along a
four-dimensional diagonal and the alternative kinematics have them
oriented along a two-dimensional diagonal, these setups will sometimes
be referred to as 4d and 2d, respectively.

After perturbatively matching the RI$'$-MOM or RI-SMOM data to the
$\MSbar$ scheme and evolving to the scale 2~GeV, there will still be
residual dependence on the nonperturbative scale $\mu^2$ due to
lattice artifacts and truncation of the perturbative series. To
control these artifacts, we perform fits including terms polynomial in
$\mu^2$ and also, following Ref.~\cite{Boucaud:2005rm}, a pole
term. Our fit function has the form $A+B\mu^2+C\mu^4+D/\mu^2$; the
constant term $A$ serves as our estimate of the relevant ratio of
renormalization factors $Z_\mathcal{O}/Z_V$. We also consider fits
without the pole term, i.e.\ with $D=0$. We use two different fit
ranges: 4 to 20~GeV$^2$ and 10 to 30~GeV$^2$.

\begin{figure*}
  \centering
  \includegraphics[width=0.495\textwidth]{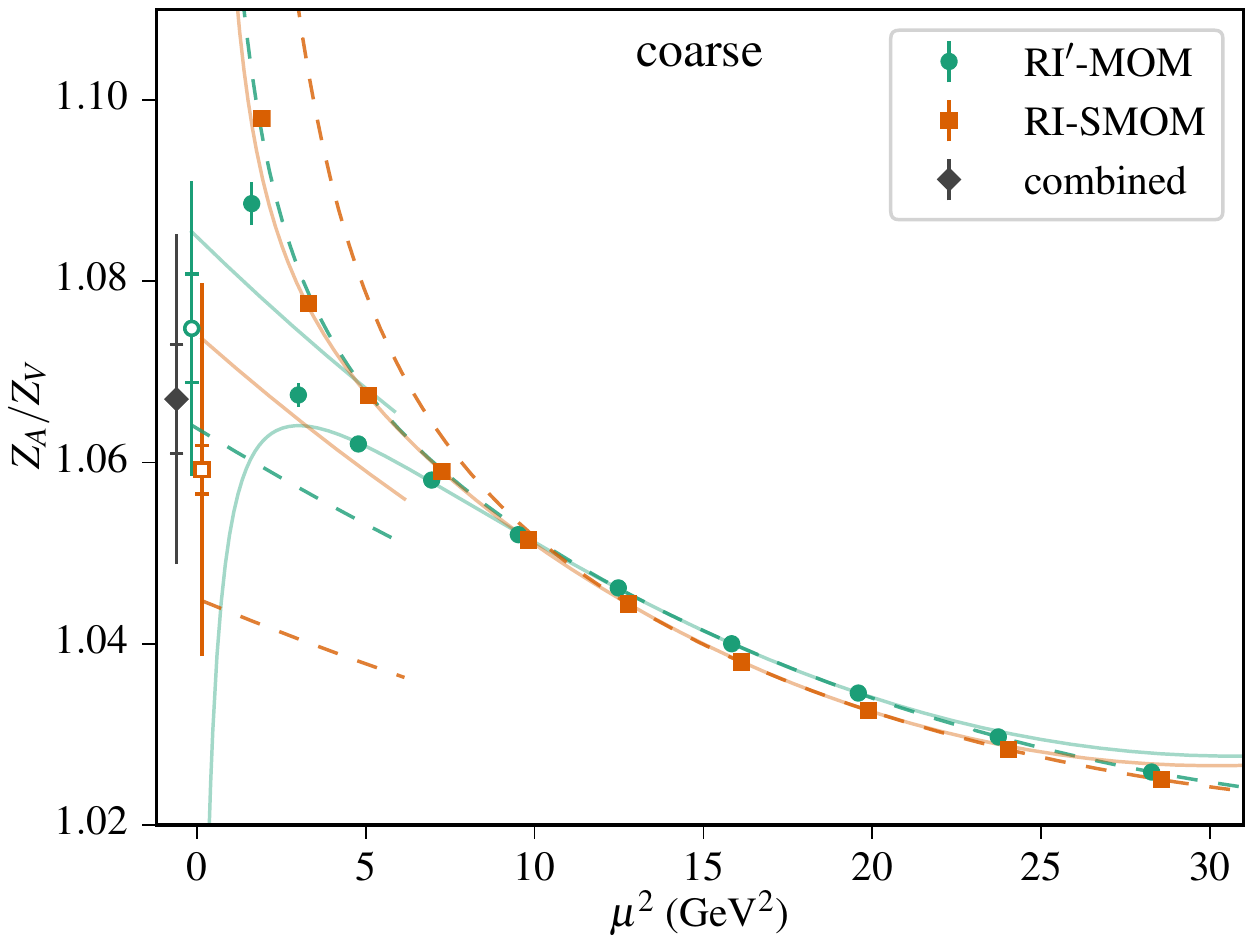}
  \includegraphics[width=0.495\textwidth]{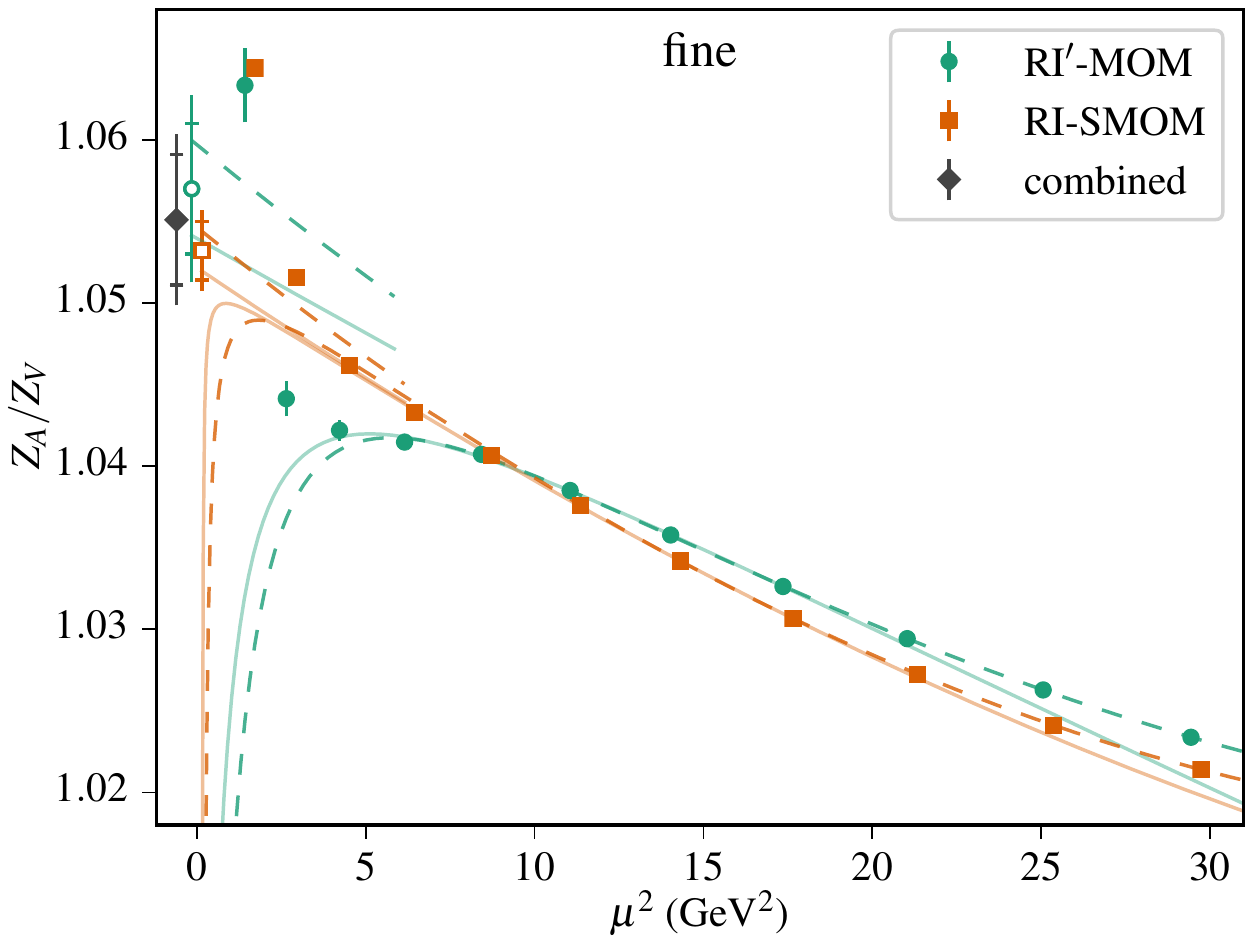}\\
  \includegraphics[width=0.495\textwidth]{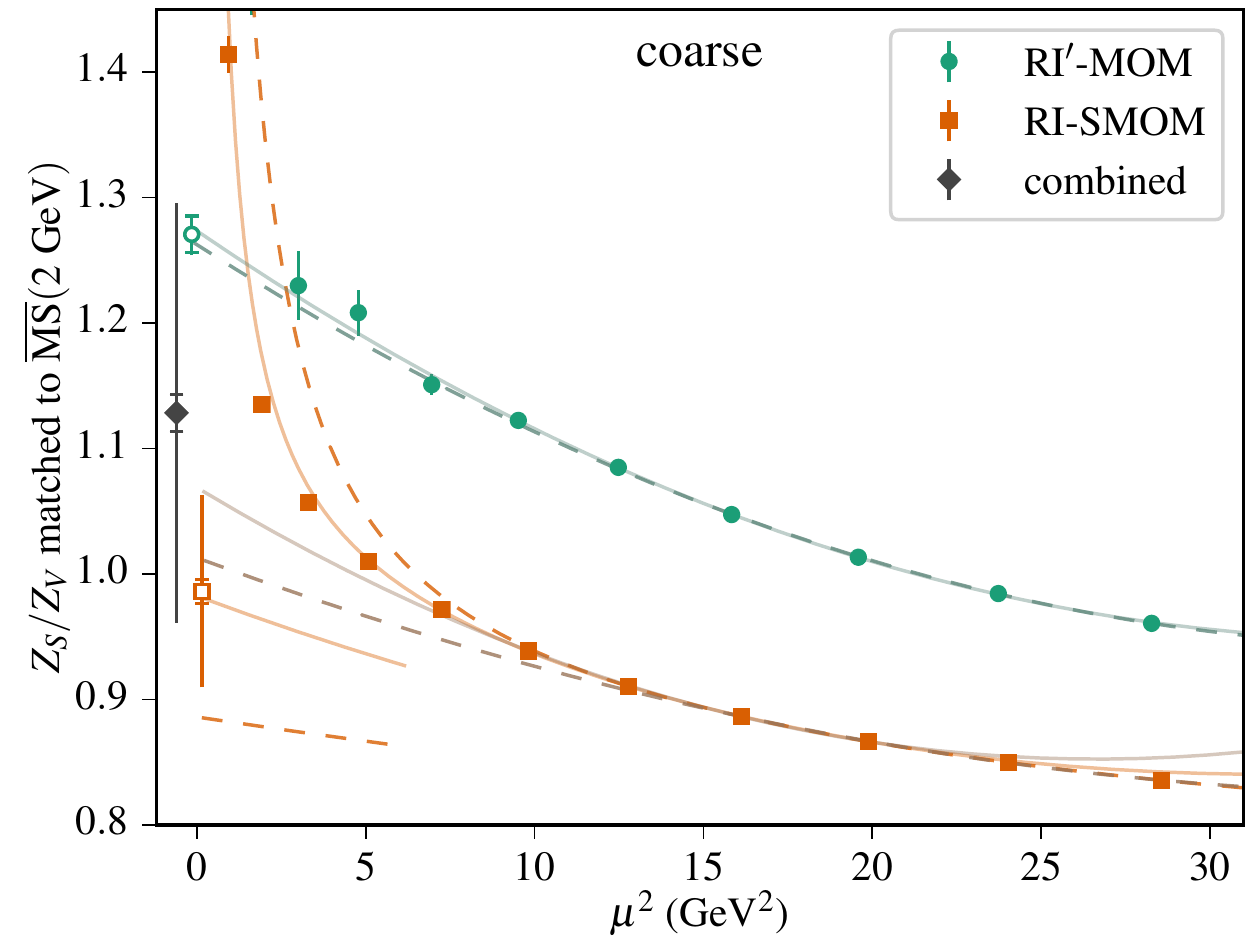}
  \includegraphics[width=0.495\textwidth]{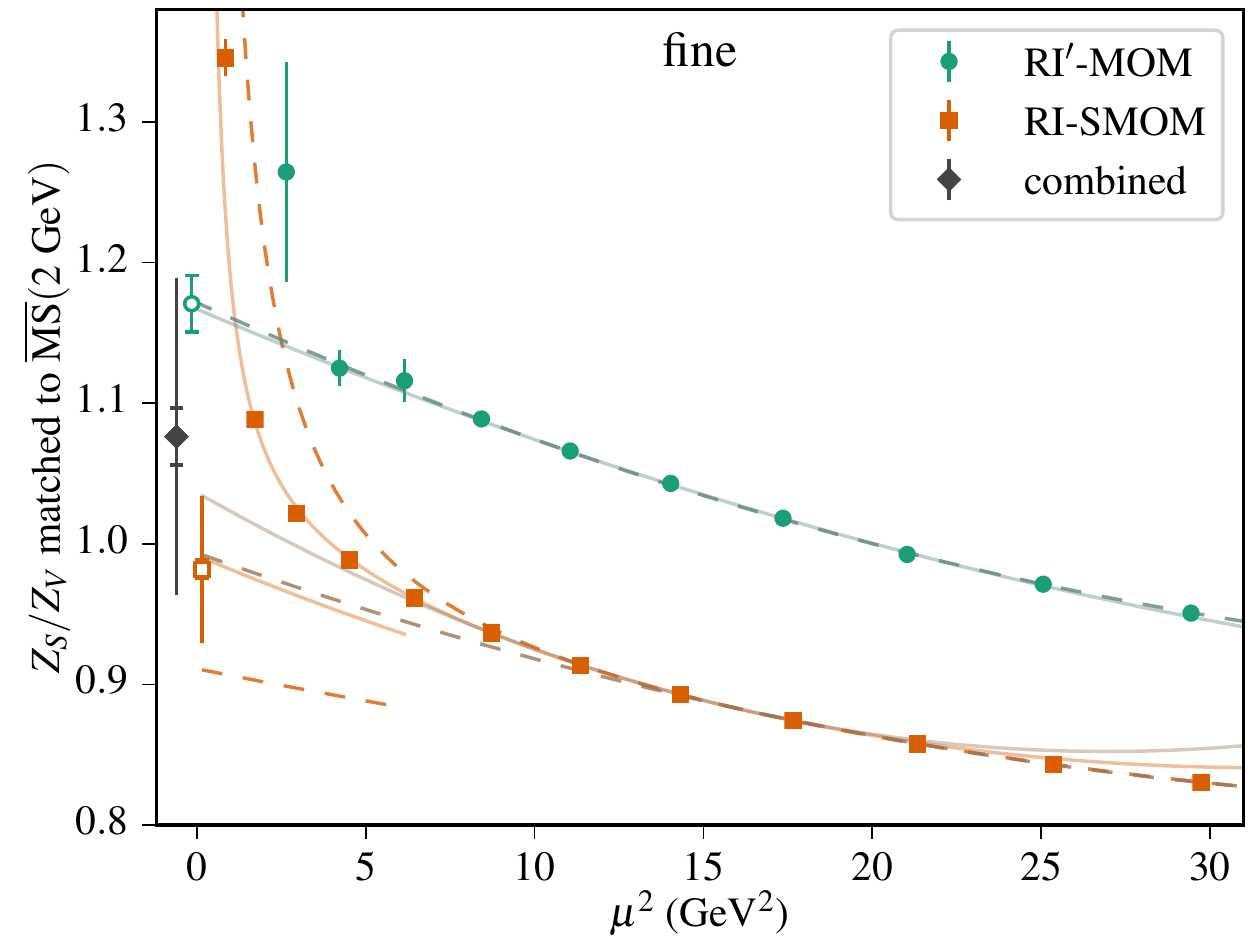}\\
  \includegraphics[width=0.495\textwidth]{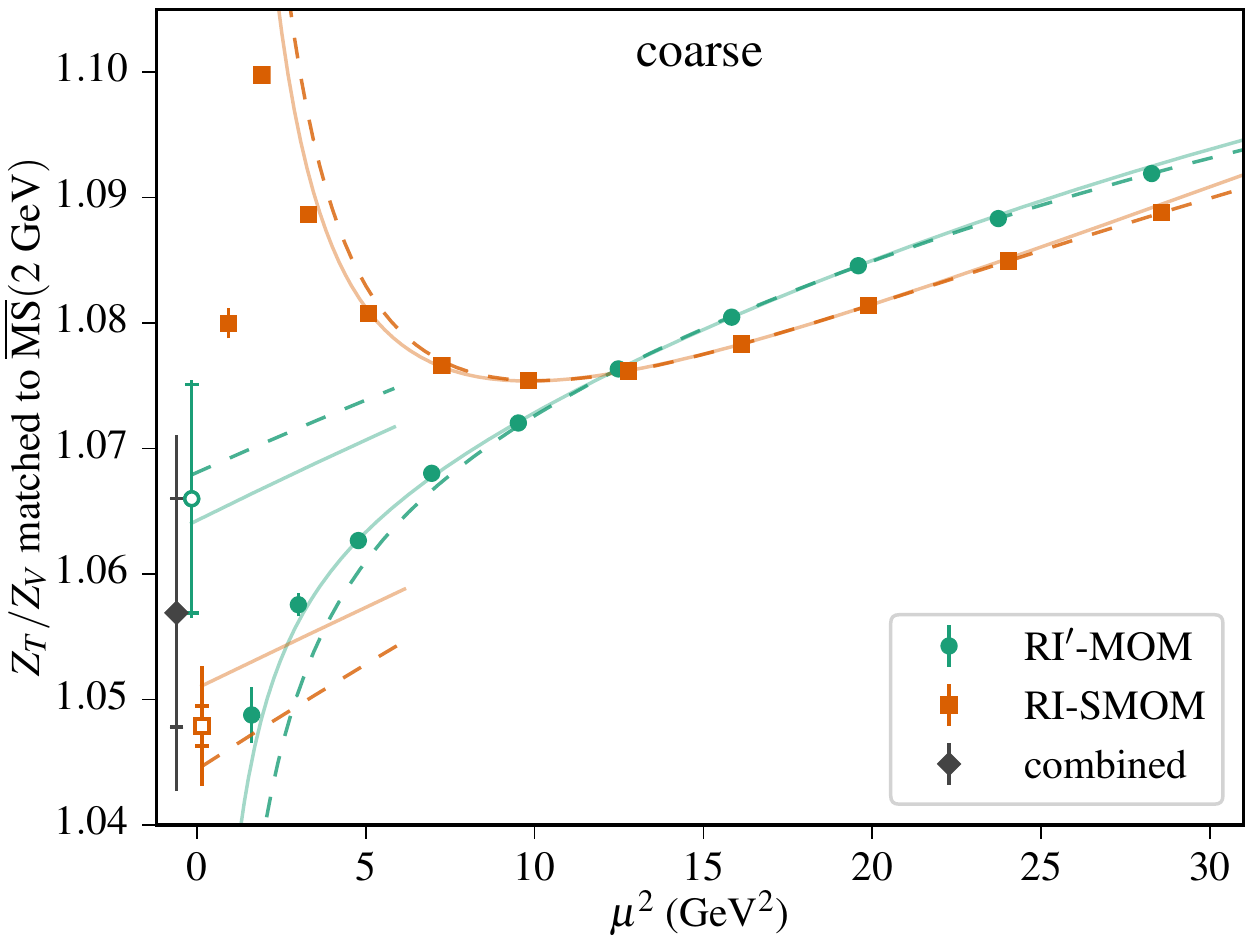}
  \includegraphics[width=0.495\textwidth]{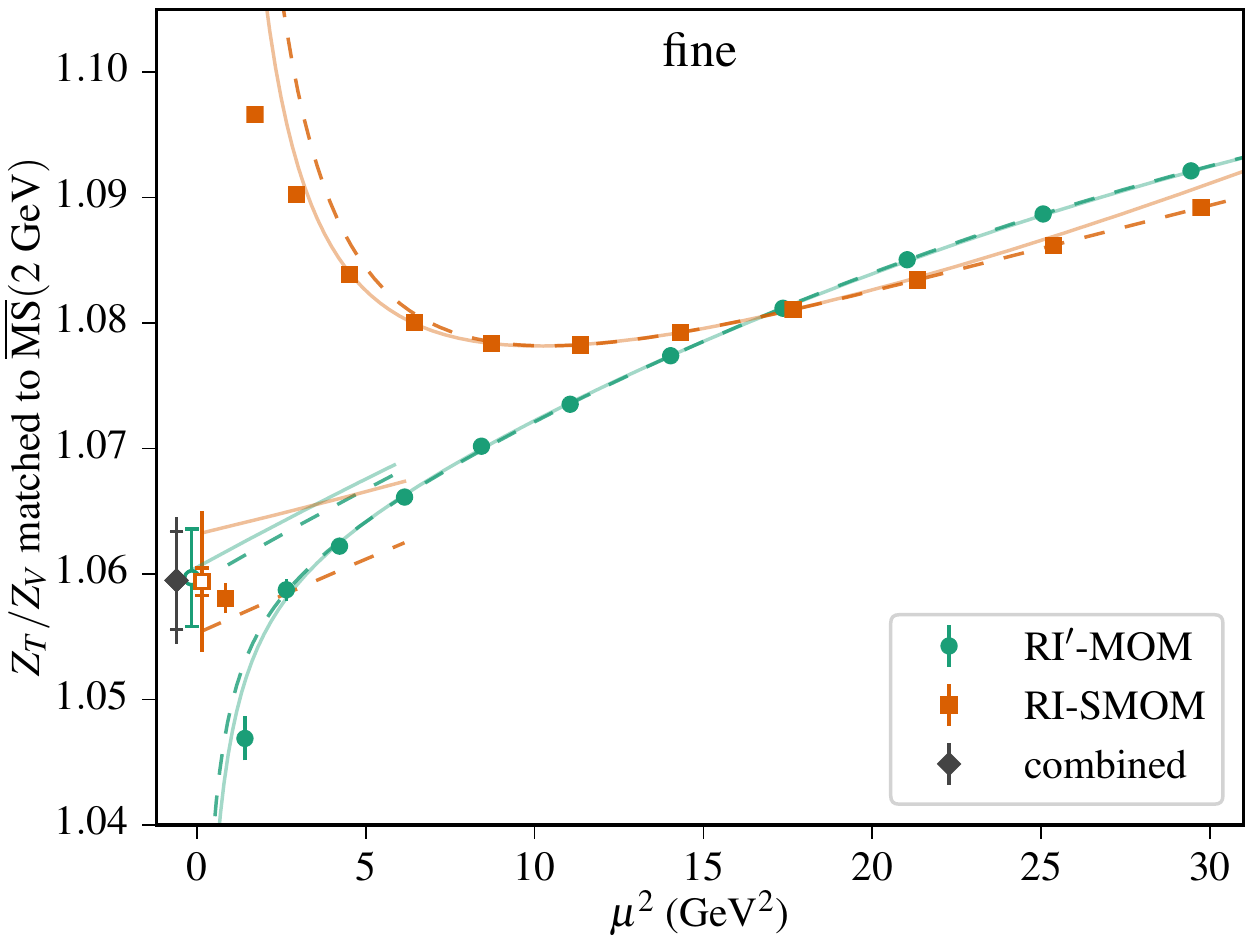}\\
  \caption{Ratios of renormalization factors $Z_A/Z_V$, $Z_S/Z_V$, and
    $Z_T/Z_V$ on the coarse (left) and fine (right) ensembles,
    determined using the RI$'$-MOM (green circles) and RI-SMOM (orange
    squares) intermediate schemes and then matched to $\MSbar$ at
    scale 2~GeV. For most points, the statistical uncertainty is
    smaller than the plotted symbol. The solid curves are fits to the
    $\mu^2$-range from 4 to 20~GeV$^2$, and the dashed curves are fits
    to the range 10 to 30~GeV$^2$. To reduce clutter, uncertainties on
    the fit curves are not shown. For the fits that include a pole
    term, the fit curve without the pole term is also plotted, in the
    range $0<\mu^2<6\text{ GeV}^2$. The fits for $Z_S/Z_V$ without a
    pole term are shown using desaturated colors. The open symbols
    near $\mu^2=0$ provide the final estimate for each intermediate
    scheme; their outer (without endcap) and inner (with endcap) error
    bars show the total and statistical uncertainties. The filled dark
    gray diamonds are the final estimates that combine both schemes.}
  \label{fig:ZX_over_ZV}
\end{figure*}

The main results on the two ensembles are shown in
Fig.~\ref{fig:ZX_over_ZV}. The RI-SMOM data are generally very precise
(more so than the RI$'$-MOM data), which makes the fit quality very
poor in many cases. If the covariance matrix from the RI$'$-MOM data
is used when fitting to the RI-SMOM data, then the fit qualities are
good except for some of the fits without a pole term for the axial and
tensor bilinears. For the RI$'$-MOM data, the fit quality is good when
using a pole term and also good for the scalar bilinear when omitting
the pole term. Therefore, we elect to always include the pole term in
our fits for $Z_A/Z_V$ and $Z_T/Z_V$. For $Z_S/Z_V$ we use fits both
with and without the pole term, however the fit with a pole term to
the RI$'$-MOM data is very noisy and therefore we exclude it.

To account for the poor fit quality for some of the RI-SMOM fits, we
scale the statistical uncertainty of the estimated ratio of
renormalization factors by $\sqrt{\chi^2/\text{dof}}$ whenever this is
greater than one. For each intermediate scheme, we take the unweighted
average of all fit results as the central value, the maximum of the
statistical uncertainties, and the root-mean-square deviation of the
fit results as the systematic uncertainty. We combine results from
both schemes in the same way to produce our final estimates, with the
constraint that both schemes are given equal weight. These estimates
are also shown in Fig.~\ref{fig:ZX_over_ZV}. For $Z_S/Z_V$ there is a
large discrepancy between the two intermediate schemes, which leads to
a large systematic uncertainty. This discrepancy is smaller on the
fine ensemble, suggesting that it is caused by lattice artifacts.

\begin{figure*}
  \centering
  \includegraphics[width=0.495\textwidth]{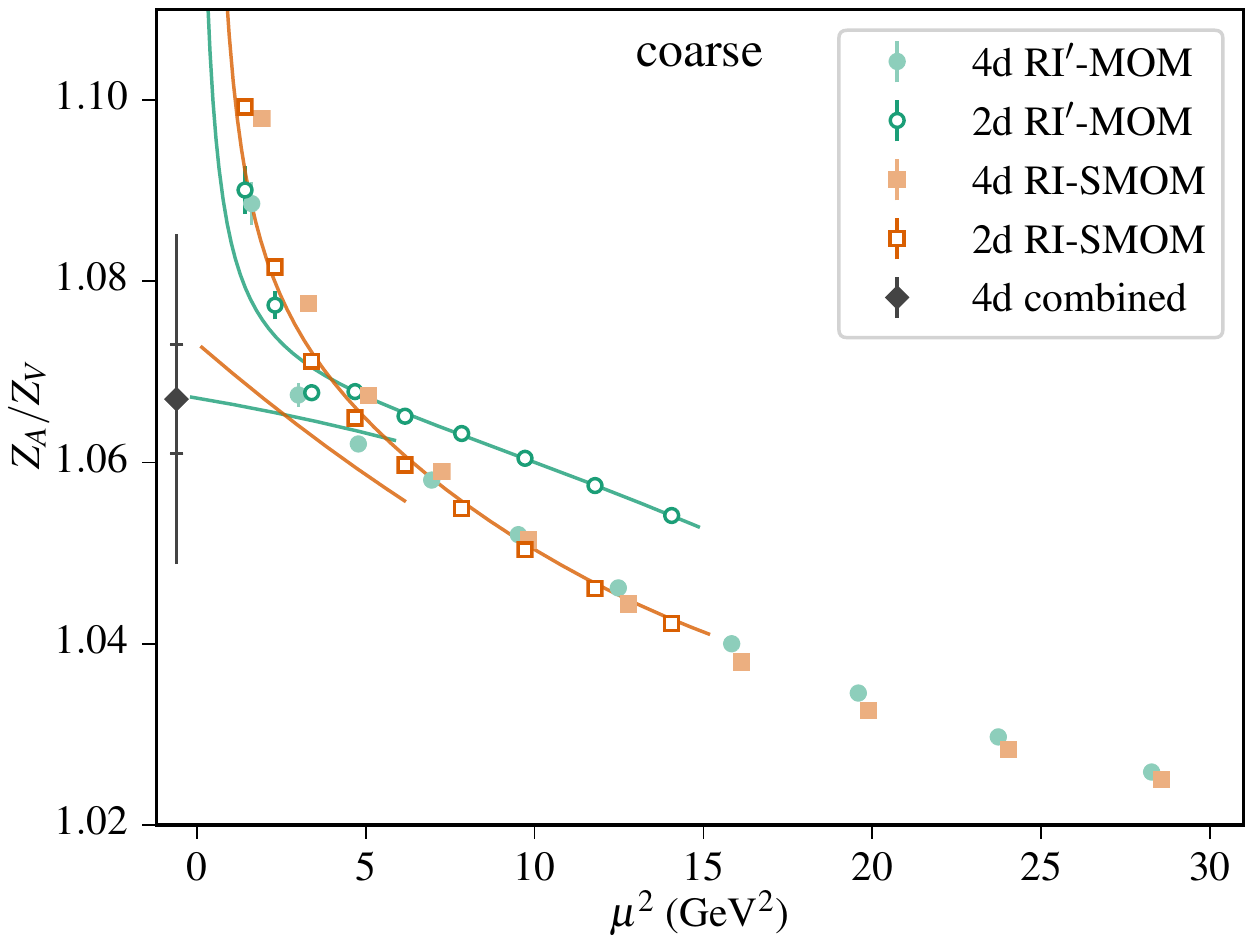}\\
  \includegraphics[width=0.495\textwidth]{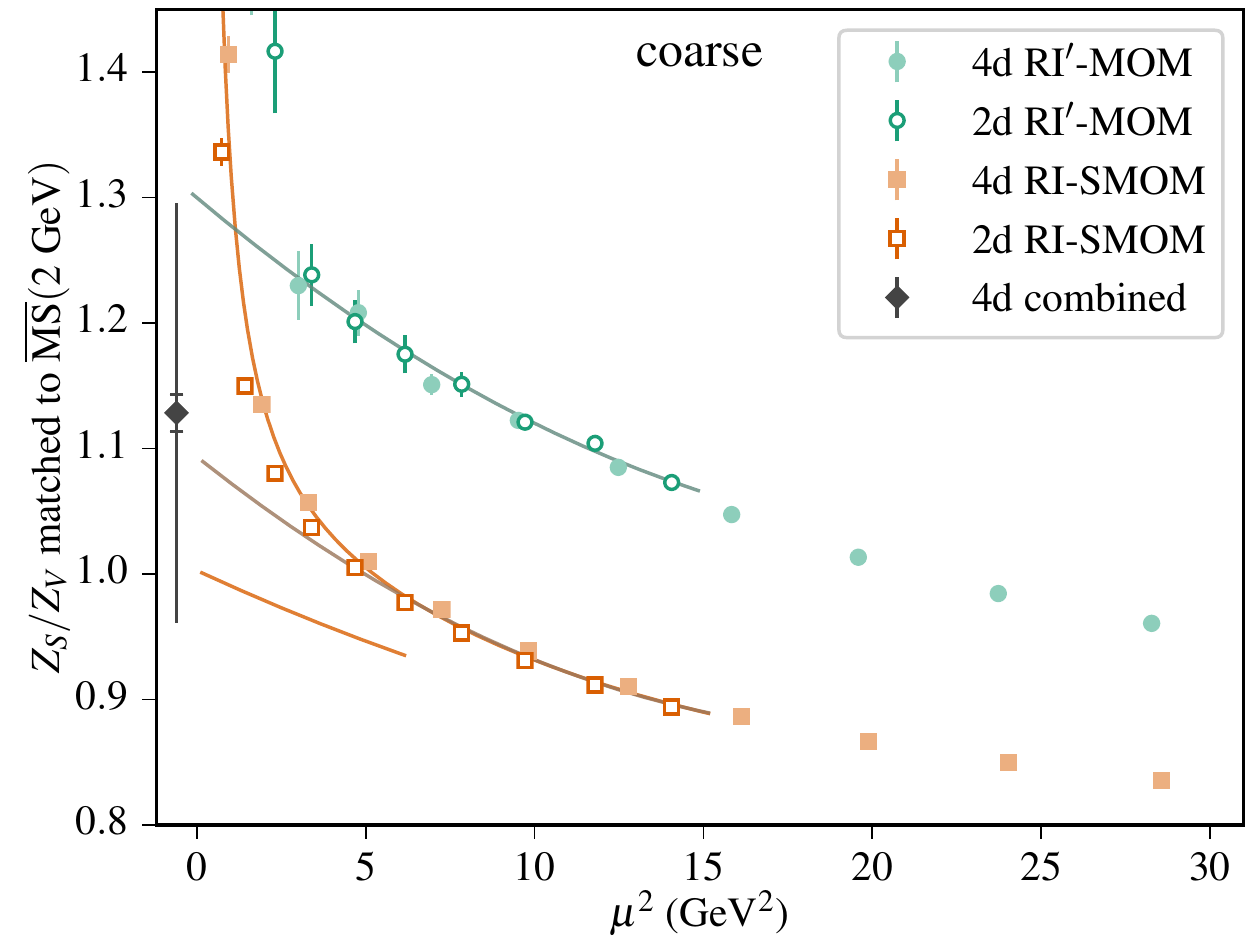}
  \includegraphics[width=0.495\textwidth]{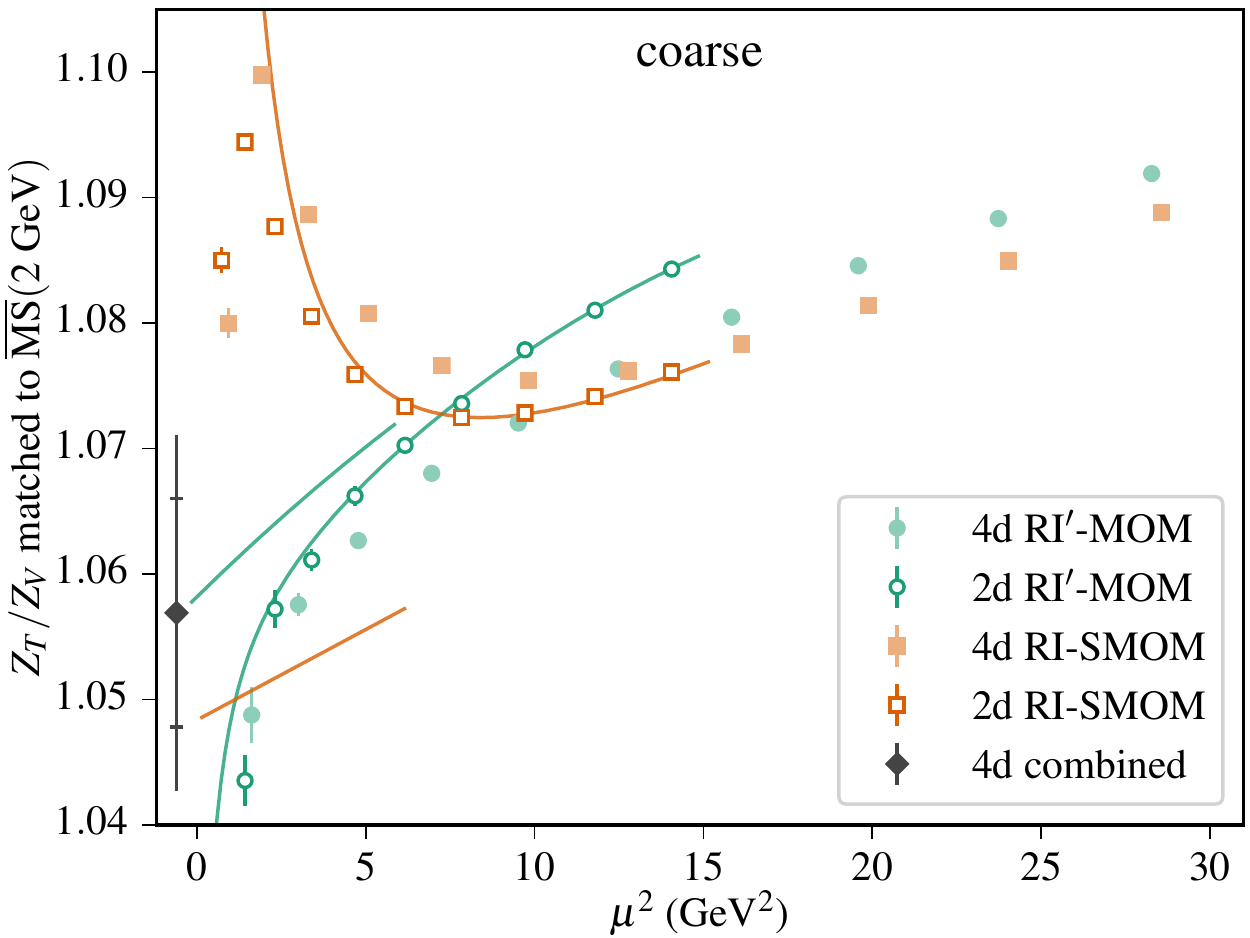}
  \caption{Check of alternative kinematics for ratios of
    renormalization factors on the coarse ensemble. The data with
    momenta along four-dimensional diagonals and the final combined
    estimates are repeated from Fig.~\ref{fig:ZX_over_ZV}. The points
    with open symbols have momenta along two-dimensional diagonals and
    the curves are fits to those points in the $\mu^2$-range from 4 to
    15~GeV$^2$. For the fits that include a pole term, the fit curve
    without the pole term is also plotted, in the range
    $0<\mu^2<6\text{ GeV}^2$. The fits for $Z_S/Z_V$ without a pole
    term are shown using desaturated colors.}
  \label{fig:ZX_check2d}
\end{figure*}

Figure~\ref{fig:ZX_check2d} shows the second set of kinematics on the
coarse ensemble. These data do not reach as high in $\mu^2$;
therefore, we choose to fit to a single range of 4 to 15~GeV$^2$.  We
use the same fit types as for the first set of kinematics, and the
results (which can seen from the values of the curves at $\mu^2=0$)
are consistent with the final estimates from the first set of
kinematics.

\begin{table}
  \centering
\begin{tabular}{r|llll}
         & $Z_V$      & $Z_A$       & $Z_S$        & $Z_T$ \\\hline
coarse   & 0.9094(36) & 0.9703(170) & 1.0262(1521) & 0.9611(134) \\
fine     & 0.9438(1)  & 0.9958(50)  & 1.0157(1065) & 0.9999(48)
\end{tabular}
  \caption{Final estimates of renormalization factors on the two ensembles.}
  \label{tab:Z_factors}
\end{table}

\begin{table}
  \centering
\begin{tabular}{l|lll}
        & $Z_A$           & $Z_S$         & $Z_T$ \\\hline
coarse  & 0.9086(21)(111) & 1.115(17)(30) & 0.9624(62) \\
fine    & 0.9468(6)(56)   & 1.107(16)(22) & 1.011(5) \\\hline
reference & \cite{Durr:2013goa,BMW_privcomm} & \cite{Durr:2010aw} & \cite{Green:2012ej}
\end{tabular}
  \caption{Previously used renormalization factors for this lattice action and these two lattice spacings.}
  \label{tab:Z_factors_prev}
\end{table}

Our final estimates of the renormalization factors, after adding
errors in quadrature, are given in Table~\ref{tab:Z_factors}. The
uncertainty on $Z_S$ is more than 10\% and we obtain percent-level
uncertainties on $Z_A$ and $Z_T$. In our previous publications using
this lattice action~\cite{Green:2012ej,Green:2012ud,Hasan:2017wwt}, we
used different values for these renormalization factors, which are
listed in Table~\ref{tab:Z_factors_prev}. These previous values were
all obtained using an RI$^{(\prime)}$-MOM type scheme. Because of our
large uncertainty, $Z_S$ is in agreement with the previous value. The
latter is also in agreement with our result from only the RI$'$-MOM
scheme. Our result for $Z_T$ is also consistent with the previous
value. However, we find that $Z_A$ is 5--7\% higher than the values
that we previously used, a discrepancy of three standard deviations on
the coarse lattice spacing and more than six on the fine one. The
previous values would imply that $Z_A/Z_V$ is within about one percent
of unity for both lattice spacings, which is very difficult to
reconcile with Fig.~\ref{fig:ZX_over_ZV}. The discrepancy in central
values of $Z_A$ is smaller for the fine lattice spacing than the
coarse one, suggesting that the two determinations could converge in
the continuum limit, although the uncertainties are large enough that
it is also possible the discrepancy has no dependence on the lattice
spacing.

\section{Renormalized charges}
\label{sec:renormalized_charges}

\begin{table}
  \centering
  \begin{tabular}{l|ccc}
    Ensemble & $g_A$ & $g_S$ & $g_T$\\\hline
    coarse & 1.244(28) & 0.759(136) & 0.989(23) \\
    fine   & 1.265(24) & 0.927(112) & 0.972(24)
  \end{tabular}
  \caption{Renormalized charges on the two ensembles.}
  \label{tab:charges_renormalized}
\end{table}

Multiplying the bare charges in Table~\ref{tab:summary_final} by the
renormalization factors in Table~\ref{tab:Z_factors} and adding the
uncertainties in quadrature, we obtain the renormalized charges on the
two ensembles, shown in Table~\ref{tab:charges_renormalized}. The
final values should be obtained at the physical pion mass, in the
continuum and in infinite volume. Since both ensembles have pion
masses very close to the physical pion mass and have large volumes, we
neglect these effects as their contribution to the overall uncertainty
is relatively small. With two lattice spacings, we are unable to fully
control the continuum limit; instead, we choose to account for
discretization effects by taking the central value from the fine
ensemble and quoting an uncertainty that covers the spread of
uncertainties on both ensembles, i.e.\ $\delta g_X = \max(\delta
g_X^f,|g_X^c-g_X^f|+\delta g_X^c)$, where $g_X^c$ and $g_X^f$ denote
the charge computed on the coarse and fine ensembles, respectively. It
should be cautioned that since discretization effects are formally
$O(\alpha_s a)$, which varies by a factor of only about $\frac{3}{4}$
between the two ensembles, it is possible that the uncertainty from
these effects is underestimated; additional calculations with finer
lattice spacings would be needed to improve this. We obtain
\begin{align}
  g_A &= 1.265(49), \label{eq:gA_final}\\
  g_S &= 0.927(303),\label{eq:gS_final}\\
  g_T &= 0.972(41). \label{eq:gT_final}
\end{align}
The overall uncertainties for the axial and tensor charges are roughly
4\%. The scalar charge has a much larger uncertainty, due to
the large uncertainty in the renormalization factor and the large difference in central values
between the two ensembles.

Results on these two ensembles can be compared with our earlier
calculations using the same lattice action and heavier pion
masses~\cite{Green:2012ej,Green:2012ud}, reevaluating those earlier
works based on the more extensive study of excited-state effects in
Section~\ref{sec:estimation_of_bare_charges} and using the
renormalization factors from Section~\ref{sec:NPR}. For $g_A$, the
summation method with $T_\text{min}\approx 0.7$~fm was found to be
acceptable; therefore, we reuse the summation-method results from
Ref.~\cite{Green:2012ud}, which had $T_\text{min}\approx 0.9$~fm. For
$g_T$, we found that the ratio method with the middle separation
($T\approx 1.2$~fm), as used in Ref.~\cite{Green:2012ej} was
inadequate; instead we will use the summation method. Finally, for
$g_S$ the large statistical uncertainty means that the source-sink
separation used in Ref.~\cite{Green:2012ej} with the ratio method was
larger than necessary, and here we will take the shortest separation
($T\approx 0.9$~fm) rather than the middle one. Some caution is also
required here, as excited-state effects can vary with the pion mass
and the choice of smearing parameters in the interpolating
operator. However, the earlier calculations are generally less
precise, which makes it more likely that excited-state effects are
small compared with the statistical uncertainty. The exception is the
fully-controlled study of finite-volume effects at
$m_\pi\approx 250$~MeV, which has a precision similar to this study;
however, it is expected that the contribution from excited states is
weakly dependent on the lattice volume in the range we
considered~\cite{Bar:2016uoj, Hansen:2016qoz}.

\begin{figure}
  \centering
  \includegraphics[width=0.495\textwidth]{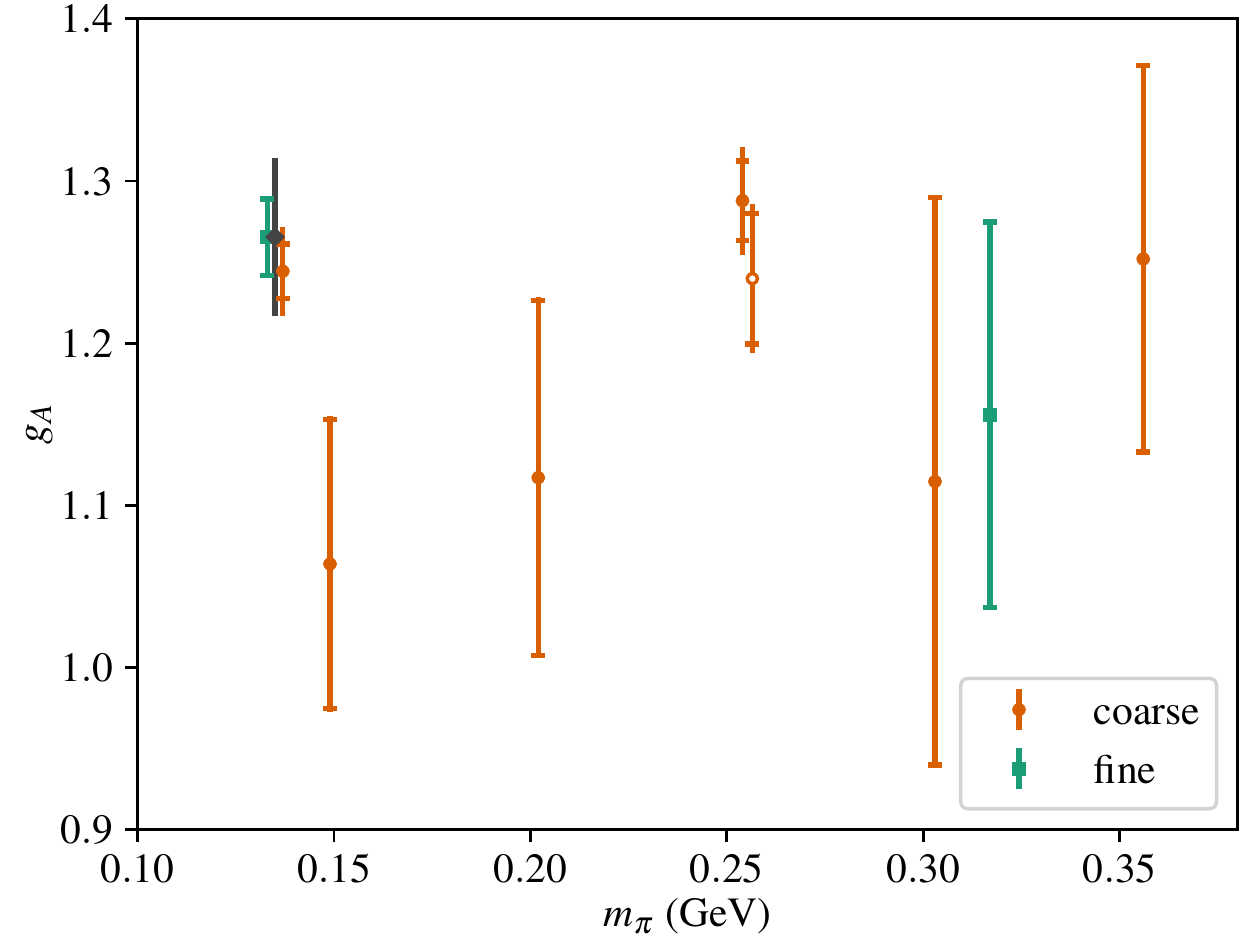}
  \includegraphics[width=0.495\textwidth]{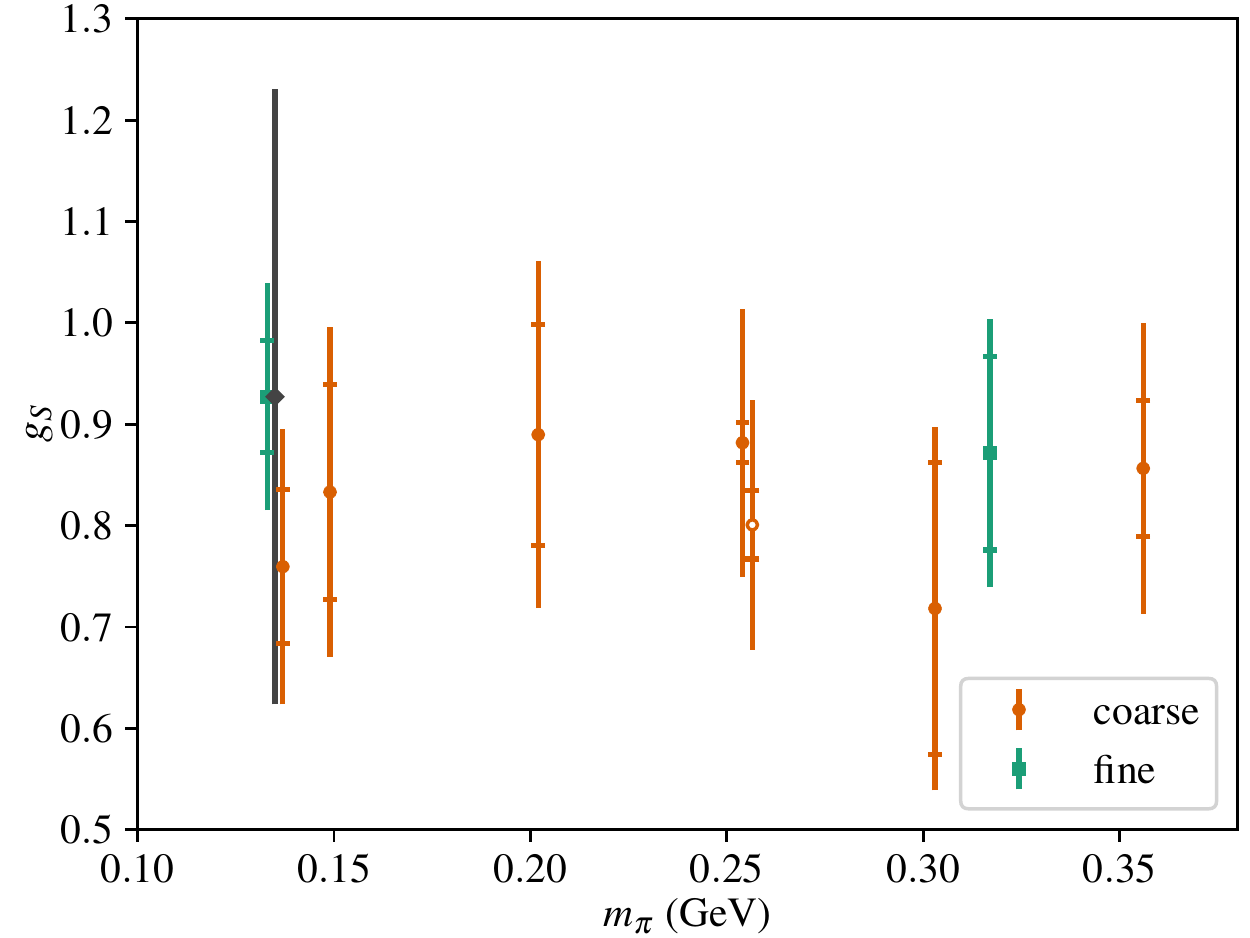}\\
  \includegraphics[width=0.495\textwidth]{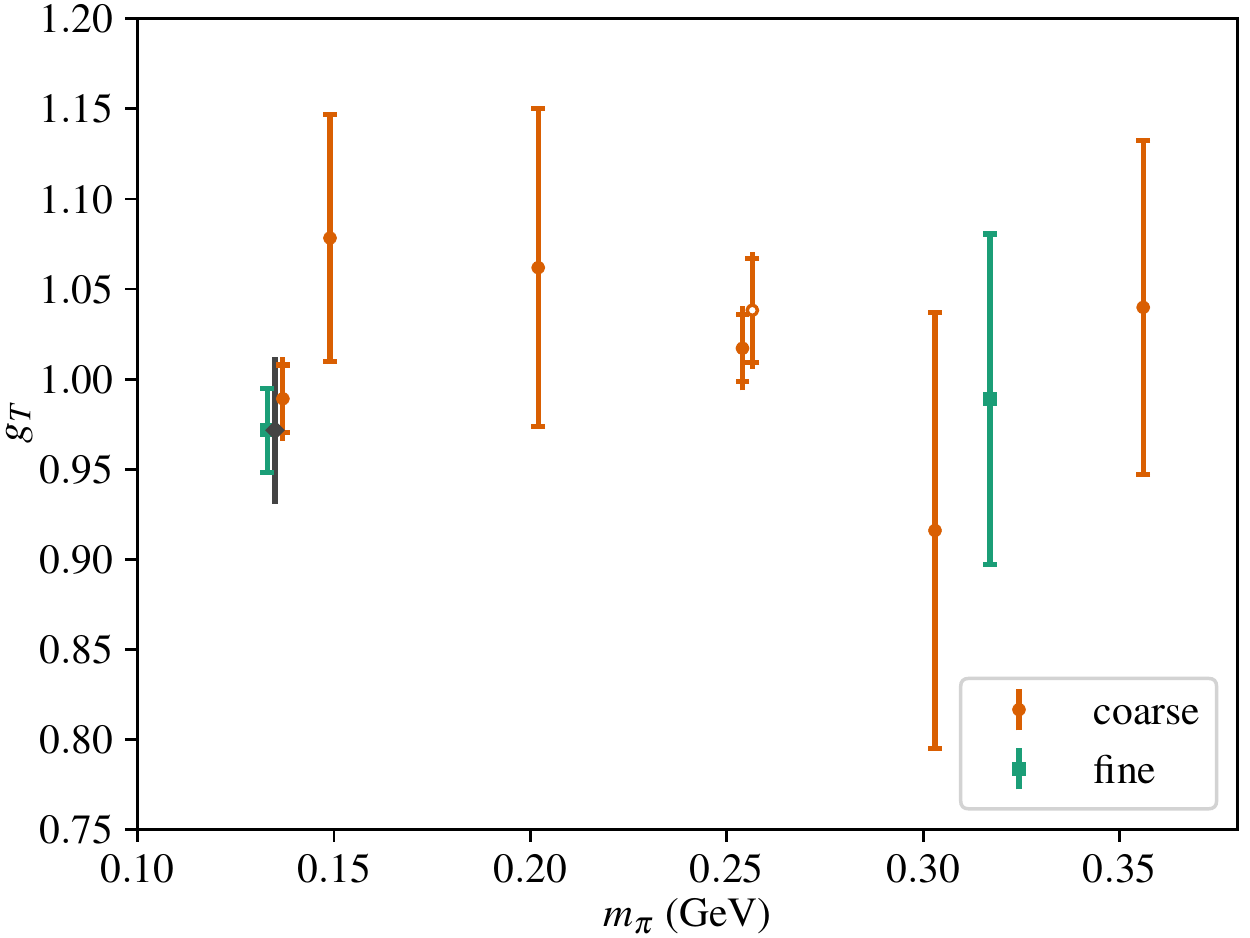}
  \caption{Isovector charges $g_A$, $g_S$, and $g_T$ versus pion
    mass. The inner error bars exclude the uncertainty on the
    renormalization factor, which is fully correlated across all
    ensembles with the same lattice spacing. The smaller of the two
    volumes at $m_\pi\approx 0.25$~GeV is displaced horizontally and
    indicated with an open symbol. The final
    estimates based on the two physical-point ensembles are indicated
    by the dark gray diamonds.}
  \label{fig:charges_vs_mpi}
\end{figure}

The comparison with our earlier results is shown in
Fig.~\ref{fig:charges_vs_mpi}. In these plots, the ensembles used for
a study of short time-extent effects are excluded and for two
ensembles at $m_\pi\approx 250$~MeV of size $32^3\times 48$ and
$24^3\times 48$, we have increased statistics.
The data show no significant dependence
on the pion mass, which justifies our neglect of this effect in the
final values of the charges. If we assume that finite-volume effects
scale as $m_\pi^2 e^{-m_\pi L}/\sqrt{m_\pi L}$ as for the axial charge
in chiral perturbation theory at large $m_\pi L$~\cite{Beane:2004rf},
then the finite-volume correction can be obtained by multiplying the
difference between the two volumes at $m_\pi\approx 250$~MeV by 0.28
and 0.23 for the coarse and fine physical-pion-mass ensembles,
respectively. One can see that this effect is also small compared
with the final uncertainties.

This comparison provides the opportunity to revisit our earlier result
for $g_A$~\cite{Green:2012ud}, which was unusually low. This was
partly caused by the lower value of $Z_A$, but the value obtained for
$m_\pi=149$~MeV is still two standard deviations below the
physical-point coarse ensemble. It appears that this is a statistical
fluctuation, since the methodology has not been significantly changed.

\newpage
\section{Summary and outlook}
\label{sec:summary}
We have computed the nucleon isovector axial, scalar, and tensor charges using two 2+1-flavor ensembles with a 2-HEX-smeared Wilson-clover action.
Both ensembles are at the physical pion mass and have lattice spacings of 0.116 and 0.093 fm.
We have demonstrated control over excited-state contamination by using eight source-sink separations in the range from roughly $0.4$ to $1.4$ fm on the coarse ensemble and three source-sink separations in the range from $0.9$ to $1.5$ fm on the fine ensemble.
The shorter source-sink separations are useful for the summation method but larger ones are needed for the ratio method. In addition, the choice of $T$ is observable-dependent: if excited-state effects are 
drowned out by noise, then shorter separations are more useful. 
We have studied a range of different fitting strategies to extract the different charges of the nucleon from ratios of correlation functions, namely
the ratio, two-state fit to the ratios, summation method, two-state fit to the summations (only on the coarse ensemble).
We have studied the stability of the different analysis methods and designed a procedure for combining the multiple estimates obtained for each observable and giving an estimate of its final value. We have observed consistency between the different analysis methods, although within larger error bars for the scalar charge.
We have determined the renormalization factors for the different observables using the nonperturbative Rome-Southampton approach and compared between the RI$'$-MOM and RI-SMOM intermediate schemes to estimate the systematic uncertainties.

Our final results are given in Eqs.~(\ref{eq:gA_final}--\ref{eq:gT_final}). The axial and tensor charges show overall uncertainties of roughly 4\%.
The obtained scalar charge, however, shows a much larger uncertainty, due to the large uncertainty in the renormalization factor and the large difference in the central values we observe between the the coarse and fine ensembles.
In this study, since both ensembles have pion masses very close to the physical pion mass and have large volumes, we neglect the pion-mass dependence and finite volume effects.
We have shown that this is justified when comparing our results to earlier calculations using the same lattice action and heavier pion masses. This calculation supersedes the earlier ones since it improves on them by working directly at the physical pion mass, using much higher statistics, and performing a more extensive study of excited-state effects.

\begin{figure*}
  \centering
  \includegraphics[width=\textwidth]{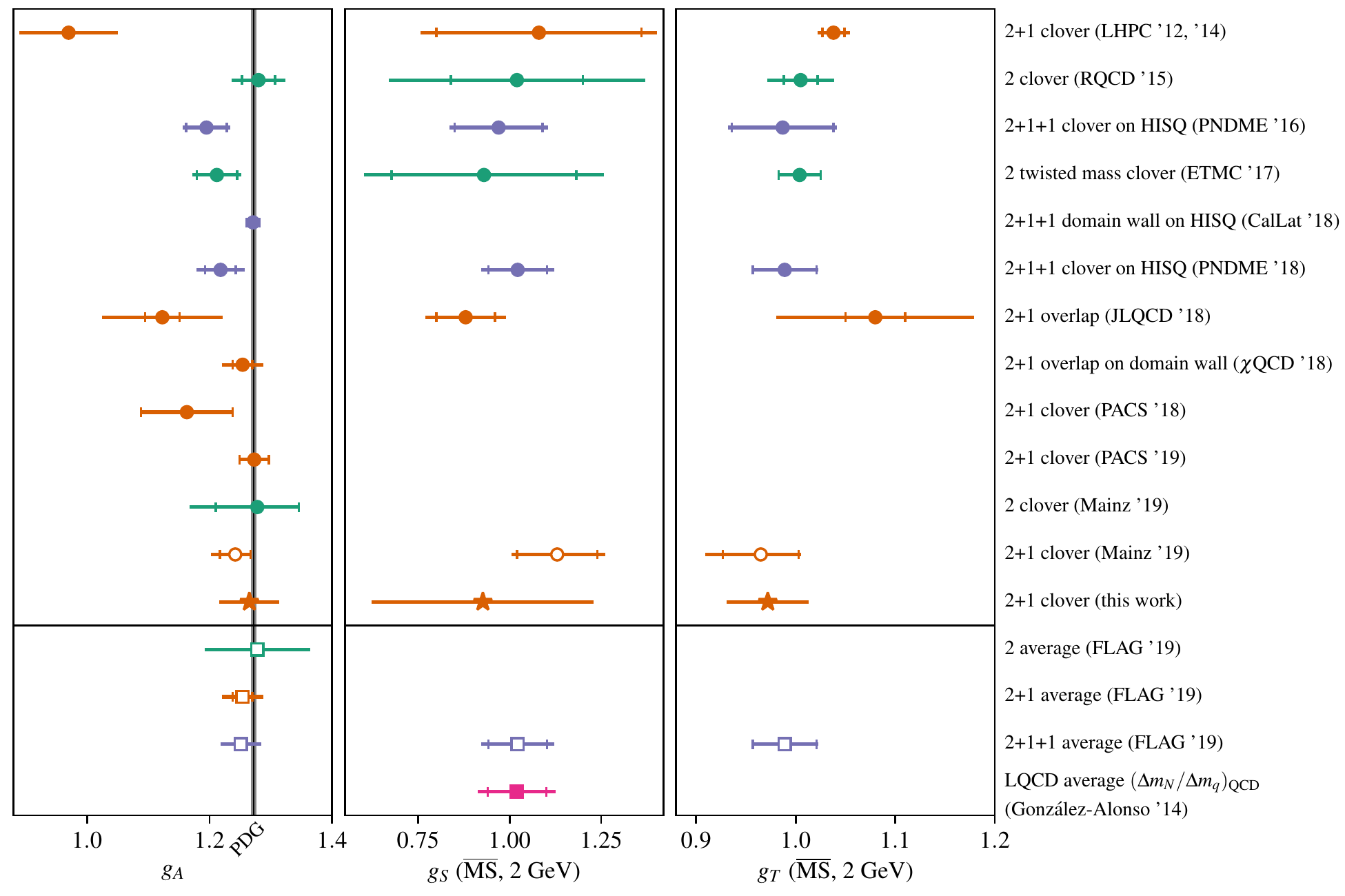}
  \caption{Recent lattice calculations of $g_A$, $g_S$, and
    $g_T$~\cite{Green:2012ej, Green:2012ud, Bali:2014nma,
      Bhattacharya:2016zcn, Alexandrou:2017qyt, Alexandrou:2017hac,
      Capitani:2017qpc, Yamanaka:2018uud, Chang:2018uxx,
      Liang:2018pis, Gupta:2018qil, Ishikawa:2018rew,
      Shintani:2018ozy, Harris:2019bih}. When separate statistical and
    systematic errors are quoted, the inner error bar (with endcap)
    indicates the statistical uncertainty and the outer one (without
    endcap) gives the quadrature sum. Open and filled symbols denote
    unpublished and published work. Green, orange, and blue denote
    calculations done with $2$, $2+1$, and $2+1+1$ dynamical quark
    flavors, which is also indicated in the legend. Circles are used
    for individual calculations and this work is indicated with
    stars. Squares are used for the averages from
    FLAG~\cite{Aoki:2019cca} and for the determination of $g_S$ using
    the conserved vector current relation and lattice QCD
    input~\cite{Gonzalez-Alonso:2013ura}. The vertical line with gray
    error band indicates the PDG value for
    $g_A$~\cite{PhysRevD.98.030001}.}
  \label{fig:global_results}
\end{figure*}

Recent lattice calculations of the isovector charges are summarized in
Fig.~\ref{fig:global_results}, although we caution that many of them
leave some sources of systematic uncertainty uncontrolled or
unestimated; see the FLAG review~\cite{Aoki:2019cca} for details. Our
results are consistent with most of these previous calculations and
also with the PDG value of $g_A$.

In our calculation, we have found a large discrepancy for $Z_S$ between the two intermediate renormalization schemes; it would be therefore useful to verify whether this goes away at finer lattice spacings, and to compare against other approaches such as the Schrödinger functional~\cite{Capitani:1998mq} or position-space~\cite{Gimenez:2004me} methods.

\acknowledgments
We thank the Budapest-Marseille-Wuppertal collaboration for making their configurations available to us.
Calculations for this project were done using the Qlua software
suite~\cite{Qlua}, and some of them made use of the QOPQDP adaptive
multigrid solver~\cite{Babich:2010qb,QOPQDP}. Fixing to Landau
gauge was done using the Fourier-accelerated conjugate gradient
algorithm~\cite{Hudspith:2014oja}.

This research used resources on the supercomputers
JUQUEEN~\cite{juqueen}, JURECA~\cite{jureca}, and JUWELS~\cite{juwels} at Jülich
Supercomputing Centre (JSC) and Hazel Hen at the High Performance
Computing Centre Stuttgart (HLRS). We acknowledge computing time
granted by the John von Neumann Institute for Computing (NIC) and by
the HLRS Steering Committee.

SM is supported by the U.S. Department of Energy (DOE), Office of Science, Office of High Energy Physics under Award Number DE-SC0009913. SM and SS are also supported by the RIKEN BNL Research Center under its joint tenure track fellowships with the University of Arizona and Stony Brook University, respectively. ME, JN, and AP are supported in part by the
Office of Nuclear Physics of the U.S. Department of Energy (DOE) under grants DE-FG02-96ER40965, DE-SC-0011090, and DE-FC02-06ER41444,respectively. SK and NH received support from Deutsche Forschungsgemeinschaft grant SFB-TRR 55.

\appendix
\section{Many-state fit}\label{app:Mstate}
In the two-state fit presented in Sec.~\ref{sec:2stateC2}, the obtained $E_1$ is much higher  
than the lowest expected $N\pi$ or $N\pi\pi$ state. 
For this reason, in addition to using the two-state fit model, we also implement a many-state model for the excited-state contributions.

Inspired by~\cite{Djukanovic:2016ocj}, 
the many-state fit models the contributions from the first few $N\pi$ noninteracting states with relative momentum $(\vec p\,)^2 < (\vec p_{\mathrm{max}})^2$.
The noninteracting levels in a finite cubic volume with periodic boundary conditions are determined by 
\begin{equation}
E_{\vec n} = \sqrt{\left(\frac{2\pi \vec n}{L}\right)^2 + m_\pi^2}  + \sqrt{\left(\frac{2\pi \vec n}{L}\right)^2 + m_N^2}\;,
\end{equation}
where $L$ denotes the spatial extent of the lattice and $\vec n$ is a three-vector of integers. We are interested in states with quantum numbers equal to that of a proton i.e.\ $I(J^P) = 1/2({1/2}^+)$. However, the state of a pion and nucleon both at rest does not contribute since its parity is opposite that of the nucleon.
The shift between free and interacting energy levels is small relative to the gap to the single nucleon state, as shown in~\cite{Hansen:2016qoz}. This justifies the use of noninteracting finite-volume spectrum for the values of $\Delta E_{\vec n} = E_{\vec n} - m_N$.

The obvious difficulty in performing such a fit comes from the many fit parameters needed to parametrize the matrix elements and the overlap of the nucleon interpolating operator onto each of the $N\pi$ states. In order to reduce the number of fit parameters involved in the many-state fit, we assume that the coefficient for ground-to-excited transitions is the same for all states, and the
off-diagonal transition matrix elements between different excited states are small
but that excited states in the two-point function in the denominator of the ratio are important. This yields a formula like the following\footnote{We thank Oliver Bär for pointing out that the ChPT prediction includes a factor of $\vec p^2/(m_\pi^2+\vec p^2)$ in the terms proportional to $b_X$ and $c_X$. However, for nonzero momenta in our lattice volumes this factor lies in the range $[0.7,1.0]$ and including it does not change the qualitative behavior of these fits.}
\begin{equation}\label{eq:Mstate}
R^X(\tau,T) = g_X^{\mathrm{bare}} + b_X \sum_{\substack{\vec n\neq 0 \\ |\vec n|^2\leq  | \vec n_\mathrm{max}|^2}} \left( e^{-\Delta E_{\vec n} \tau} + e^{-\Delta E_{\vec n}
(T-\tau)} \right) + c_X \sum_{\substack{\vec n\neq 0 \\ |\vec n|^2\leq |\vec n_\mathrm{max}|^2}} e^{-\Delta E_{\vec n} T}.
\end{equation}
where the three parameters are $g_X^\mathrm{bare}$, $b_X$, and $c_X$. Also, the above $\vec n$ fulfills $|\vec n|^2 \leq |\vec n_{\mathrm{max}}|^2$ and we exclude the state where the nucleon and pion are at rest, $\vec n=0$.
\begin{figure}
\begin{center}
\includegraphics[width=0.4\textwidth]{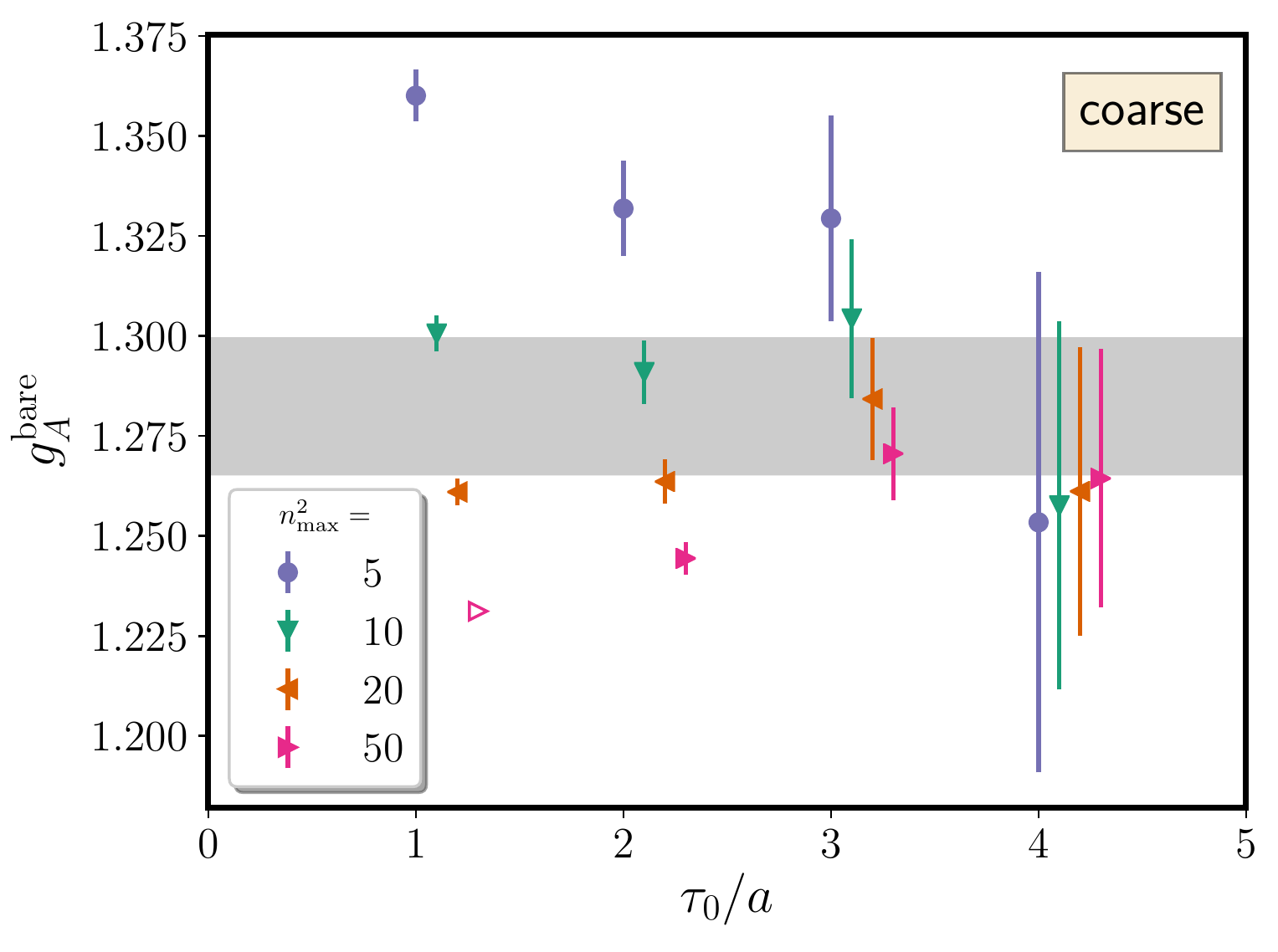}
\hspace{0.001\textwidth}
\includegraphics[width=0.4\textwidth]{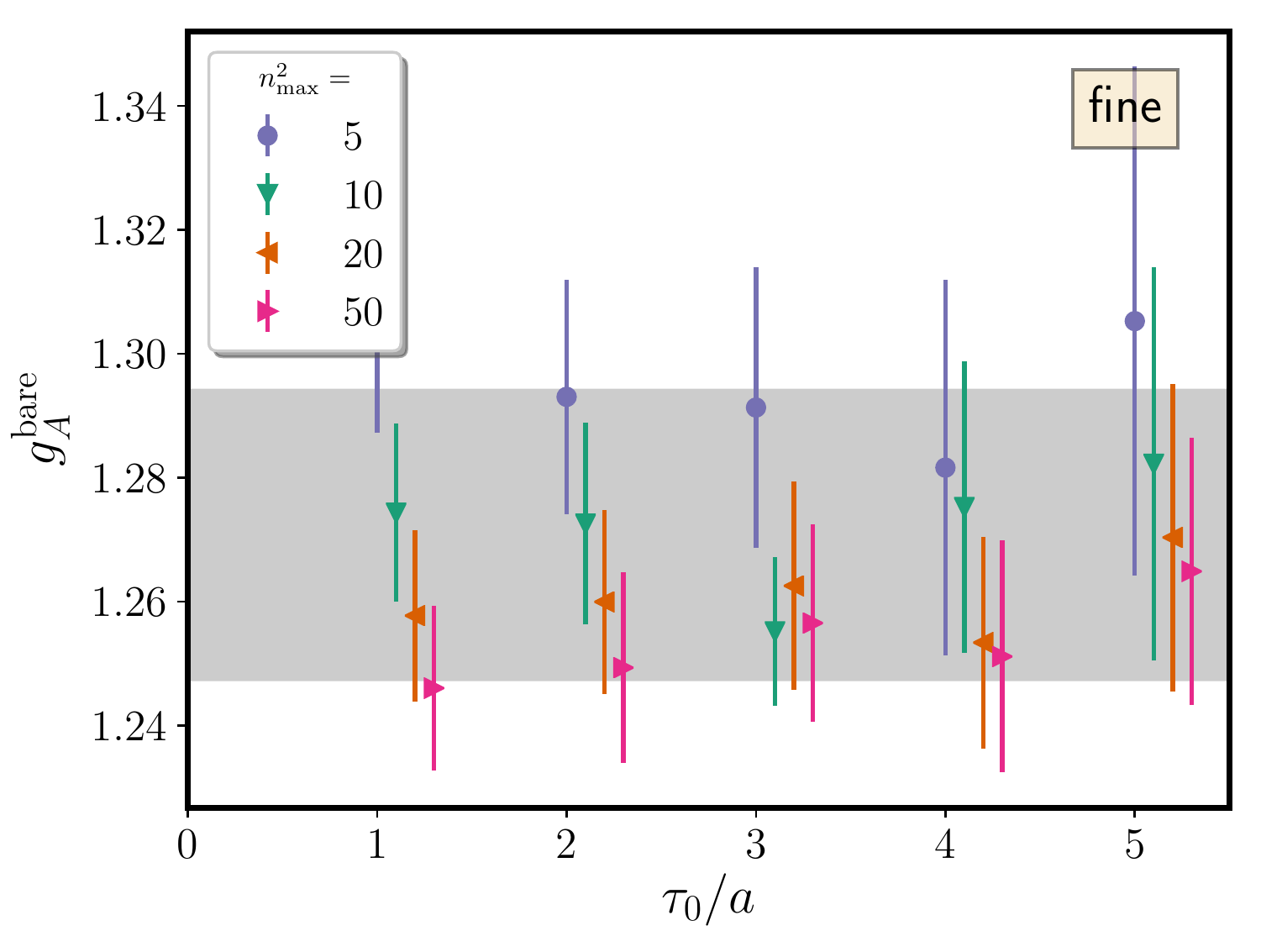}
\\
\includegraphics[width=0.4\textwidth]{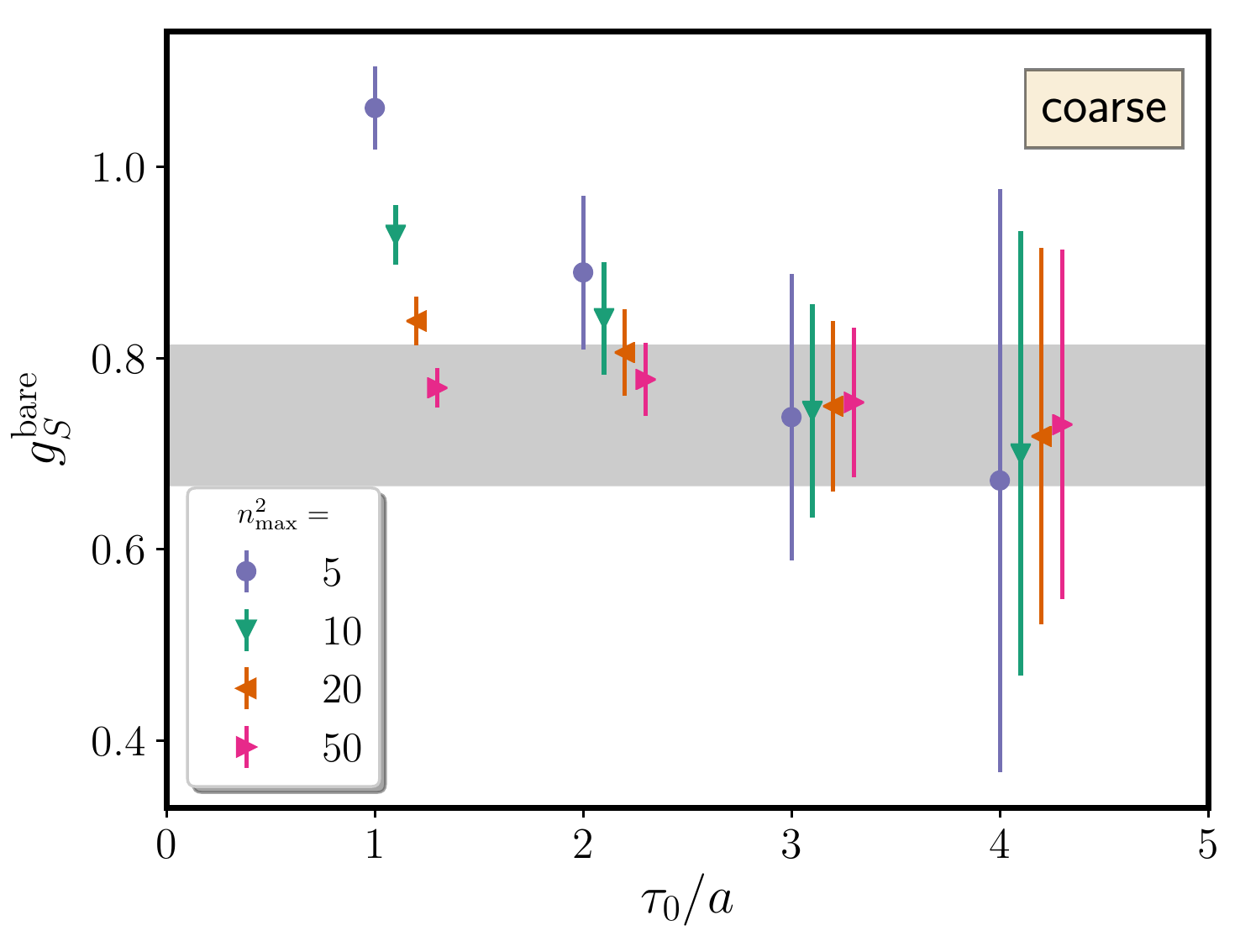}
\hspace{0.001\textwidth}
\includegraphics[width=0.4\textwidth]{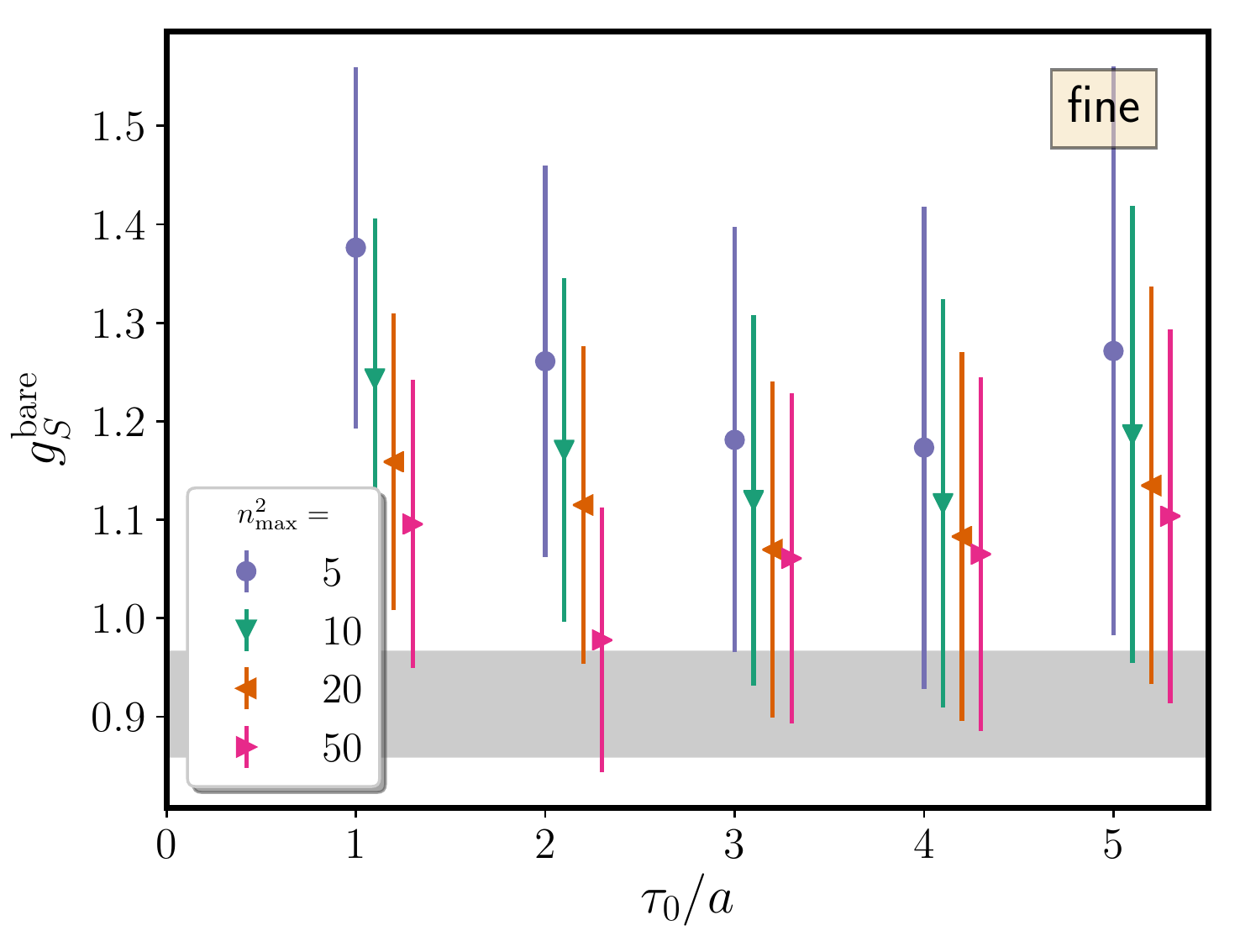}
\\
\includegraphics[width=0.4\textwidth]{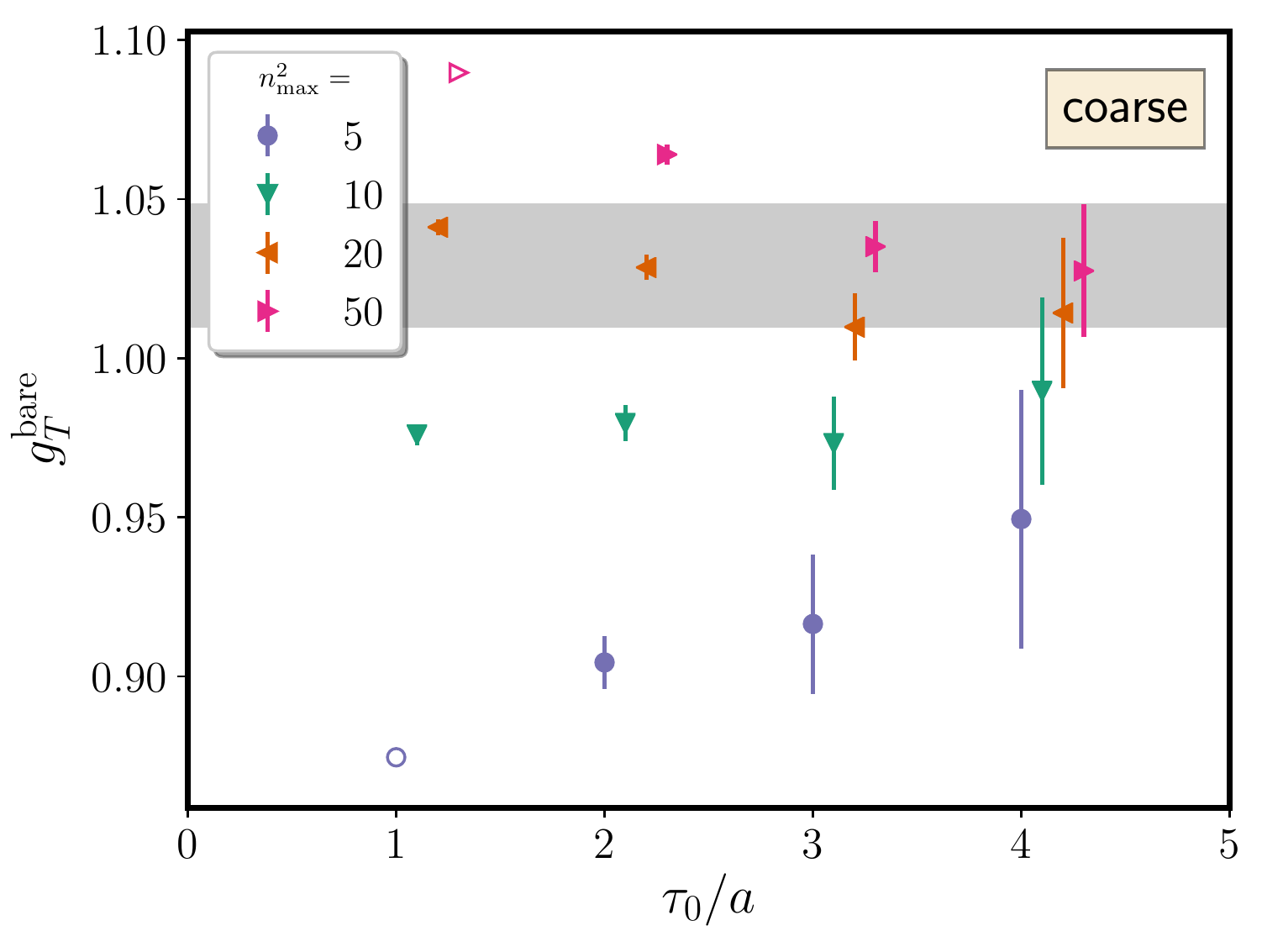}
\hspace{0.001\textwidth}
\includegraphics[width=0.4\textwidth]{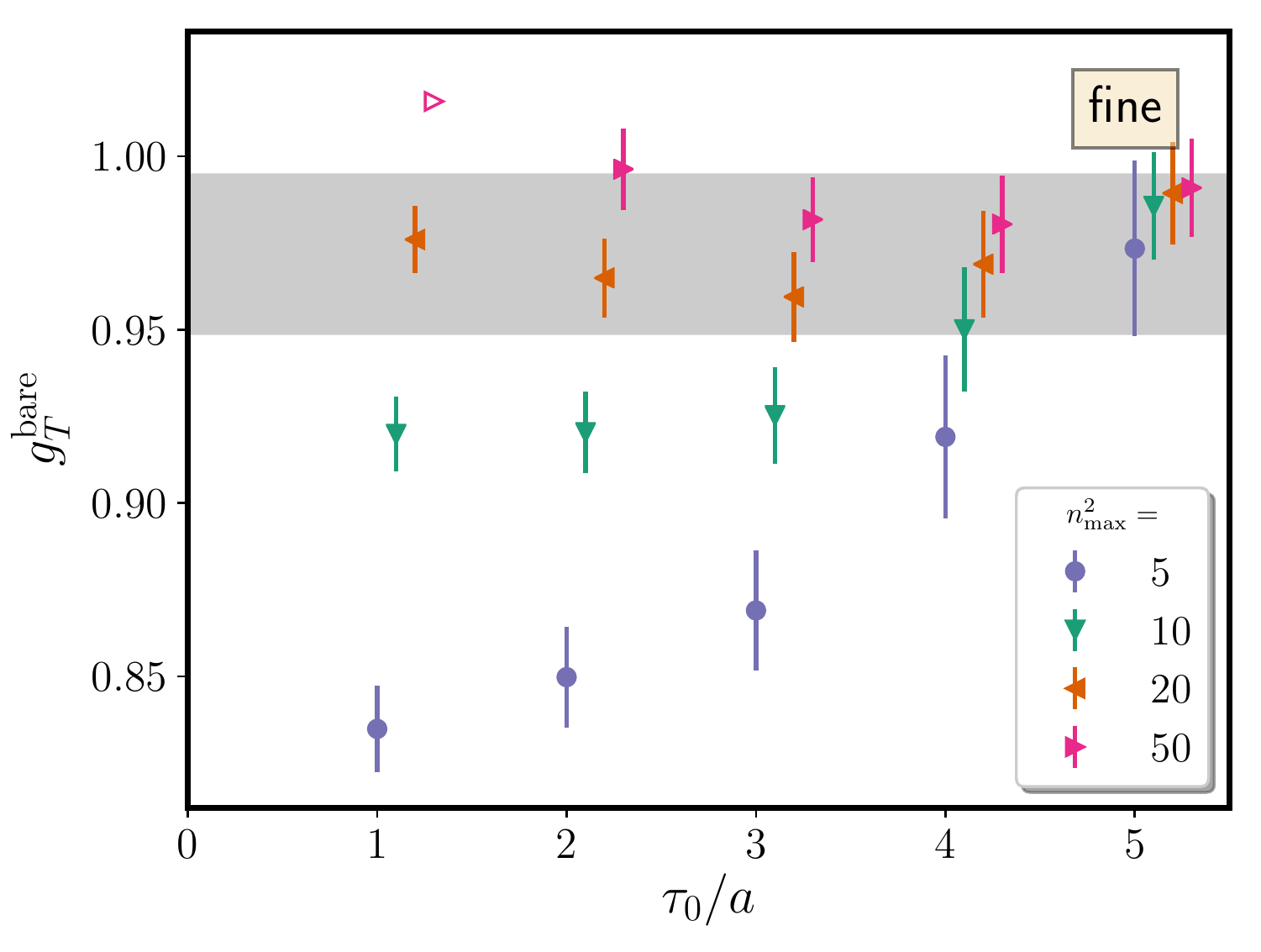}
\end{center}
  \caption{The bare axial, scalar, and tensor charges obtained using the many-state fit to $R^X(\tau,T)$ for the coarse and fine ensembles. The open symbols refer to fits with $p$-value $<0.02$. The gray bands denote the final estimates of the charges from Tab.~\ref{tab:summary_final}.}
  \label{fig:Mstate_summary}
\end{figure}
We perform the many-state fit using four different values of $n_\mathrm{max}^2$, $n_\mathrm{max}^2 \in \{5,10,20,50\}$.

Figure~\ref{fig:Mstate_summary} shows a summary of the estimated unrenormalized isovector axial, scalar, and tensor charges as functions of $\tau_0$ using this approach applied on both the coarse (left column) and fine (right column) ensembles.
We notice that the estimated charges at short $\tau_0$ depend significantly on $n_\mathrm{max}^2$ and that increasing $n_\mathrm{max}^2$ results in decreasing the statistical uncertainties of the estimated charges. In addition, Fig~\ref{fig:Mstate_summary} shows that the obtained charges for different $n_\mathrm{max}^2$ values tend to be consistent at the largest $\tau_0$. 
When comparing to our final estimates of the bare charges in Tab.~\ref{tab:summary_final} (gray bands), we notice that the estimates from the many-state fit approach are consistent within error bars and that the many-state fit leads to larger statistical uncertainties for $g_A^\mathrm{bare}$ and $g_S^\mathrm{bare}$ compared to other analysis methods.
The strong dependence on $n_\mathrm{max}^2$ at short $\tau_0$ suggests that this method may be relatively unreliable and that a more sophisticated model such as the one in Ref.~\cite{Hansen:2016qoz} is needed to extend into the resonance regime.

\section{Table of bare charges}\label{app:table}

Table~\ref{tab:bare_charges} contains the bare charges used (after
renormalizing) in Fig.~\ref{fig:charges_vs_mpi}, based on this work
(physical-point ensembles), data from previous
publications~\cite{Green:2012ej,Green:2012ud}, and increased
statistics at $m_\pi\approx 250$~MeV.

\begin{table}
  \centering
  \begin{tabular}{cccc|cc|lll}
    $\beta$ & Size & $am_{ud}$ & $am_s$ & $m_\pi$ [MeV] & $m_\pi L$ & \multicolumn{1}{c}{$g_A^\text{bare}$} & \multicolumn{1}{c}{$g_S^\text{bare}$} & \multicolumn{1}{c}{$g_T^\text{bare}$} \\\hline
3.31 & $48^3\times 48$ & $-0.09933$ & $-0.04$ & 137(2) & 3.9 & 1.282(17) & 0.740(74) & 1.029(20) \\
3.5  & $64^3\times 64$ & $-0.05294$ & $-0.006$& 133(1) & 4.0 & 1.271(24) & 0.913(54) & 0.972(23) \\\hline
3.31 & $48^3\times 48$ & $-0.09900$ & $-0.04$ & 149(1) & 4.2 & 1.096(92) & 0.812(103)& 1.122(71) \\
3.31 & $32^3\times 48$ & $-0.09756$ & $-0.04$ & 202(1) & 3.8 & 1.151(113)& 0.867(106)& 1.105(92) \\
3.31 & $32^3\times 48$ & $-0.09530$ & $-0.04$ & 254(1) & 4.8 & 1.327(25) & 0.859(19) & 1.058(19) \\
3.31 & $24^3\times 48$ & $-0.09530$ & $-0.04$ & 254(1) & 3.6 & 1.278(42) & 0.780(33) & 1.080(30) \\
3.31 & $24^3\times 48$ & $-0.09300$ & $-0.04$ & 303(2) & 4.3 & 1.149(180)& 0.700(141)& 0.953(126)\\
3.31 & $24^3\times 48$ & $-0.09000$ & $-0.04$ & 356(2) & 5.0 & 1.290(123)& 0.835(65) & 1.082(96) \\
3.5  & $32^3\times 64$ & $-0.04630$ & $-0.006$& 317(2) & 4.8 & 1.161(119)& 0.858(94) & 0.989(92)
  \end{tabular}
  \caption{Bare charges used in Fig.~\ref{fig:charges_vs_mpi}.}
  \label{tab:bare_charges}
\end{table}

\bibliography{gAST}
\bibliographystyle{utphys-noitalics}

\end{document}